\documentclass[aip,cha,reprint,groupedaddress,nofootinbib, superscriptaddress]{revtex4-1}

\usepackage[utf8]{inputenc}

\usepackage{graphicx,amsmath,amssymb}

\usepackage{longtable}
\usepackage{booktabs}
\usepackage{soul}
\usepackage{tabularx}
\usepackage{ltablex}
\usepackage{xcolor}
\usepackage{braket}
\usepackage{footnote}
\usepackage{xr}
\usepackage{prettyref}
\usepackage{url}

\newcommand{\msm}[1]{\textcolor{blue}{msm says: #1}}
\newcommand{\mm}[1]{\textcolor{green}{mm says: #1}}
\newcommand{\hao}[1]{\textcolor{magenta}{hao says: #1}}
\newcommand{\minyang}[1]{\textcolor{purple}{minyang says: #1}}

\newcommand{\vek}[1]{\boldsymbol{#1}}
\newcommand{\avg}[1]{\langle #1\rangle}
\newcommand{\abs}[1]{\lvert #1\rvert}
\newcommand{\mat}[1]{\mathsf{#1}}
\newcommand{\TT}{^{\intercal}}
\newcommand{\dd}{\mathrm{d}}
\newcommand{\ee}{\mathrm{e}}
\newcommand{\ie}{\emph{i.e.}}
\newcommand{\eg}{\emph{e.g.}}

\usepackage{lineno}

\setcitestyle{square,comma,numbers,sort&compress}

\begin{document}


\title{Ranking in evolving complex networks}



\author{Hao Liao}
\affiliation{Guangdong Province Key Laboratory of Popular High Performance Computers, College of Computer Science and Software Engineering, Shenzhen University, Shenzhen 518060, PR China}

\author{Manuel Sebastian Mariani}
\email[]{manuel.mariani@unifr.ch}
\thanks{Corresponding author}
\affiliation{Department of Physics, University of Fribourg, 1700 Fribourg, Switzerland}
\affiliation{Guangdong Province Key Laboratory of Popular High Performance 
Computers, College of Computer Science and Software Engineering, Shenzhen University, Shenzhen 518060, PR China}

\author{Mat{\'u}{\v{s}} Medo}
\email[]{matus.medo@unifr.ch}
\thanks{Corresponding author}
\affiliation{Institute of Fundamental and Frontier Sciences,
University of Electronic Science and Technology of China, Chengdu 610054, PR China}
\affiliation{Department of Physics, University of Fribourg, 1700 Fribourg, Switzerland}
\affiliation{Department of Radiation Oncology, Inselspital, Bern University Hospital and University of Bern, 3010 Bern, Switzerland}

\author{Yi-Cheng Zhang}
\affiliation{Department of Physics, University of Fribourg, 1700 Fribourg, Switzerland}

\author{Ming-Yang Zhou}
\affiliation{Guangdong Province Key Laboratory of Popular High Performance Computers, College of Computer Science and Software Engineering, Shenzhen University, Shenzhen 518060, PR China}

\begin{abstract}
Complex networks have emerged as a simple yet powerful framework to represent and analyze a wide range of complex systems. 
The problem of ranking the nodes and the edges in complex networks is critical for a broad range of 
real-world problems because it affects how we access online information and products, how success and 
talent are evaluated in human activities, and how scarce resources are allocated by companies and policymakers, among others. 
This calls for a deep understanding of how existing ranking algorithms perform, and which are their possible biases that may impair their effectiveness.
Well-established ranking algorithms (such as the popular Google's PageRank) are static in nature and, as a consequence, 
they exhibit important shortcomings when applied to real networks that rapidly evolve in time. 
The recent advances in the understanding and modeling of evolving networks have enabled the development 
of a wide and diverse range of ranking algorithms that take the temporal dimension into account.
The aim of this review is to survey the existing ranking algorithms, both static and time-aware, and their applications to evolving networks.
We emphasize both the impact of network evolution on well-established static algorithms and the benefits from including 
the temporal dimension for tasks such as prediction of real network traffic, prediction of future links, and identification of highly-significant nodes. 
\end{abstract}


\maketitle

\tableofcontents

\section{Introduction}


In a world where we need to make many choices and our attention is inherently limited,
we often rely on, or at least use the support of, automated scoring and ranking algorithms to orient ourselves in the copious amount of available information.
One of the major goals of ranking algorithms is to \emph{filter} \cite{hanani2001information} the large amounts of available data in order to \emph{retrieve} \cite{baeza1999modern} the most relevant information for our purposes. 
For example, quantitative metrics for assessing the relevance of websites are motivated by the impossibility, for an individual web user, to browse manually large portions of the Web to find the information that he or she needs. 

The resulting rankings influence our choices of items to purchase \cite{chen2004impact,fleder2009blockbuster,zeng2015modeling},
the way we access scientific knowledge \cite{feenberg2016s} and online content \cite{cho2004impact,fortunato2006topical, pan2007google, murphy2006primacy,zhou2010impact}.
At a systemic level, ranking affects how success and excellence in human activity are recognized \cite{hirsch2005index,radicchi2009diffusion,radicchi2011best,spitz2014measuring,wasserman2015cross,waltman2016review}, how funding is allocated in scientific research \cite{wilsdon2016metric}, how to characterize and counteract the outbreak of an infectious disease \cite{brockmann2013hidden,iannelli2016effective, wang2016statistical}, and how undecided citizens vote for political elections \cite{epstein2015search}, for example. 
Given the strong influence of rankings on various aspects of our lives, achieving a deep understanding
of how respective ranking methods work, what are their basic assumptions and limitations, and designing
improved methods is of critical importance.
Rankings' disadvantages, such as the biases that they can induce, can otherwise outweigh the benefits that they bring.
Achieving such understanding becomes more urgent as a huge and ever-growing amount of data is created and 
stored every day and, as a result, automated methods to extract relevant information are poised to further gain importance in the future.


An important class of ranking algorithms is based on the \emph{network representation} of the input data.
Complex networks have recently emerged as one of the leading frameworks to analyze a wide range 
of complex social, economic and information systems~\cite{boccaletti2006complex, newman2010networks, jackson2010social, barabasi2016network}. 
A network is composed of a set of \emph{nodes} (also referred to as \emph{vertexes}) and a set of
relationships (\emph{edges}, or \emph{links}) between the nodes.
A network representation of a given system greatly reduces the complexity of the original system and,
if the representation is well-designed, it allows us to gain fundamental insights into the structure
and the dynamics of the original system. Fruitful applications of the complex networks approach
include prediction of spreading of infectious diseases \cite{balcan2009seasonal, brockmann2013hidden}, characterization 
and prediction of economic development of countries \cite{hidalgo2009building, tacchella2012new,cristelli2015heterogeneous}, 
characterization and detection of early signs of financial crises \cite{battiston2012debtrank,kenett2012evolvement, battiston2016complexity}, and 
design of robust infrastructure and technological networks \cite{schneider2011mitigation, reis2014avoiding}, among many others. 

\emph{Network-based ranking algorithm}s use network representations of the data to compute
the inherent value, relevance, or importance of individual nodes in the system \cite{duhan2009page,medo2013network}.
The idea of using the network structure to infer node status is long known in social sciences \cite{katz1953new,bonacich1987power}, 
where node-level ranking algorithms are typically referred to as \emph{centrality metrics} \cite{borgatti1995centrality}; in
this review, we use the labels ``ranking algorithms'' and ``centrality metrics'' interchangeably.
Today, research on ranking methods is being carried out by scholars from several fields (such
as physicists, mathematicians, social scientists, economists, computer scientists). Network-based ranking 
algorithms have drawn conspicuous attention from the physics community due to the direct applicability
of statistical-physics concepts such as
random walk and diffusion
\cite{brin1998anatomy, rosvall2014memory, rocha2014random, masuda2016random, zhang2016dynamics}, as well as
percolation \cite{piraveenan2013percolation, morone2015influence,radicchi2016leveraging}.
Real-world applications of network-based ranking algorithms cover a vast spectrum of problems, including the design of 
search engines \cite{brin1998anatomy, duhan2009page,langville2011google} and recommender systems \cite{zhang2010personalized, zhang2007heat,zhou2010solving}, evaluation of academic research \cite{chen2007finding, radicchi2009diffusion, ding2009pagerank}, 
and identification of the influential spreaders \cite{ren2014iterative, pei2014searching, morone2015influence,radicchi2016leveraging}. 
A paradigmatic example of a network-based ranking algorithm is
the Google's popular PageRank algorithm. While originally devised to rank web pages \cite{brin1998anatomy}, it has since then
found applications in a vast range of real systems \cite{franceschet2011pagerank, gleich2015pagerank, ermann2015google,
jiang2008self, mao2015quantifying, walker2007ranking, ding2009pagerank,
bollen2006journal, jing2008visualrank, ivan2011web, radicchi2011best}.



While there are excellent reviews and surveys that address network-based ranking 
algorithms \cite{duhan2009page, lu2012recommender, medo2013network, ermann2015google, gleich2015pagerank} and random walks 
in complex networks \cite{masuda2016random}, the impact of network evolution on ranking algorithms is discussed there only marginally.
This gap deserves to be filled because real networks usually evolve
in time \cite{barabasi1999emergence,bianconi2001competition, dorogovtsev2002evolution,albert2002statistical,medo2011temporal, papadopoulos2012popularity,wang2013quantifying}, and temporal effects have been shown to strongly influence properties and effectiveness 
of ranking algorithms \cite{newman2009first,rosvall2014memory, perra2012random, scholtes2014causality,lambiotte2015effect,delvenne2015diffusion,mariani2015ranking,vidmer2016role}.


The aim of this review is to fill this gap and thoroughly present the temporal aspects of network-based 
ranking algorithms in social, economic and information systems that evolve in time.
The motivation for this work is multifold. First, we want to present in a cohesive fashion information that is scattered across literature from diverse fields, providing scholars from diverse communities with a reference point on the temporal aspects of network-based ranking. Second, we want to highlight the essential role of the temporal dimension, whose omission can produce misleading results when it comes to evaluating the centrality of a node.
Third, we want to highlight open challenges and suggest new research directions.
Particular attention will be devoted to network-based ranking algorithms that build on classical statistical physics concepts such as random walks and diffusion. 


In this review, we will survey both \emph{static} and \emph{time-aware} network-based ranking algorithms.
\emph{Static} algorithms take into account the only network's adjacency matrix $\mat{A}$ (see paragraph \ref{sec:basic}) to compute node centrality, and they constitute the topic of Section \ref{sec:static}. The dynamic nature of most of the real networks can cause static algorithms to fail in a number of real-world systems, as outlined in Section \ref{sec:failure}.
\emph{Time-aware} algorithms take as input the network's adjacency matrix at a given time $t$ and/or the list of node and/or edge time-stamps. Time-aware algorithms for growing unweighted networks are be presented in Section \ref{sec:time}. 
When multiple interactions occur among the network's nodes, algorithms based on \emph{temporal-network representations} of the data are often needed; this class of ranking algorithms is presented in Section \ref{sec:temporal} alongside a brief introduction to the temporal network representation of time-stamped datasets.
Network-based time-aware recommendation methods and their applications are covered by Section \ref{sec:recommender}; the impact of recommender systems on network evolution is also discussed in this section.

Due to fundamental differences between the datasets where network centrality metrics are applied, no universal and all-encompassing ranking algorithm can exist.
The best that we can do is to identify the specific ranking task (or tasks) that a metric is able to address.
As it would be impossible to cover all the applications of network centrality metrics studied in the literature, we narrow our focus to applications where the temporal dimension plays an essential role (section \ref{sec:applications}). For example, we discuss how time-aware metrics can be used to early identify significant nodes (paragraph \ref{sec:academic}), how node importance can be used as a predictor of future GDP of countries (paragraph \ref{prediction_gdp}), and how time-aware metrics outperform static ones in the classical link-prediction problem (paragraph \ref{sec:link_prediction}). 
We provide a general discussion on the problem of ranking in Section \ref{sec:perspectives}. In this discussion, we focus on possible ways of validating the existing centrality metrics, and on the importance of detecting and suppressing the biases of the rankings obtained with various metrics.

The basic notation used in this review is provided in paragraph \ref{sec:basic}. 
While this review covers a broad range of interwoven topics, the
chapters of this review are as self-contained as possible, and in principle, they can be read independently.
We do discuss the underlying mathematical details only occasionally and refer to the corresponding references instead where the interested reader can find more details.

\section{Static centrality metrics}
\label{sec:static}
Centrality metrics aim at identifying the most important, or central, nodes in the network.
In this section, we review the centrality metrics that only take into account the topology of the network, encoded in the network's adjacency matrix $\mat{A}$. These metrics neglect the temporal dimension and are thus referred to as \emph{static} metrics in the following.
Static centrality metrics have a long-standing tradition in social network analysis -- suffice it to say that
Katz centrality (paragraph \ref{sec:eigenvector}) was introduced in the 50s \cite{bonacich1987power}, 
betweeness centrality (paragraph \ref{sec:shortestpath}) in the 70s \cite{freeman1977set},
and indegree (citation count, paragraph \ref{sec:degree}) is used to rank scholarly publications since the 70s.
The crucial role played by the PageRank algorithm
in the outstanding success of the web search engine Google was one of the main motivating factors for the new wave of interest in centrality metrics starting in the late 90s.
Today, we are aware that static centrality metrics may exhibit severe shortcomings when applied to dynamically evolving systems
-- as will be examined in depth in the next sections -- and, for this reason, they should never be applied without 
considering the specific properties of the system in hand.
Static metrics are nevertheless important both as basic heuristic tools of network analysis,
and as building blocks for time-aware metrics.

\subsection{Getting started: basic language of complex networks}
\label{sec:basic}
Since excellent books \cite{caldarelli2007scale, jackson2010social,
newman2010networks,barabasi2016network} and reviews \cite{dorogovtsev2002evolution,albert2002statistical,newman2003structure,boccaletti2006complex} on complex networks and their applications have already been written, in the following we only introduce the network-analysis concepts and tools indispensable for the presented ranking methods.
We represent a network (usually referred to as graph in mathematics literature) 
composed by $N$ nodes and $E$ edges by the symbol $\mathcal{G}(N,E)$.
The network is completely described by its \emph{adjacency matrix} $\mat{A}$.
In the case of an \emph{undirected} network, matrix element $A_{ij}=1$ if an edge exists that connects $i$ and $j$, zero otherwise;
the \emph{degree} $k_{i}$ of node $i$ is defined as the number of connections of node $i$.
In the case of a \emph{directed} network, $A_{ij}=1$ if an edge exists that connects $i$ and $j$, zero otherwise.
The \emph{outdegree} $k^{out}_{i}$ (\emph{indegree} $k^{in}_i$) of node $i$ is defined as the number of outgoing (incoming) edges
that connect node $i$ to other nodes.

We introduce now four structural notions that will be useful in the following:
\emph{path, path length, shortest path}, and \emph{geodesic distance between two nodes}.
The definitions provided below apply to undirected networks, but they can be straightforwardly generalized to directed networks (in that case,
we speak of directed paths). 
A network path is defined as a sequence of nodes such that consecutive nodes in the path are connected by an edge.
The \emph{length of a path} is defined as the number of edges that compose the path.
The \emph{shortest path} between two nodes $i$ and $j$ is the path of the smallest possible lenght that start in node $i$ and end up in $j$
-- note that multiple distinct shortest paths between two given nodes may exist.
The \emph{(geodesic) distance} between two nodes $i$ and $j$ is the length of the shortest paths between $i$ and $j$.
We will use these basic definitions in paragraph \ref{sec:shortestpath} to define the closeness and betweeness centrality.

\subsection{Degree and other local centrality metrics}
\label{sec:degree}

We start by centrality metrics that are \emph{local} in nature.
By local metrics, we mean that only the immediate neighborhood of a given node is taken into account for the calculation
of the node's centrality score, as opposed to, for example, eigenvector-based metrics that consider all network's paths
(paragraph \ref{sec:eigenvector}) and metrics that consider all network's shortest paths that pass through a given node to determine its score (paragraph \ref{sec:shortestpath}).

\subsubsection{Degree} The simplest structural centrality metric is arguably degree. For an undirected network, the degree of a node $i$ is defined as
\begin{equation}
k_i=\sum_{j}A_{ij}.
\end{equation}
That is, the degree of a node is given by the number of edges attached to it.
Using degree as a centrality metric assumes that a node is important if it has received many connections.
Even though degree neglects nodes' positions in the network, its simple basic assumption may be reasonable enough for some systems.
Even though the importance of the neighbors of a given node is presumably relevant to determine its centrality \cite{katz1953new,franceschet2011pagerank}, and this aspect is ignored by degree, one can argue that node degree still captures some part of node importance.
In particular, degree is highly correlated with more sophisticated centrality metrics for uncorrelated networks \cite{fortunato2006approximating, pastor2016topological}, and can perform better than more sophisticated network-based metrics in specific tasks such as identifying influential spreaders \cite{klemm2012measure,lu2016vital}.

In the case of a directed network, one can define node degree either by counting the number of incoming edges or the number of outgoing edges -- the outcomes of the two different counting procedures are referred to as node indegree and outdegree, respectively.
In most cases, incoming edges can be interpreted as a recommendation, or a positive endorsement, received from the node from which the edge comes from.
While the assumption that edges represent positive endorsements is not always true\footnote{In a signed social network, for example, an edge between two individuals can be either positive (friendship) or negative (animosity). See \cite{kunegis2009slashdot}, among others, for generalizations of popular centrality metrics to signed networks.}, it is reasonable enough for a broad range of real systems and, unless otherwise stated, we assume it throughout this review.
For example, in networks of academic publications, a paper's incoming edges represent its received citations, and their number can be considered
as a rough proxy for the paper's impact\footnote{This assumption neglects the multi-faceted nature of the scientific citation process \cite{bornmann2008citation}, and the fact that citations can be both positive and negative -- a citation can be considered as "negative" if the citing paper points out flaws of the cited paper, for example \cite{catalini2015incidence}. We describe in the following sections some of the shortcomings of citation count and possible ways to counteract them.}.
For this reason, indegree is often used as a centrality metric in directed networks
\begin{equation}
k^{in}_i=\sum_{j}A_{ij}.
\end{equation} 
Indegree centrality is especially relevant in the context of quantitative evaluation of academic research, where 
the number of received citations is often used to gauge scientific impact (see \cite{waltman2016review} for a review of bibliometric indices based on citation count).

\subsubsection{H-index}
A limitation of the degree centrality is that it only considers the node's number of received edges, regardless of the centrality of 
the neighbors.
By contrast, in many networks it is plausible to assume that a node is important if it is connected to other important nodes in the network.
This thesis is enforced by eigenvector-based centrality metrics (section \ref{sec:eigenvector}) by taking into account the whole network topology to
determine the node's score.
A simpler way to implement this idea is to consider nodes' neighborhood in the node score computation,
without including the global network structure. 

The $H$-index is a centrality metric based on this idea.
The $H$-index for unipartite networks \cite{lu2016h} is based on the metric of the same name introduced by Hirsch for bipartite networks \cite{hirsch2005index} to assess researchers' scientific output.
Hirsch's index of a given researcher is defined as the maximum integer $h$ such that at least $h$
publications authored by that researcher received at least $h$
citations. In a similar way, Lu et al. \cite{lu2016h} define the $H$-index of a given node in a  monopartite network as the maximum integer $h$ such
that there exist at least $h$ neighbors of that node with at least $h$ neighbors.
In this way, a node's centrality score is determined by the neighbors' degrees, without the need for inspecting the whole network topology. 
The $H$-index brings some advantage with respect to degree in identifying the most influential spreaders in real networks \cite{lu2016h,lu2016vital}, yet the correlation between the two metrics is large for uncorrelated networks (\ie, networks with negligible degree-degree correlations) \cite{pastor2016topological}.
More details on the $H$-index and its relation with other centrality metrics can be found in paragraph \ref{kshell}.

\subsubsection{Other local centrality metrics}
Several other methods have been proposed in the literature to build local centrality metrics.
The main motivation to prefer local metrics to ``global'' metrics (such as eigenvector-based metrics) is that
local metrics are much less computationally expensive and thus can be evaluated in relatively short time even in huge graphs.
As paradigmatic examples, we mention here the local centrality metric introduced by Chen et al. \cite{chen2012identifying} which considers information up to distance
four from a given node to compute the node's score, and the graph expansion technique introduced by Chen et al. \cite{chen2004local} to estimate a website's
PageRank score, which will be defined in paragraph \ref{sec:eigenvector}.
As this review is mostly concerned with time-aware methods, a more detailed description of static 
local metrics goes beyond its scope. An interested reader can refer to a recent review article by Lu et al. \cite{lu2016vital} for more information.

\subsection{Metrics based on shortest paths}
\label{sec:shortestpath}
This paragraph presents two important metrics -- closeness and betweenness centrality -- that are based on shortest paths in the network defined in paragraph \ref{sec:basic}.

\subsubsection{Closeness centrality}
The basic idea behind closeness centrality is that a node is central if it is ``close''
(in a network sense) to many other nodes. The first way to implement this idea is to define
the closeness centrality~\cite{sabidussi1966centrality} score of 
node $i$ as the reciprocal of its average geodesic distance from the other nodes
\begin{equation}
c_i=\frac{N-1}{\sum\limits_{j\neq i}d_{ij}},
\label{closeness1}
\end{equation}
where $d_{ij}$ denotes the geodesic distance between $i$ and $j$. 
A potential trouble with Eq.~\eqref{closeness1} is that if only one node cannot reached from node $i$, node $i$'s score as determined
by Eq. \eqref{closeness1} is identically zero.
This makes Eq. \eqref{closeness1} useless for networks composed of more than one component.
A common way to overcome this difficulty \cite{rochat2009closeness, boldi2014axioms} is to define node $i$'s centrality as a harmonic mean of its distance from the other nodes
\begin{equation}
c_i=\frac{1}{N-1}\sum_{j\neq i}\frac{1}{d_{ij}}.
\label{closeness2}
\end{equation}
With this definition, a given node $i$ receives zero contribution from nodes that are not reachable with a path that includes node $i$, whereas it receives a large contribution from close nodes -- node $i$'s neighbors give contribution one to $i$'s closeness score,
for example.

The main assumption behind closeness centrality deserves some attention. If we are interested in node centrality as an indicator of the expected 
hitting time for an information (or a disease) flowing through the network, closeness centrality implicitly assumes
that information only flows through the shortest paths \cite{borgatti1995centrality, borgatti2005centrality}.
As pointed out by Borgatti \cite{borgatti2005centrality}, this assumption makes the metric useful only for assessing node importance
in situations where flow from any source node has a prior global knowledge of the network and knows how to reach the target node through the shortest possible path -- this could be the case for a parcel delivery process
or a passenger trip, for example.
However, this assumption is unrealistic for real-world spreading processes where information flows typically have no prior knowledge of 
the whole network topology (think of an epidemic disease) and can often reach target nodes through longer paths 
\cite{iannelli2016effective}.
Empirical experiments reveal that closeness centrality typically (but not for all datasets) underperforms with respect to local metrics in identifying influential spreaders 
for SIS and SIR diffusion processes \cite{lu2016h, lu2016vital}.

\subsubsection{Betweenness centrality} The main assumption of betweenness centrality is that 
a given node is central if many shortest paths pass through it.
To define the betweenness centrality score, we denote by $\sigma_{st}$ the number of shortest paths between nodes
$s$ and $t$, and by $\sigma_{st}^{(i)}$ the number of these paths that pass through node $i$.
The betweenness centrality score of node $i$ is given by
\begin{eqnarray}\label{eqn2}
b_i=\frac{1}{(N-1)(N-2)}\sum_{s\neq t} \frac{\sigma_{st}^{(i)}}{\sigma_{st}},
\end{eqnarray}
Similarly to closeness centrality, betweenness centrality implicitly assumes that only the shortest paths matter 
in the process of information transmission (or traffic) between two nodes.
While a variant of betweenness centrality which accounts for all network paths has also been considered, it
often gives similar results as the original betweenness centrality \cite{newman2005measure}, and no evidence has been found that
these metrics accurately reproduce information flows in real networks \cite{newman2010networks}.
In addition, betweenness centrality performs poorly with respect to other centrality metrics in identifying influential 
spreaders in numerical simulations with the standard SIR and SIS models \cite{lu2016h, lu2016vital}.
Another drawback of closeness and betweenness centrality is their high computational 
complexity mostly due to the calculation of the shortest paths between all pairs of nodes in the network.

\subsection{Coreness centrality and its relation with degree and $H$-index}
\label{kshell}
The coreness centrality (also known as the $k$-shell centrality \cite{NP6888}) is based on the idea of decomposing the network~\cite{Carmi200701} to distinguish among the nodes that form the core of the network and the peripheral nodes.
The decomposition process is usually referred to as the $k$-shell decomposition \cite{NP6888}.
The $k$-shell decomposition starts by removing all nodes with only one connection (together with their edges), until no more such nodes 
remain, and assigns them to the $1$-shell. 
For each remaining node, the number of edges connecting to the other remaining nodes 
is called its residual degree. 
After having assigned the $1$-shell, all the nodes with residual degree $2$ are recursively removed and the $2$-shell is created. This procedure continues as the residual degree increases until all nodes in the networks have been
assigned to one of the shells. 
The index $c_i$ of the shell to which a given node $i$ is assigned to is referred to as the node coreness (or $k$-shell) centrality. Coreness centrality can also be generalized to weighted networks \cite{garas2012k}.

There is an interesting mathematical connection between $k$-shell centrality, degree and $H$-index.
In paragraph \ref{sec:degree}, we have introduced the $H$-index 
of a node $i$ as
the maximum value $h$ such that there exist $h$ neighbors of $i$, each of them 
having at least $h$ neighbors. 
This definition can be formalized by introducing 
an operator $\mathcal{H}$ which acts on a finite number of real numbers $(x_1,x_2,...,x_n)$ by returning the maximum integer $y=\mathcal{H}(x_1,x_2,...,x_n)>0$ such that there exist at least $y$ elements in $\{x_1,\dots,x_n\}$
each of which is larger or equal than $y$ \cite{lu2016h}.
Denoting by $(k_{j_1},k_{j_2},...,k_{j_{k_i}})$ the degrees of the neighbors of a given node $i$, the H-index of node $i$ can be written in terms of $\mathcal{H}$ as
\begin{equation}
h_i=\mathcal{H}(k_{j_1},k_{j_2},...,k_{j_{k_i}}).
\end{equation}
Lu et al. \cite{lu2016h} generalize the $h$ index by recursively defining the $n$th-order $H$-index of node $i$ as
\begin{equation}
 \label{eq:hindex}
h_i^{(n)}=\mathcal{H}(k_{j_1}^{(n-1)},k_{j_2}^{(n-1)},...,k_{j_{k_i}}^{(n-1)}),
\end{equation}
where $h_i^{(0)}=k_i$ and $h_{i}^{(1)}$ is the $H$-index of node $i$.
For arbitrary undirected networks, it can be proven \cite{lu2016h} that as $n$ grows, $h_i^{(n)}$ converges to coreness centrality $c_i$,
\begin{equation}
\lim_{n\rightarrow \infty}h_i^{(n)}=c_i
\end{equation}
The family of $H$-indexes defined by Eq.~\eqref{eq:hindex} thus bridges from node degree ($h_i^{(0)}=k_i$) to coreness that is from a local centrality metric to a global one.

In many real-world networks, the $H$-index 
outperforms both degree and coreness in identifying influential spreaders as defined by classical SIR and SIS network 
spreading models. 
Recently, Pastor-Satorras and Castellano \cite{pastor2016topological} performed an extensive analytic and numerical investigation of the mathematical 
properties of the $H$-index, finding analytically that the $H$-index is expected to be strongly correlated with degree in uncorrelated networks. The same strong correlation is found also in real networks, and Pastor-Satorras and Castellano \cite{pastor2016topological} point out that the $H$-index is a poor indicator of node spreading influence as compared
to the non-backtracking centrality \cite{kawamoto2016localized}.
While the comparison of different metrics with respect to their ability to identify influential spreaders 
is not one of the main focus of the present review, we remind the interested
reader to \cite{lu2016vital,radicchi2016leveraging, pastor2016topological} for recent reports on this important problem.

\subsection{Eigenvector-based centrality metrics}
\label{sec:eigenvector}
Differently from the local metrics presented in paragraph \ref{sec:degree} and the metrics based on shortest paths presented
in paragraph \ref{sec:shortestpath},
eigenvector-based metrics take into account all network paths to determine node importance.
We refer to the metrics presented in this paragraph as eigenvector-based metrics because their respective score
vectors can be interpreted as the principal eigenvector of a matrix which only depends on the network's adjacency matrix $\mat{A}$.

\subsubsection{Eigenvector centrality}
Often referred to as Bonacich centrality \cite{bonacich1987power}, the eigenvector centrality assumes that
a node is important if it is connected to other important nodes.
To build the eigenvector centrality, we start from a uniform score value
$s_{i}^{(0)}=1$ for all nodes $i$.
At each subsequent iterative step, we update the scores according to the equation
\begin{equation}
\label{iterative}
s_{i}^{(n+1)}=\sum_{j}A_{ij}\,s_{j}^{(n)}=(\mat{A}\vek{s})_i.
\end{equation}
The vector of eigenvector centrality scores is defined as the fixed point of this equation. 
The iterations converge to a vector proportional to the principal eigenvector $\vek{v}_{1}$ of the adjacency matrix $\mat{A}$.
To prove this, we first observe that for connected networks, $\vek{v}_1$ is unique \footnote{For networks with more than one component, there can exist more than one eigenvector associated with the principal eigenvalue of $\mat{A}$. However, it can be shown \cite{newman2010networks} that only one of them can have all components larger than zero, and thus it is the only one that can be obtained by iterating Eq.~\eqref{iterative} since we assumed $s_i^{(1)}=1$ for all $i$.} due to the Perron-Frobenius theorem \cite{perron1907theorie}.

We then expand the initial condition $\vek{s}^{(0)}$ in terms of the eigenvectors $\{\vek{v}_{\alpha}\}$ of $\mat{A}$ as
\begin{equation}
\vek{s}^{(0)}=\sum_{\alpha}c_{\alpha}\vek{v}_{\alpha}.
\end{equation}
Denoting the eigenvalue associated with the eigenvector $\vek{v}_{\alpha}$ as $\lambda_{\alpha}$ and assuming that the eigenvalues are ordered and thus $\abs{\lambda_{\alpha+1}}<\abs{\lambda_{\alpha}}$, Eq.~\eqref{iterative} implies
\begin{equation}
\vek{s}^{(n)} = \mat{A}^{n}\vek{s}^{(0)} =
c_1 \lambda_1^{n} \vek{v}_1+\sum_{\alpha>1}c_{\alpha}\lambda_{\alpha}^{n}\vek{v}_{\alpha}.
\end{equation}
When $n$ grows, the second term on the r.h.s. of the previous equation is exponentially small with respect to the r.h.s.'s first term, which results in
\begin{equation}
\vek{s}^{(n)} / \lambda_{1}^{n} \to c_{1}\vek{v}_{1}.
\end{equation}
By iterating Eq.~\eqref{iterative} and continually normalizing the score vector, the scores converge to the adjacency matrix's principal eigenvector $\vek{v}_1$.
Consequently, the vector of eigenvector centrality scores $\vek{v}_1$ satisfies the self-consistent equation
\begin{equation}
s_{i}=\lambda_{1}^{-1}\sum_{j}A_{ji}\,s_{j},
\end{equation}
which can be rewritten in a matrix form as
\begin{equation}
\vek{s}=\lambda_{1}^{-1}\mat{A}\,\vek{s}.
\label{eigenvector}
\end{equation}
While the assumptions of the eigenvector centrality seem plausible -- high centrality nodes give  high contribution to the centrality of the nodes they connect to. However, it has important shortcomings when we try to apply it to directed networks.
If a node has received no incoming links, it has zero score according to Eq.~\eqref{eigenvector} 
and consequently gives zero contribution to the scores of the nodes it connects to.
This makes the metric particularly useless for directed acyclic graphs (such as the network of scientific papers) where all papers have are assigned zero score. The Katz centrality metric, that we introduce below, solves this issue by adding a constant term to the r.h.s. of Eq.~\eqref{eigenvector}, and thus achieves that the score of all nodes is positive.

\subsubsection{Katz centrality}
Katz centrality metric \cite{katz1953new} builds on the same premise of eigenvector centrality -- a node is important if it
is connected to other important nodes -- but, differently from eigenvector centrality, it assigns certain minimum score to each node. Different variants of the Katz centrality have been proposed in the literature \cite{katz1953new,hubbell1965input, bonacich1987power, bonacich2001eigenvector} -- we refer to the original papers the reader interested in the subtleties associated with the different possible definitions of Katz centrality, and we present here the variant proposed in \cite{bonacich2001eigenvector}.
The vector of Katz-centrality (referred to as \emph{alpha-centrality} in \cite{bonacich2001eigenvector}) scores is defined as \cite{bonacich2001eigenvector}
\begin{equation}
\vek{s}=\alpha\,\mat{A}\,\vek{s}+\beta\,\vek{e}.
\end{equation}
where $\vek{e}$ is the $N$-dimensional vector whose elements are all equal to one.
The solution $\vek{s}$ of this equation exists only if $\alpha<1/\lambda_1$ where $\lambda_1$ is the principle eigenvalue of $\mat{A}$. So we restrict the following analysis to this range of $\alpha$ values.
The solution reads
\begin{equation}
\vek{s}=\beta\,(\mat{1}_N-\alpha\mat{A})^{-1}\,\vek{e},
\label{katz}
\end{equation}
where $\mat{1}_N$ denotes the $N\times N$ identity matrix. This solution can be written in the form of a geometric series
\begin{equation}
\vek{s}=\beta\,\sum_{k=0}^{\infty}\alpha^{k}\,\mat{A}^{k}\,\vek{e}.
\label{katz1}
\end{equation}
Thus the score $s_i$ of node $i$ can be expressed as
\begin{equation}
\begin{split}
s_i&=\beta\,\Biggr(1+\alpha\,\sum_j A_{ij}+\alpha^2 \,\sum_{j,l}A_{ij}\,A_{jl}\\
&+\alpha^{3}\,\sum_{j,l,m}A_{ij}\,A_{jl}\,A_{lm}+\mathcal{O}(\alpha^4)\Biggr).
\end{split}
\label{katz2}
\end{equation}
Eq. \eqref{katz1} (or, equivalently, \eqref{katz2}) shows that the Katz centrality score of a node is determined by all network paths that pass through node $i$. 
Note indeed that $\sum_{j}A_{ij}$, $\sum_{j,l}A_{ij}\,A_{jl}$ and $\sum_{j,l,m}A_{ij}\,A_{jl}\,A_{lm}$ represent the paths of length one, two and three, respectively, that have node $i$ as the last node.
The parameter $\alpha$ determines how contributions of various paths attenuate with path length. For small values of $\alpha$, long paths are strongly penalized and node score is mostly determined by the shortest paths of length one (i.e., by node degree). By contrast, for large values of $\alpha$ (but still $\alpha<\lambda_{1}^{-1}$, otherwise the series in Eq.~\eqref{katz2} does not converge) long paths to node score.

Known in social science since the 50s \cite{katz1953new}, Katz centrality has been used in a number of applications -- for example, a variant of the Katz centrality has been recently shown to substantially outperform other static centrality metrics in predicting neuronal activity \cite{fletcher2016structure}.
While it overcomes the drawbacks of eigenvector centrality and produces a meaningful ranking also when applied to directed acyclic graphs or disconnected networks, the Katz centrality may still provide unsatisfactory results in networks where the outdegree distribution
is heterogeneous. An important node is then able to manipulate the other nodes' importance by simply creating many edges towards
a selected group of target nodes, and thus improving their score. Google's PageRank centrality overcomes this limitation by weighting less the edges that are created by nodes with many outgoing connections. Before introducing PageRank centrality, we briefly discuss a simple variant of Katz centrality that has been applied to networks of sport professional teams and players.

\subsubsection{Win-lose scoring systems for ranking in sport}
\label{sec:winlose}
Network-based win-lose scoring systems aim to rank competitors (for simplicity, we refer to players in this paragraph) in sports based on one-versus-one matches (like tennis, football, baseball).
In terms of complex networks, each player $i$ is represented by a node and the weight of the directed edge $j\to i$ represents the number of wins of player $j$ against player $i$.
The main idea of a network-based win-lose scoring scheme is that a player
is strong if it is able to defeat other strong players (i.e., players that have defeated many opponents~\cite{park2005network}).
This idea led Park and Newman~\cite{park2005network} to define the
vector of win scores $\vek{w}$ and the vector of lose scores $\vek{l}$ through a variant of the Katz centrality equation
\begin{equation}
\begin{aligned}
\vek{w}&=(\mat{1}-\alpha\,\mat{A}\TT)^{-1}\,\vek{k}^{out},\\
\vek{l}&=(\mat{1}-\alpha\,\mat{A})^{-1}\,\vek{k}^{in},\\
\end{aligned}
\end{equation}
where $\alpha\in(0,1)$ is a parameter of the method. The vector $\vek{s}$ of player scores is defined as the win-loss differential $\vek{s}=\vek{w}-\vek{l}$.
To understand the meaning of the win score, we use again the geometric series of matrices and obtain
\begin{equation}
w_{i}=\sum_{j}A_{ji}+\alpha\,\sum_{j,k}A_{kj}\,A_{ji}+\alpha^{2}\,\sum_{j,k,l}A_{kj}\,A_{jl}\,A_{li} +\mathcal{O}(\alpha^3).
\end{equation}
From this equation, we realize that the first contribution to $w_{i}$ is simply the total number of wins of player $i$.
The successive terms represent ``indirect wins'' achieved by player $i$.
For example, the $\mathcal{O}(\alpha)$ term represents the total number of wins of players beaten by player $i$.
An analogous reasoning applies to the loss score $\vek{l}$.
Based on similar assumptions as the win-lose score, PageRank centrality metric has been applied to rank sport competitors by Radicchi~\cite{radicchi2011best}, leading to the Tennis
Prestige Score [\url{http://tennisprestige.soic.indiana.edu/}]. A time-dependent generalization of the win-lose score will be presented in paragraph \ref{sec:dynwinlose}.

\subsubsection{PageRank}
\label{sec:pagerank}
PageRank centrality metric has been introduced by Brin and Page \cite{brin1998anatomy} with the aim to rank web pages in the Web graph,
and there is general agreement on the essential role played by this metric in determining the outstanding success of Google's Web search engine \cite{franceschet2011pagerank,langville2011google}.
For a directed network, the vector $\vek{s}$ of PageRank scores is defined by the equation
\begin{equation}
\vek{s}=\alpha\,\mat{P}\,\vek{s}+(1-\alpha)\,\vek{v}
\label{pagerank}
\end{equation}
where $P_{ij}=A_{ij}/k^{out}_j$ is the network's transition matrix,
$\vek{v}$ is called the teleportation vector and $\alpha$ is called the teleportation parameter. A uniform vector ($v_i=1$ for all nodes $i$) is the original choice by Brin and Page, and it is arguably the most common choice for $\vek{v}$, although benefits and drawbacks of other teleportation vectors $\vek{v}$ have been also explored in the literature~\cite{lambiotte2012ranking}.
This equation is conceptually similar to the equation that defines Katz centrality,
with the important difference that the adjacency matrix $\mat{A}$ has been replaced by the transition matrix $\mat{P}$.
From Eq.~\eqref{pagerank}, it follows that the vector of PageRank scores can be interpreted as the leading eigenvector of the matrix
\begin{equation}
\mat{G}=\alpha \,\mat{P}+(1-\alpha)\,\vek{v}\vek{e}\TT
\end{equation}
where $\vek{e} = (1,\dots,1)_N$. $\mat{G}$ is often referred to as the \emph{Google matrix}.
We refer to the review article by Ermann et al. \cite{ermann2015google} for a detailed review of the mathematical properties of the spectrum of $\mat{G}$.

Beyond its interpretation as an eigenvalue problem, there exist also an illustrative physical interpretation of the PageRank equation.
In fact, Eq.~\eqref{pagerank} represents the equation that defines the stationary state of a stochastic process on the network
where a random walker located at a certain node can make two moves: (1) with probability $\alpha$, jump to another node by following a randomly chosen edge starting at the node where the walker is located; (2) with probability $1-\alpha$, ``teleport'' to a randomly chosen node. In other words, Eq.~\eqref{pagerank} can be seen as the stationary equation of the process described by the following master equation
\begin{equation}
\vek{s}^{(n+1)}=\alpha\,\mat{P}\,\vek{s}^{(n)}+(1-\alpha)\,\vek{v}.
\label{pagerank_master}
\end{equation}
The analogy of PageRank with physical diffusion on the network is probably one of the properties that have most stimulated physicists' interest in studying the PageRank's properties.

It is important to notice that in most real-world directed networks (such as the Web graph),
there are nodes (referred to as \emph{dangling nodes} \cite{berkhin2005survey}) without outgoing edges. These nodes with zero out-degree receive score from other pages, but do not redistribute it to other nodes. In Equation~\eqref{pagerank}, the existence of dangling nodes makes the transition matrix ill-defined as its elements corresponding to transitions from the dangling nodes are infinite. There are several strategies how to deal with dangling nodes. For example, one can remove them from the system --
this removal is unlikely to significantly affect the ranking of the remaining nodes as by definition they receive no score from the dangling nodes. One can also artificially set $k^{out}=1$ for the dangling nodes which does not affect the score of the other nodes \cite{newman2010networks}.
Another strategy is to replace the ill-defined elements of the transition matrix with uniform entries;
in other words, the transition matrix is re-defined as
\begin{equation}
P_{ij}=\begin{cases}
       A_{ij}/k^{out}_j & \text{if } k_{j}^{out}>0,\\
	   1/N		  & \text{if } k_{j}^{out}=0.\\
       \end{cases}
\label{transition}
\end{equation}
We direct the interested reader to the survey article by Berkhin \cite{berkhin2005survey} for a presentation of possible alternative strategies to account for the dangling nodes.

Using the Perron-Frobenius theorem, one can show that the existence and uniqueness of the solution of Eq.~\eqref{pagerank} is guaranteed if $\alpha\in[0,1)$ \cite{berkhin2005survey,gleich2015pagerank}.
We assume $\alpha$ to be in the range $(0,1)$ in the following.
The solution of Eq.~\eqref{pagerank} reads
\begin{equation}
\vek{s}=(1-\alpha)\,(1-\alpha\,\mat{P})^{-1}\,\vek{v}.
\end{equation}
Similarly as we did for Eq.~\eqref{katz}, $\vek{s}$ can be expanded by using the geometric series to obtain
\begin{equation}
\vek{s}=(1-\alpha)\,\sum_{k=0}^{\infty}\alpha^{k}\,\mat{P}^{k}\,\vek{v}.
\label{pr_paths}
\end{equation}
As was the case for the Katz centrality, also the PageRank score of a given node is determined by all the paths that pass through that node. The teleportation parameter $\alpha$ controls the exponential damping of longer paths. For small $\alpha$ (but larger than zero), long paths give a negligible contribution and -- if outdegree fluctuations are sufficiently small -- the ranking by PageRank
approximately reduces to the ranking by indegree \cite{chen2007finding, perra2008spectral}.
When $\alpha$ is close to (but smaller than) one, long paths give a substantial contribution to node score. While there is no universal criterion for choosing $\alpha$, a number of studies (see \cite{boldi2005pagerank,bianchini2005inside,avrachenkov2008singular} for example) have pointed out that large values of $\alpha$ might lead to a ranking that is highly sensible to small perturbations on the network's structure, which suggests \cite{avrachenkov2008singular} that values around $0.5$ should be preferred to the original value set by Brin and Page ($\alpha=0.85$).

An instructive interpretation of $\alpha$ comes from viewing the PageRank algorithm as a stochastic process on the network (Eq.~\eqref{pagerank_master}). At each iterative step, the random walker has to decide whether to follow a network edge or whether to teleport to a randomly chosen node. On average, the length of the network paths covered by the walker before teleporting to a randomly chosen node is given by \cite{medo2013network}
\begin{equation}
\braket{l}_{RW}= (1-\alpha)\,\sum_{k=0}^{\infty}k\,\alpha^{k}=\alpha / (1-\alpha),
\end{equation}
which is exactly the ratio between the probability of following a link and the probability of random teleportation.
For $\alpha=0.85$, we obtain $\braket{l}_{RW}=5.67$ which corresponds to following six edges before teleporting to a random node.
Some researchers ascribe the PageRank's success to the fact that the algorithm provides a reasonable model of the real behavior of Web users, who surf the Web both by following the hyperlinks that they find in the webpages,
and -- perhaps when they get bored or stuck -- by restarting their search (teleportation).
For $\alpha=0.5$, we obtain $\braket{l}_{RW}=1$.
This choice of $\alpha$ may better reflect the behavior of researchers on 
the academic network of papers (one may feel compelled to check the references in a work of interest but rarely to check the references in one of the referred papers). Chen et al.~\cite{chen2007finding} use $\alpha=0.5$ for a citation network motivated by the finding that half of the entries in the reference list of a typical publication cite at least one other article
in the same reference list, which might further indicate that researchers tend to cover paths up to the length of one when browsing the academic literature.

\subsubsection{PageRank variants}
\label{sec:pr_variants}
PageRank is built on three basic ingredients: the transition matrix $\mat{P}$, the teleportation parameter $\alpha$, and
the teleportation vector $\vek{v}$.
With respect to the original algorithm, the majority of PageRank variants are still based on Eq.~\eqref{pagerank}, and modify one or more than one of these three elements. In this paragraph, we focus on PageRank variants that do not explicitly depend on edge time-stamp. 
We present three such variants; the reader interested in further variations is referred to the review article by Gleich~\cite{gleich2015pagerank}. Time-dependent variants of PageRank and Katz centrality will be the focus of Section \ref{sec:time}.

\paragraph{Reverse PageRank} The reverse PageRank \cite{fogaras2003start} score vector 
for a network described by the adjacency matrix $\mat{A}$ 
is defined as the PageRank score vector for the network described by the transpose $\mat{A}\TT$ of $\mat{A}$.
In other words, reverse-PageRank scores flow in the opposite edge direction with respect to PageRank scores.
Reverse PageRank has been also referred to as CheiRank in the literature \cite{zhirov2010two,ermann2015google}.
In general, PageRank and reversed PageRank provide different information on the node role in a network. Large reverse PageRank values mean that the nodes can reach many other nodes in the network.
An interested reader can find more details in the review article by Ermann et al. \cite{ermann2015google}.

\paragraph{LeaderRank} In the LeaderRank algorithm \cite{lu2011leaders}, a ``ground-node'' is added to the original network and connected with each of the network's $N$ nodes by an outgoing and incoming link.
A random walk is performed on the resulting network; the LeaderRank score of a given node is given by the fraction of time the random walker spends on that node.
With respect to PageRank, LeaderRank keeps the idea of performing a random walk on network's links, but it does not feature any teleportation mechanism and, as a consequence, it is parameter-free. By adding the ground node, the algorithm effectively enforces a degree-dependent teleportation probability\footnote{From a node of the original degree $k_i$, the probability to follow one of the actual links is $k_i / (k_i + 1)$ as opposed to the probability $1 / (k_i + 1)$ to move to the ground node and then consequently to a random one of the network's $N$ nodes.}.
Lu et al. \cite{lu2011leaders} analyzed online social network data to show that the ranking by LeaderRank is less sensitive than that by PageRank to perturbations
 of the network structure and to malicious manipulations on the system. Further variants based on degree-dependent weights have been used to detect influential and robust nodes \cite{li2014identifying}. Furthermore, the ``ground-node'' notion of LeaderRank has also been generalized to bipartite networks by network-based recommendation systems \cite{zhou2013power}.
 
\paragraph{PageRank with ``smart'' teleportation} In the original PageRank formulation, 
the teleportation vector $\vek{v}$ represents the amount of score
that is assigned to each node by default, independently of network structure.
The original  PageRank algorithm implements the simplest possible teleportation vector, $\vek{v}=\vek{e}/N$, where $e_i=1$ for all components $i$.
This is not the only possible choice, and one might instead assume that higher-degree nodes should be given higher baseline score than obscure nodes.
This thesis has been referred to as ``smart'' teleportation by Lambiotte and Rosvall \cite{lambiotte2012ranking}, 
and it can be implemented by setting $v_{i}\propto k^{in}_{i}$.
Smart-teleportation PageRank score can still be expressed in terms of network paths through Eq.~\eqref{pr_paths}; differently from PageRank, the zero-order contribution to a given node's smart-teleportation PageRank score is its indegree.
The ranking by smart-teleportation PageRank results to be remarkably more stable than original PageRank with respect to variations
in the teleportation parameter $\alpha$ \cite{lambiotte2012ranking}. A detailed comparison of the two rankings as well as a discussion of alternative teleportation strategies can be found in~\cite{lambiotte2012ranking}.

\paragraph{Pseudo-PageRank}
It is worth mentioning that sometimes PageRank variants are built on a different definition of node score, based on
a column sub-stochastic matrix\footnote{A matrix $\mat{Q}$ is \emph{column sub-stochastic} if and only if $\sum_{i}Q_{ij}\leq 1$.} $\mat{Q}$ and a non-normalized additive vector $\vek{f}$ ($f_i \geq 0$).
The corresponding equation that define the vector $\vek{s}$ of nodes' score is
\begin{equation}
\vek{s}=\alpha\,\mat{Q}\,\vek{s}+\vek{f}.
\label{pseudopagerank}
\end{equation}
In agreement with the review by Gleich \cite{gleich2015pagerank}, we refer the problem of solving Eq.~\eqref{pseudopagerank} as the pseudo-PageRank problem.

One can prove that the pseudo-PageRank problem is equivalent to a PageRank problem (theorem 2.5 in~\cite{gleich2015pagerank}). More precisely, let $\vek{y}$ be the solution of a pseudo-PageRank system with parameter $\alpha$, column sub-stochastic matrix $\mat{Q}$ and additive vector $\vek{f}$.
If we define  $\vek{v}: = \vek{f} /(\vek{e}\TT \vek{f})$ and $\vek{x} := \vek{y} / (\vek{e}\TT \vek{y})$, then $\vek{x}$ is the
solution of a PageRank system with teleportation parameter $\alpha$, stochastic matrix $\mat{P} = \mat{Q} + \vek{v} \vek{c}\TT$, and teleportation vector $\vek{v}$, where $\vek{c}\TT = \vek{e}\TT  - \vek{e}\TT \mat{Q}$ is the correction vector needed to make $\mat{Q}$ stochastic. 
The (unique) solution $\vek{y}$ of Eq.~\eqref{pseudopagerank} thus inherits the properties of the corresponding PageRank solution $\vek{x}$ -- the two solutions $\vek{y}$ and $\vek{x}$ only differ by a uniform normalization factor. Since PageRank and pseudo-PageRank problem are equivalent, we use the two descriptions interchangeably in the following.


\subsubsection{HITS algorithm}
HITS (Hyperlink-Induced Topic Search) algorithm~\cite{kleinberg1999authoritative} is a popular eigenvector-based ranking method originally aimed at ranking webpages in the WWW. In a unipartite directed network, the HITS algorithm assigns two scores to each node in a self-consistent fashion. Score $h$ -- referred to as node hub-centrality score -- is large for nodes that point to many authoritative nodes.
The other score $a$ -- referred to as node authority-centrality score -- is large for nodes that are pointed by many hubs.
The corresponding equations for node scores $a$ and $h$ read
\begin{equation}
\begin{split}
\vek{a} &=\alpha \,\mat{A}\,\vek{h},\\
\vek{h} &=\beta\, \mat{A}\TT\,\vek{a},\\
\end{split}
\end{equation}
where $\alpha$ and $\beta$ are parameters of the method.
The previous equations can be rewritten as
\begin{equation}
\begin{split}
\mat{A}\,\mat{A}\TT\,\vek{a}&=\lambda\, \vek{a},\\
\mat{A}\TT\,\mat{A}\,\vek{h}&=\lambda\, \vek{h},\\
\end{split}
\end{equation}
where $\lambda=(\alpha\beta)^{-1}$. The resulting score vectors $\vek{a}$ and $\vek{h}$ are thus eigenvectors of the matrices $\mat{A}\,\mat{A}\TT$ and $\mat{A}\TT\,\mat{A}$, respectively.
The HITS algorithm has been used, for example, to rank publications in citation networks by the literature search engine Citeseer~\cite{li2006citeseer}. In the context of citation networks, it is natural to identify topical reviews as hubs, since they contain many references to influential papers in the literature.
A variant of the HITS algorithm where a constant additive term is added to both node scores is explored by Ng et al.~\cite{ng2001stable} and results in a more stable ranking of the nodes with respect to small perturbations in network structure.
An extension of HITS to bipartite networks \cite{deng2009generalized} is presented in paragraph~\ref{sec:cohits}.

\begin{figure*}[t]
\centering
\includegraphics[width=0.8\textwidth]{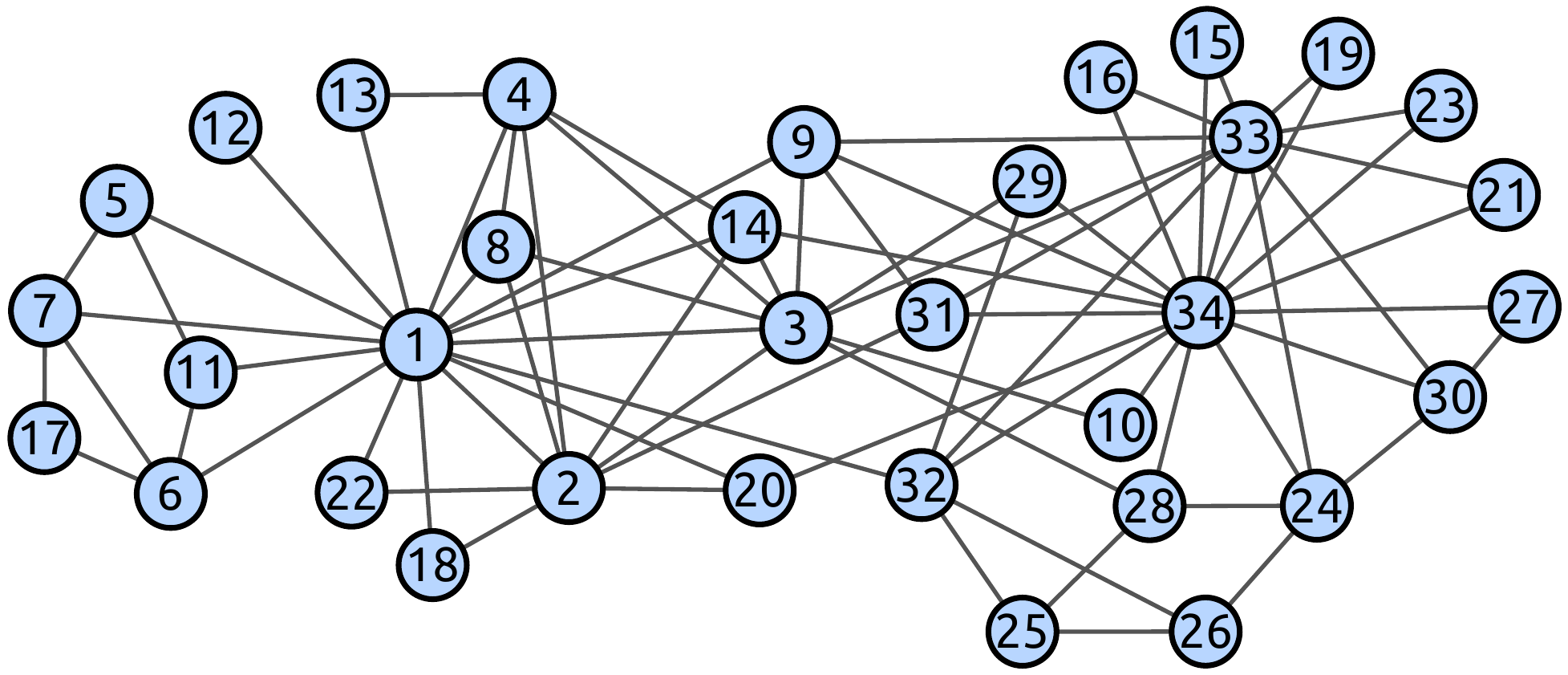}\\[8pt]
\includegraphics[scale=0.7]{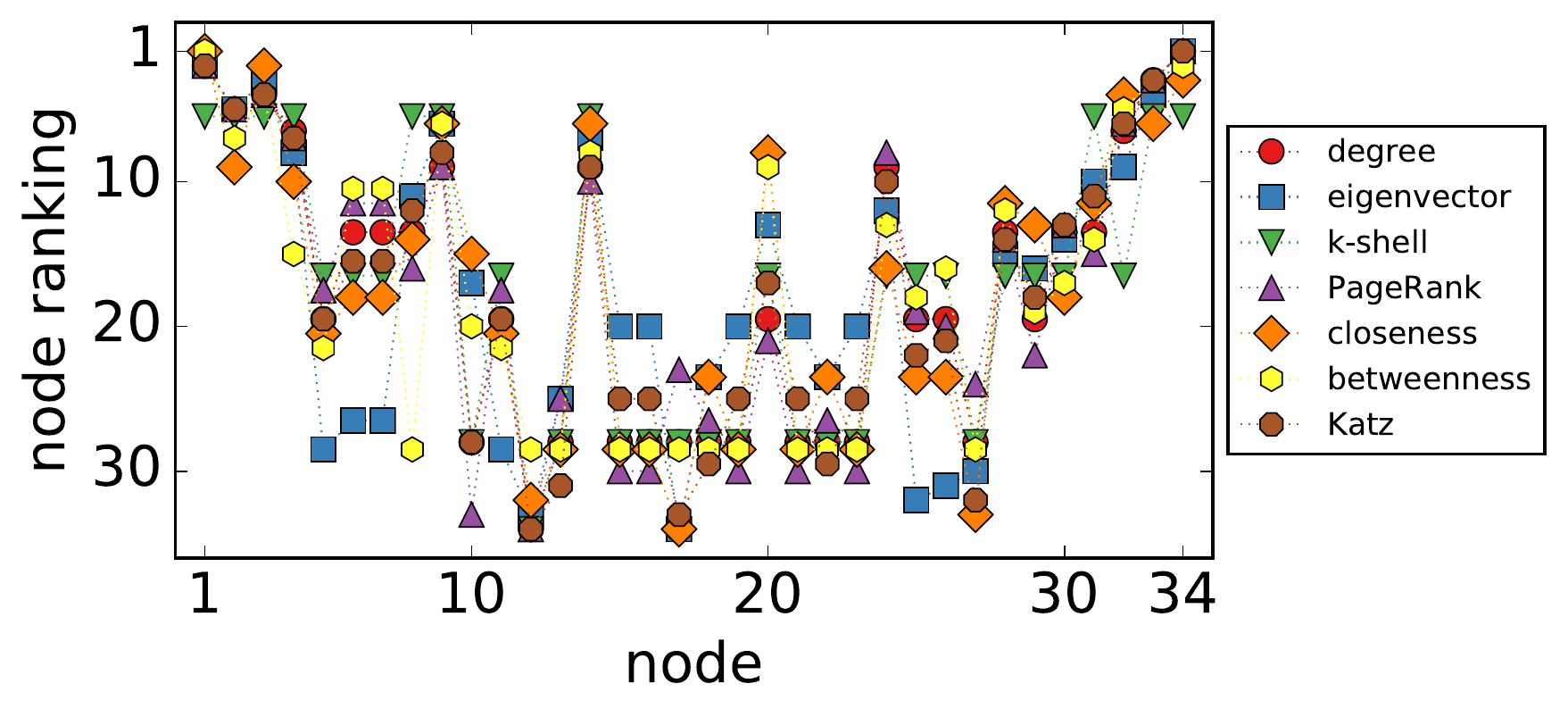}
\caption{\emph{Top:} The often-studied Zachary's karate club network has $34$ nodes and $78$ links (visualized with the Gephi software).
\emph{Bottom:} Ranking of the nodes in the Zachary karate club network by the centrality metrics described in this section.}
\label{fig:zachary}
\end{figure*}

\subsection{A case study: node centrality in the Zachary's karate club network}
In this subsection, we illustrate the differences of the above-discussed node centrality metrics on the example of the small Zachary karate club network~\cite{Zachary1977}. The Zachary Karate Club network captures friendships between the members of a US karate club (see Figure~\ref{fig:zachary} for its visualization). Interestingly, the club has split in two parts as a result of a conflict between its the instructor (node 1) and the administrator (node 34); the network is thus often used as one of a test cases for community detection methods~\cite{fortunato2010community,fortunato2016community}. The rankings of the network's nodes produced by various node centrality metrics are shown in Figure~\ref{fig:zachary}. The first thing to note here that nodes 1 and 34 are correctly identified as being central to the network by all metrics. The $k$-shell metric produces integer values distributed over a limited range (in this network, the smallest and largest $k$-shell value are 1 and 4, respectively) which leads to many nodes obtaining the same rank (a highly degenerate ranking).

While most nodes are ranked approximately the same by all metrics, there are also nodes for which the metrics produced widely disparate results. Node 8, for example, is ranked high by $k$-shell but low by betweenness. An inspection of the node's direct neighborhood reveals why this is the case: node 8 is connected high-degree nodes 1, 2, 4, and 14, which directly assures its high $k$-shell value of 4. At the same time, those neighboring nodes themselves are connected, which creates shortcuts circumventing node 8 and implies that node 8 lies on very few shortest paths. Node 10 has low degree and thus also low PageRank score (among the evaluated metrics, these two have the highest Spearman's rank correlation). By contrast, its closeness is comparatively high as the node is centrally located (\eg, the two central nodes 1 and 34 can be reached from node 10 in two and one steps, respectively). Finally, while PageRank and eigenvector centrality are closely related and their values are quite closely correlated (Pearson's correlation 0.89), they still produce rather dissimilar rankings (their Spearman's correlation 0.68 is the lowest among the evaluated metrics).

\subsection{Static ranking algorithms in bipartite networks}
There are many systems that are naturally represented by \emph{bipartite networks}: users are connected with the content that they consumed online, customers are connected with the products that they purchased, scientists are connected with the papers that they authored, and many others. In a bipartite network, two groups of nodes are present (such as the customer and product nodes, for example) and links exist only between the two groups, not within them. The current state of a bipartite network can be captured by the network's adjacency matrix $\mat{B}$, whose element $B_{i\alpha}$ is one if node $i$ is connected to node $\alpha$ and zero otherwise. Note that to highlight the difference between the two groups of nodes, we use Latin and Greek letters, respectively, to label them. Since a bipartite network's adjacency matrix is a different mathematical object than a monopartite network's adjacency matrix, we label them differently as $\mat{B}$ and $\mat{A}$, respectively. While $\mat{A}$ is by definition a square matrix, $\mat{B}$ in general is not. Due to this difference, the adjacency matrix of a bipartite network is sometimes referred to as the \emph{incidence matrix} in graph theory~\cite{newman2010networks}.

Since bipartite networks are composed of two kinds of nodes, one might be interested in scoring and rankings the two groups of nodes separately. Node-scoring metrics for bipartite networks thus usually give two vectors of scores, one for each kind of nodes, as output. In this paragraph, we present three ranking algorithms (coHITS, method of reflections, and the fitness-complexity algorithm) specifically aimed at bipartite networks, together with their possible interpretation in real socio-economic networks.

Before presenting the metrics, we stress that neglecting the connections between nodes of the same kind can lead to a considerable loss of information. For example, many bipartite networks are embedded in a social network of the participating users. In websites like \url{Digg.com} and \url{Last.fm}, users can consume content and at the same time they can also select other users as their friends and potentially be influenced by their friends' choices \cite{vidmer2015unbiased}. Similarly, the scholarly network of papers and their authors is naturally influenced by personal acquaintances and professional relationships among the scientists~\cite{sarigol2014predicting}. 
In addition, often more than two layers are necessary to properly account for different types of interactions between the nodes \cite{kivela2014multilayer,boccaletti2014structure}.
We focus here on the settings where a bipartite network representation of the input data provides sufficiently insightful results, and refer to \cite{sole2014centrality,de2015ranking,taylor2015eigenvector} for representative examples of ranking algorithms for multilayer networks.

\subsubsection{Co-HITS algorithm}
\label{sec:cohits}
Similarly to the HITS algorithm for monopartite networks, co-HITS algorithm~\cite{deng2009generalized} works with two kinds of scores. Unlike HITS, co-HITS assigns one kind of scores to nodes in one group and the other kind of scores to nodes in the other group; each node is thus given one score. Denoting the two groups of nodes as $1$ and $2$, respectively, and their respective two score vectors as $\vek{x}=\{x_i\}$ and $\vek{y}=\{y_{\alpha}\}$, the general co-HITS equations take the form
\begin{equation}
\label{eq.HITS}
\begin{split}
x_i &= (1-\lambda_1)x_i^0 + \lambda_1 \sum_{\alpha} w_{i\alpha}^{21} y_{\alpha},\qquad \\
y_{\alpha} &= (1-\lambda_2)y_{\alpha}^0 + \lambda_2 \sum_i w_{\alpha i}^{12} x_i.
\end{split}
\end{equation}
Here $\lambda_1$ and $\lambda_2$ are analogs of the PageRank's teleportation parameter and $x_i^0$ and $y_{\alpha}^0$ are provide baseline, in general node-specific, scores that are given to all nodes regardless of the network topology. The summation terms represent the redistribution of scores from group 2 to group 1 (through $w_{i\alpha}^{21}$) and vice versa (through $w_{\alpha i}^{12}$). The transition matrices $\mat{w}^{21}$ and $\mat{w}^{12}$ as well as the baseline score vectors $\vek{x}^0$ and $\vek{y}^0$ are column-normalized, which implies that the final scores can be obtained by iterating the above-specified equations without further normalization.

Similarly to HITS, equations for $x_i$ and $y_{\alpha}$ can be combined to obtain a set of equations where only node scores of one node group appear. Connections between the original iterative framework and various regularization schemes, where node scores are defined through optimization problems motivated by the co-HITS equations, are explored by Deng et al.~\cite{deng2009generalized}.
One of the studied regularization frameworks is eventually found to yield the best performance in a query suggestion task when applied to real data.

\subsubsection{Method of reflections}
\label{sec:mr}
The method of reflections (MR) was originally devised by Hidalgo and Hausmann to quantify the competitiveness of countries and the complexity of products based on the network of international exports~\cite{hidalgo2009building}. Although the metric can be applied to any bipartite network, in order to present the algorithm, we use the original terminology where countries are connected with the products that they export. The method of reflections is an iterative algorithm and its equations read
\begin{equation}
\label{eq:MR}
k_i^{(n)}=\frac{1}{k_i}\sum_{\alpha}B_{i\alpha}\,k_\alpha^{(n-1)},\qquad
k_\alpha^{(n)}=\frac{1}{k_\alpha}\sum_{i} B_{i\alpha}\,k_i^{(n-1)}
\end{equation}
where $k_i^{(n)}$ is the score of country $i$ at iteration step $n$, and $k_\alpha^{(n)}$ is the score of product $\alpha$ at step $n$. Both scores are initialized with node degree ($k_i^{(0)}=k_i$ and $k_\alpha^{(0)}=k_\alpha$). In the original method, a threshold value is set, and when the total change of the scores is smaller than this value, the iterations stop. 
The choice of the threshold is important because the scores converge to a trivial fixed point \cite{caldarelli2012network}. The threshold has to be big enough so that rounding errors do not exceed the differences among the scores, as discussed in \cite{caldarelli2012network, cristelli2013measuring}. 
An eigenvector-based definition was given later by Hausmann et al.~\cite{hausmann2014atlas}, which yields the same results as the original formulation and it has the advantage of making the threshold choice unnecessary.

Another drawback of the iterative procedure is that while one might expect that subsequent iterations ``refine'' the information provided by this metric, Cristelli et al. \cite{cristelli2013measuring} point out that the iterations of the metric shrink information instead of refining it. Furthermore, Mariani et al.  \cite{mariani2015measuring} noticed that stopping
the computation after two iterations maximizes the agreement between the country ranking by the MR score and the country ranking by their importance for the structural stability of the system.
While the country scores by the MR have been shown to provide better predictions of the economical growth of countries as compared to traditional economic indicators \cite{hausmann2014atlas}, both the economical interpretation of the metric and its use as a predictive tool have raised some criticism \cite{tacchella2012new, cristelli2013measuring}. 
The use of this metric as a predictor of GDP will be discussed in paragraph \ref{prediction_gdp}.

\subsubsection{Fitness-complexity metric}
\label{sec:fcm}
Similarly to the method of reflections, the Fitness-Complexity (FC) metric aims to simultaneously measure the competitiveness of countries' (referred to as country fitness) and the products' level of sophistication (referred to as product complexity) based on international trade data~\cite{tacchella2012new}. The basic thesis of the algorithm is that competitive countries tend to diversify their export basket, whereas sophisticated products tend to be only exported by the most diversified (and thus the most fit) countries. The rationale of this thesis is that the production of a given good requires a certain set of diverse technological capabilities. More complex products require more capabilities to be produced and, for this reason, they are less likely to be produced by developing countries whose productive system has a limited range of capabilities. This idea is described more in detail and supported by a toy model in~\cite{cristelli2013measuring}.

The metric's assumptions can be represented by the following iterative set of equations
\begin{equation}
\label{eq:FC}
\begin{split}
F_i^{(n)} &= \sum_{\alpha}B_{i\alpha} Q_\alpha^{(n-1)},\\
Q_\alpha^{(n)} &= \frac{1}{\sum_{i} B_{i\alpha}/F_i^{(n-1)}} 
\end{split}
\end{equation}  
where $F_i$ and $Q_{\alpha}$ represent the fitness of country $i$ and the complexity of product $\alpha$, respectively. After each iterative step $n$, both sets of scores $\{F_{i}^{(n)}\}$ and $\{Q_{\alpha}^{(n)}\}$ are further normalized by their average value $\overline{F^{(n)}}$ and $\overline{Q^{(n)}}$, respectively. 

To understand the effects of non-linearity on product score, consider a product $\alpha_1$ with two exporters $i_1$ and $i_2$ whose fitness values are $0.1$ and $10$, respectively. Before normalizing the scores, product $\alpha_1$ achieves the score of $0.099$ which is largely determined by the score of the least-fit country. By contrast, a product $\alpha_2$ that is only exported by country $i_2$ achieves
the score of $10$ which is much higher than the score of product $\alpha_1$.
This simple example shows well the economic interpretation of the metric. First, if there is a low-score country that can export a given product, the complexity level of this product is likely to be low. By contrast, if only high-score countries are able to export the product, the product is presumably difficult to be produced and it should be thus assigned a high score. By replacing the $1/F_i$ terms in \eqref{eq:FC} with $1/F_i^{\gamma}$ ($\gamma>0$), one can tune the role of the least-fit exporter in determining the product score. As shown in~\cite{mariani2015measuring}, when the exponent $\gamma$ increases, the ranking of the nodes better reproduces their structural importance, but at the same time it is more sensitive to noisy data.

Variants of the algorithm have been proposed in order to penalize more heavily products that are exported by low-fitness countries \cite{mariani2015measuring, wu2016mathematics} and to improve the convergence properties of the algorithm \cite{stojkoski2016impact}. Importantly, the metric has been shown \cite{dominguez2015ranking,mariani2015measuring} to outperform other existing centrality metrics (such as
degree, method of reflections, PageRank, and betweenness centrality) in ranking the nodes according to
their importance for the structural stability of economic and ecological networks. In particular, in the case of ecological plant-pollinator networks \cite{dominguez2015ranking}, the ``fitness score'' $F$ of pollinators can be interpreted as their importance, and the ``complexity score'' $Q$ of plants as their vulnerability. When applied to plant-pollinator networks, the fitness-complexity algorithm (referred to as \emph{MusRank} by Dominguez et al. \cite{dominguez2015ranking}) reflects the idea that important insects pollinate many plants, whereas vulnerable plants are only pollinated by the most diversified -- ``generalists'' in the language of biology -- insects.

\subsection{Rating-based ranking algorithms on bipartite networks}
\label{sec:static_reputation}
In the current computerized society where individuals separated by hundreds or thousands kilometers who have never met each other can easily engage in mutual interactions, reputation systems are crucial in creating and maintaining the level of trust among the participants that is necessary for the system to perform well~\cite{resnick2000reputation,josang2007survey,pinyol2013computational}. For example, buyers and sellers in online auction sites are asked to evaluate each others' behavior and the obtained information is used to obtain their trustworthiness scores. User feedback is typically assumed in the form of ratings in a given rating scale, such as the common 1-5 star system where 1 star and 5 stars represent the worst and best possible rating, respectively (this scale is employed by the important e-commerce such as Amazon and Netflix, for example).

Reputation systems have been shown to help the buyers avoid fraud~\cite{gregg2006role} and the sellers with high reputation fare on average better than the sellers with low reputation~\cite{mcdonald2002reputation}. Note that the previously discussed PageRank algorithm and other centrality metrics discussed in Section~\ref{sec:eigenvector} also represent particular ways of assessing the reputation of nodes in a network. In that case, however, no explicit evaluations are required because links between nodes of the network are interpreted as implicit endorsements of target nodes by the source nodes.

The most straightforward way to aggregate the ratings collected by a given object is to compute their arithmetical mean (this method is referred to as \emph{mean} below). However, such direct averaging is rather sensitive to noisy information (provided by the users who significantly differ from the prevailing opinion about the object) and manipulation (ratings intended to obtain the resulting ranking of objects)~\cite{IEEEDM12422011,ACM6202009}. To enhance the robustness of results, one usually introduces a reputation system where reputation of users is determined along with the ranking of objects. Ratings by little reputed users are then assumed to have potentially adverse affect on the final results, and they are thus given corresponding low weights which helps to limit their impact on the system~\cite{masum2004manifesto,resnick2000reputation,josang2007survey}.


A particularly simple reputation system, \emph{iterative refinement} (IR), has been introduced by Laureti et al.~\cite{laureti2006information}. In IR, a user's reputation score is inversely proportional to the difference between the rating given by this user and the estimated quality values of the corresponding objects. The estimates of user reputation and object quality are iterated until a stable solution is found. We denote the rating given by user $i$ to object $\alpha$ as $r_{i\alpha}$ and the number of ratings given by user $i$ as $k_i=\sum_{\alpha} A_{i\alpha}$ where $A_{i\alpha}$ are elements of the bipartite network's adjacency matrix. The above-mentioned arithmetic mean can thus be represented as $Q_{\alpha}= \sum_i r_{i\alpha}A_{i\alpha} / k_{\alpha}$ where the ratings of all users are assigned the same weight. The iterative refinement algorithm~\cite{laureti2006information} assigns users weights $w_i$ and estimates the quality of object $\alpha$ as
\begin{equation}
\label{IR-Q}
Q_{\alpha}= \frac{\sum_i w_i r_{i\alpha} A_{i\alpha}}{\sum_i w_iA_{i\alpha}}
\end{equation}
where the weights are computed as
\begin{equation}
\label{IR-w}
w_i = \left(\frac{1}{k_i} \sum_{\alpha} A_{i\alpha}(r_{i\alpha} - Q_{\alpha})^2 + \varepsilon\right)^{-\beta}.
\end{equation}
Here $\beta\geq0$ controls how strongly are the users penalized for giving ratings that differ from the estimated object quality and $\varepsilon$ is a small parameter that prevents the user weight from diverging in the (unlikely) case when $r_{i\alpha}=Q_{\alpha}$ for all objects evaluated by user $i$.

Equations \eqref{IR-Q} and \eqref{IR-w} represent an interconnected system of equations that can be solved, similarly to HITS and PageRank, by iterations (the initial user weights are usually assumed to be identical, $w_i=1$). When $\beta=0$, IR simplifies to arithmetic mean. As noted by Yu et al.~\cite{yu2006decoding}, while $\beta = 1/2$ provides better numerical stability of the algorithm as well as translational and scale invariance, $\beta = 1$ is equivalent to a maximum likelihood estimate of object quality assuming that the individual ``rating errors'' of a user are normally distributed with unknown user-dependent variance. As shown by Medo and Wakeling~\cite{medo2010effect}, the IR's efficacy is importantly limited by the fact that most real settings use a limited range of integer values. Yu et al.~\cite{yu2006decoding} generalize the IR by assigning a weight to each individual rating.


There are two further variants of IR. The first variant is based on computing a user's weight as the correlation between the user's ratings and the quality estimates of the corresponding objects~\cite{Yanbo}. The second variant~\cite{liao2014ranking} is the same but it suppresses the weight of users who have rated only a few items, and further multiplies the estimated object quality with $\max_{i\in\mathcal{U}_i} w_i$ (\ie, the highest weight among the users who have rated a given object). The goal of this modification is to suppress the objects that have been rated by only a few users (if an item is rated well by one or two users, it can appear at the top of the object ranking which is in most circumstances not a desired feature).

\section{The impact of network evolution on static centrality metrics}
\label{sec:failure}
Many systems grow (or otherwise change) in time and it is natural to expect that this growth leaves an imprint on the metrics that are computed on static snapshots of these systems.
This imprint often takes form of a time bias that can significantly influence the results obtained with a metric: nodes may score low or high largely due to the time at which they entered the system.
For example, a paper published a few months ago is destined to rank badly by citation count because it had not have the time yet to show its potential by attracting a corresponding number of citations.
In this section, we discuss a few examples of systems where various forms of time bias exist and, importantly, discuss how the bias can be dealt with or even entirely removed.

\subsection{The first-mover advantage in preferential attachment and its suppression}
\label{sec:first-mover}
Node in-degree, introduced in paragraph \ref{sec:degree}, is the simplest node centrality metric in a directed network. It relies on the assumption that if a node has been of interest to many nodes that have  linked to it, the node itself is important. Node out-degree, by contrast, reflects the activity of a node and thus it cannot be generally interpreted as the node's importance. Nevertheless, node in-degree too becomes problematic as a measure of node importance when not all the nodes are created equal. Perhaps the best example of this is a growing network where nodes appear gradually. Early nodes then enjoy a two-fold advantage over late nodes: (1) they have more time to attract incoming links, (2) in the early phase of the network's growth, nodes face less competition because there are fewer nodes to chooser from. Note that any advantage given to a node is often further magnified by preferential attachment~\cite{barabasi1999emergence}, also dubbed as cumulative advantage, which is in effect in many real systems~\cite{jeong2003measuring,redner2005citation}.

\begin{figure*}
\centering
\includegraphics[scale = 0.7]{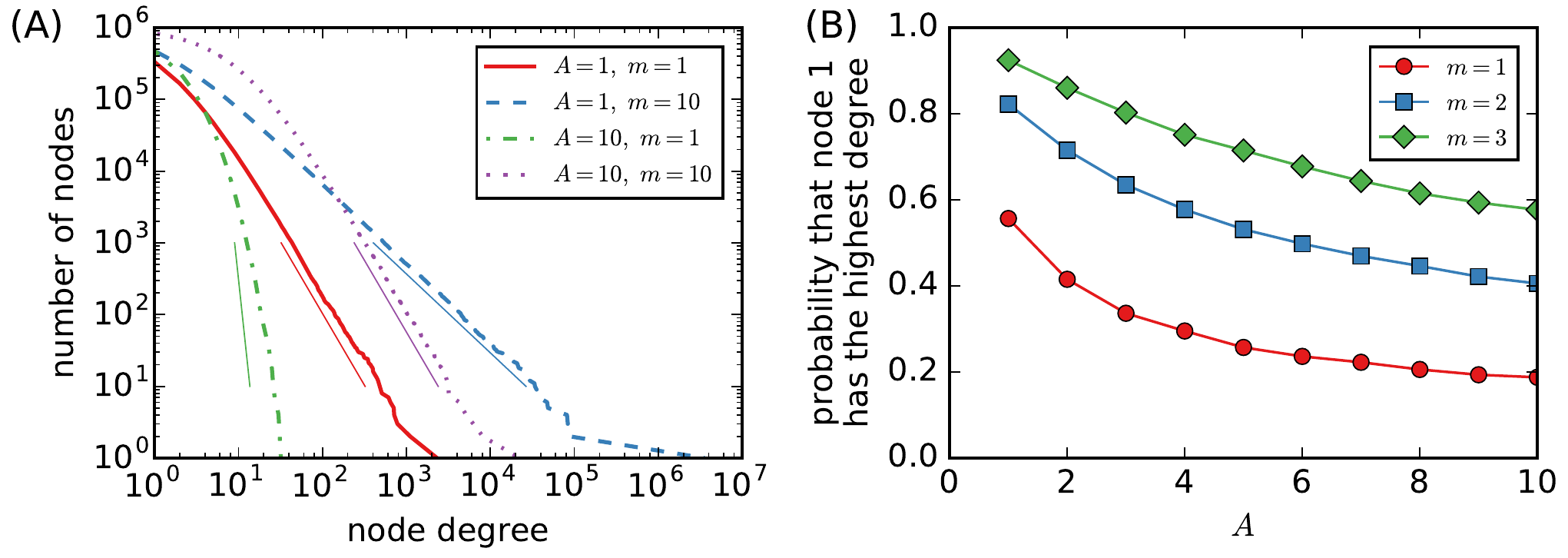}
\caption{(A) Cumulative node degree distributions of directed preferential attachment networks grown from a single node to $10^6$ nodes at different parameter values. Indicative straight lines have slope $1 + C/m$ predicted for the cumulative degree distribution by theory. (B) The probability that the first node has the largest in-degree in the network (estimated from 10,000 model realizations).}
\label{fig:PA}
\end{figure*}

We assume now the usual preferential attachment (PA) setting where at every time step, one node is introduced in the system and establishes directed links to a small number $m$ of existing nodes. The probability of choosing node $i$ with in-degree $k_i$ is proportional to $k_i + C$. Nodes are labeled with their appearance time; node $i$ has been introduced at time $i$.
The continuum approximation (see \cite{krapivsky2001organization} and Section VII.B in~\cite{albert2002statistical}) is now the simplest approach to study the evolution of the node mean in-degree $\avg{k}_i(n)$ [here $n$ is the time counter that equals the current number of nodes in the network, $n$]. The change of $\avg{k}_i(n)$ in time step $t$ is equal to the probability of receiving a link at time $t$, implying
\begin{equation}
\label{cont_approach}
\frac{\dd\avg{k}_i(n)}{\dd n} = m\,\frac{\avg{k}_i(n) + C}{\sum_j k_j(n) + C}
\end{equation}
where the multiplication with $m$ is due to $m$ links being introduced in each time step. Since each of those links increases the in-degree of one node by one, the denominator follows a deterministic trajectory and we can write $\sum_j [k_j(n) + C] = mn + Cn$. The resulting differential equation for $\avg{k}_i(n)$ can be then easily solved to show that the expected in-degree of node $i$ after introducing $n$ nodes, $\overline{k_i(n)}$, is proportional to $C[(i/n)^{-1/(1+C/m)}-1]$. Here $t:=i/n$ can be viewed as the ``relative birth time'' of node $i$. Newman~\cite{newman2009first} used the master equation formalism~\cite{dorogovtsev2002evolution} to find the same result as well as to compute higher moments of $k_i(n)$, and the divergence of $\overline{k_i(n)}:=\overline{k(t)}$ for $t\to0$ (corresponding to early papers in the thermodynamic limit of system size) has been dubbed as the ``first-mover advantage''. The generally strong dependence of $\overline{k(t)}$ on $t$ for any network size $n$ implies that only the first handful of nodes have the chance to become exceedingly popular (see Figure~\ref{fig:PA} for an illustration).

To counter the strong advantage of early nodes in systems under influence of the preferential attachment mechanism, Newman \cite{newman2009first} suggests to quantify a node's performance by computing the number of standard deviations by which its in-degree exceeds the mean for the node's appearance 
time (a quantity of this kind is commonly referred to as the $z$-score (see paragraph~\ref{sec:rescaled} for a general discussion of this approach).
The $z$-score can be thus used to identify the nodes that ``beat'' PA and attract more links
than one would expect based on their time of arrival. According to the intermediate  results provided by Newman \cite{newman2009first}, the score of a paper that appeared at time $t=i/n$ and attracted $k$ links reads
\begin{equation}
\label{z-theory}
z(k, t) = \frac{k - \mu(t)}{\sigma(t)} = \frac{k - C(t^{-\beta}-1)}{\big[Ct^{-2\beta}(1-t^{\beta})\big]^{1/2}}
\end{equation}
where $\beta = m / (m + C)$. Parameters $m$ and $C$ are to be determined from the data (in the case of the APS citation data analyzed by Newman~\cite{newman2009first}, the best fit of the overall degree distribution is obtained with $C=6.38$ and $m=22.8$, leading to $\beta=0.78$; note that the value of $C$ is substantially larger than the often-assumed $C=1$). While this is not clear from the published manuscript~\cite{newman2009first}, the preprint version available at \url{https://arxiv.org/abs/0809.0522} makes it clear that paper $z$ scores are actually obtained by evaluating $\mu(t)$ and $\sigma(t)$ empirically by considering ``a Gaussian-weighted window of width 100 papers around the paper of interest''. 
The use of empirically observed values of $\mu(t)$ and $\sigma(t)$ rather than the expected value under the basic preferential attachment model is preferable due to the limitation of the model in describing the growth of real citation networks. 
In particular, the growth of a network under preferential attachment has been shown to depend heavily and permanently on the initial condition~\cite{berset2013effect,fotouhi2013network}, and the original preferential attachment model has been extended by introducing node fitness~\cite{bianconi2001competition,caldarelli2002scale} and node
aging in particular~\cite{amaral2000classes,adamic2000power,hajra2005aging,medo2011temporal} as two essential additional driving forces that shape the network growth. Differently from Eq.~\eqref{z-theory}, which relies on rather simplistic model assumptions, the $z$ score built on empirical quantities is  independent of the details of the network's evolution. There is just one parameter to determine: the size of the window used to compute $\mu(t)$ and $\sigma(t)$.

The resulting $z$ score was found to perform well in the sense of selecting the nodes with arrival times rather uniformly distributed over the system's lifespan, and the existence of those papers is said to provide ``a hopeful sign that we as scientists do pay at least some attention to good papers that come along later'' \cite{newman2009first}. The selected papers have been revised by Newman~\cite{newman2014prediction} five years later  and they have been found to outperform randomly drawn control groups with the same prior citation counts. Since a very recent paper can score high on the basis of a single citation, a minimum citation count of five has been imposed in the analysis. An analogous approach is used in the following paragraph to correct the bias of PageRank. A general discussion of rescaling techniques for static centrality metrics is presented in Section~\ref{sec:rescaled}.

\subsection{PageRank's temporal bias and its suppression}
\label{sec:pr_bias}
As explained in Section~\ref{sec:eigenvector}, the PageRank score of a node
can be interpreted as the probability that a random walker is found at 
the node during its hopping on the directed network. The network's
evolution strongly influences the resulting score values. For example, 
a recent node that had little time to attract incoming links is
likely to achieve a low score that improves once the node is 
properly acknowledged by the system. The  corresponding bias of 
PageRank against recent nodes has been documented in the World Wide
Web data, for example~\cite{baeza2004web}. Such bias can potentially 
override the useful information generated by the algorithm and render 
the resulting scores useless or even misleading. Understanding PageRank's
biases and the situations in which they appear is therefore essential for the use of the algorithm in practice.

While PageRank's bias against recent nodes can be understood as 
a natural consequence of these nodes still lacking the links that they will have the chance to collect in the future, structural features of the network can further pronounce the bias. The network of citations among scientific papers provides a particularly striking 
example because a paper can only cite papers that have been published in 
the past. All links in the citation network thus aim back in time. This particular network feature makes it hard for random walkers on the network to reach recent nodes (the only way this can happen is through the PageRank's teleportation mechanism) and ultimately results in a  strong bias towards old nodes that has been reported in the literature~\cite{chen2007finding,maslov2008promise}.

\begin{figure}
\centering
\includegraphics[scale=0.67]{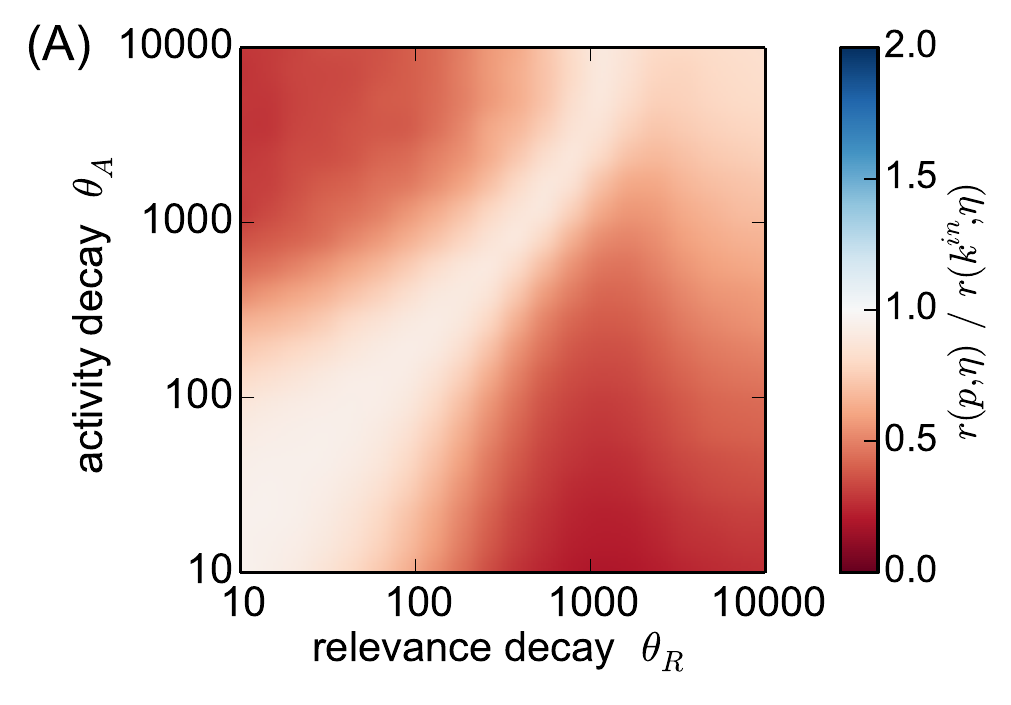}\quad\includegraphics[scale=0.67]{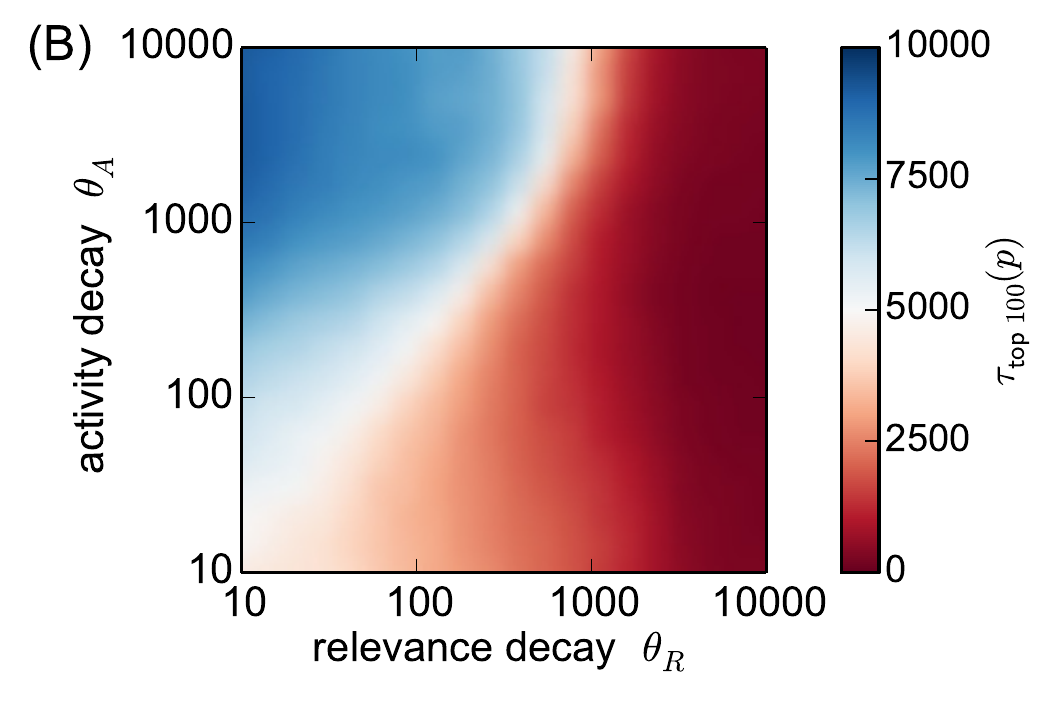}
\caption{Performance of PageRank in growing preferential attachment networks with 10,000 nodes where both node relevance and node activity decay. (A) After computing PageRank for the resulting networks, one can evaluate the correlation between node PageRank score and node fitness as well as the correlation between node in-degree and node fitness. This panel shows that when $\theta_A\gg\theta_R$ or $\theta_A\ll\theta_R$, PageRank correlates with node fitness less than in-degree (\emph{i.e.}, PageRank uncovers node properties worse than in-degree). (B) The average appearance time of top 100 nodes in the ranking by PageRank, $\tau_{\text{top 100}}(p)$, shows that PageRank strongly favors recent nodes when $\theta_A\gg\theta_R$ and it favors old nodes when $\theta_A\ll\theta_R$. The parameter regions where PageRank is strongly biased are those where it is outperformed by in-degree. (Adapted from~\cite{mariani2015ranking}.)}
\label{fig:PR_bias}
\end{figure}

Mariani et al.~\cite{mariani2015ranking} parametrized the space of network features 
by assuming that the decay of node relevance (which determines the rate at 
which a node receives new in-coming links) and the decay of node activity (which
determines the rate at which a node creates new out-going links) occur at 
generally different time scales $\varTheta_R$ and $\varTheta_A$, respectively.
Note that this parametrization is built on the fitness model with aging
that has been originally developed to model citation data~\cite{medo2011temporal}. 
The main finding by Mariani et al.~\cite{mariani2015ranking} is that when the two time scales are
of the similar order, the average time span the links that aim forward and backward in time is approximately the same and the random walk that underlies the PageRank algorithm can thus uncover
some useful information (see Figure~\ref{fig:PR_bias} for an illustration). 
When $\varTheta_R\gg\varTheta_A$, the aforementioned bias towards old nodes 
emerges and PageRank captures the intrinsic node fitness worse than the
simple benchmark in-degree count. PageRank is similarly outperformed by
node in-degree when $\varTheta_R\ll\varTheta_A$ for the very opposite reason as before: PageRank than favors recent nodes over the old ones.

Having found that the scores produced by the PageRank algorithm may
be biased by node age, the natural question that emerges is
as to whether the bias can be somehow removed. While several modifications of the PageRank algorithm have been proposed
to attenuate the advantage of old nodes (they will be the main subject of section \ref{sec:time}),
a simple method to effectively suppress the bias is to compare each node's score with the scores of nodes of similar age, 
in the same spirit as the $z$-score for citation count introduced in the previous paragraph.
This has been addressed by~\cite{mariani2016identification} where the authors study the
directed network of citations among scientific papers---a system
where the bias towards the old nodes is particularly strong\footnote{In the context of the model discussed in the previous paragraph, $\varTheta_A = 0$ in the citation network because a paper can establish links to other papers only at the moment of its appearance; it thus holds that $\varTheta_R\gg\varTheta_A$ and the bias towards old nodes is indeed expected.}.
Mariani et al.~\cite{mariani2016identification} propose to compute the PageRank score of all papers and then rescale the 
scores by comparing the score of each paper with the scores of other papers 
published short before and short after. In agreement with the findings
presented by Parolo et al.~\cite{parolo2015attention}, the window of papers included in the 
rescaling of a paper's score is best defined on the basis of the number of 
papers in the window (the authors use the window of 1000 papers).
Denoting the PageRank score of paper $i$ as $p_i$ and mean and 
standard deviation of PageRank scores in the window around 
paper $i$ as $\mu_i(p)$ and $\sigma_i(p)$, respectively, the proposed rescaled score of node $i$ reads
\begin{equation}
R_i(p) = \frac{p_i - \mu_i(p)}{\sigma_i(p)}.
\end{equation}
Note that unlike the input PageRank scores, the rescaled score can also be negative;
this happens when a paper's PageRank score is lower than the average for the 
papers published in the corresponding window. Quantities $\mu_i(p)$ and $\sigma_i(p)$ depend 
on the choice of window size; this dependence is however rather weak.

\begin{figure}
\centering
\includegraphics[scale=0.67]{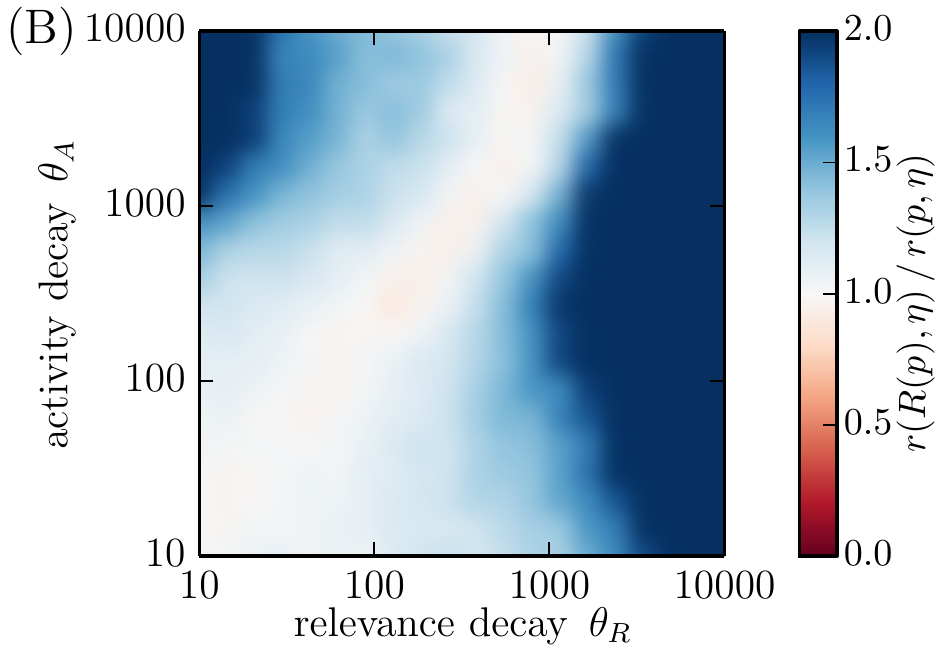}\quad\includegraphics[scale=0.67]{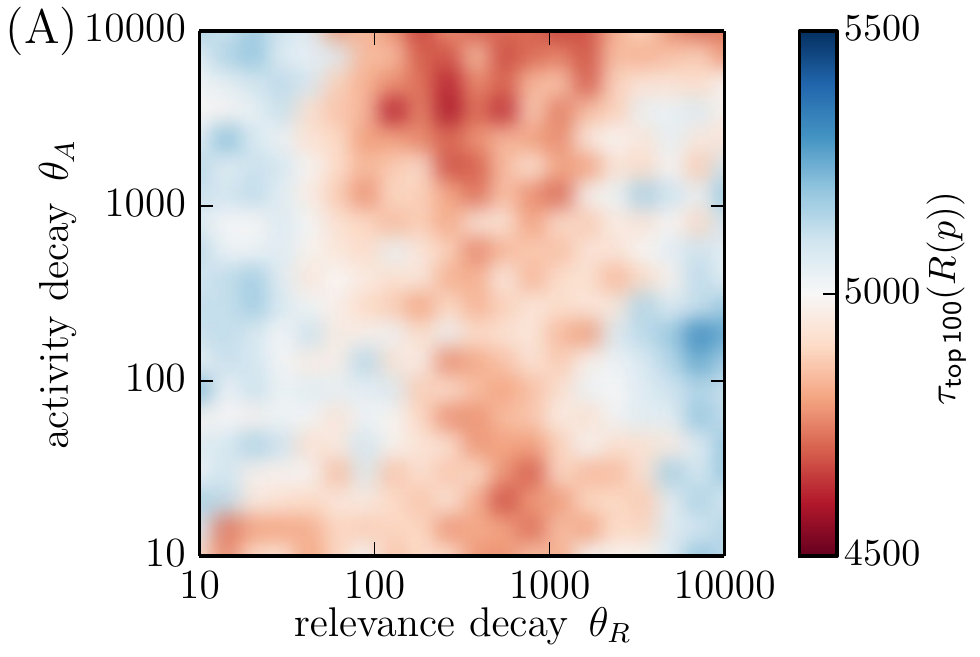}
\caption{Performance of rescaled PageRank $R(p)$ in artificial growing networks with 10,000 nodes where both node relevance and node activity decay.
(A) The average age of the top 100 nodes as ranked by rescaled PageRank, $\tau_{\text{top 100}}(R(p))$. The small range of values (note the color scale range and compare with Figure~\ref{fig:PR_bias}B) confirms that the age bias of PageRank is effectively removed by rescaling.
(B) Rescaled PageRank outperforms PageRank in uncovering node fitness in most of the investigated parameter space.}
\label{fig:PR_bias_removal}
\end{figure}

\begin{figure}
\centering
\includegraphics[scale=0.8]{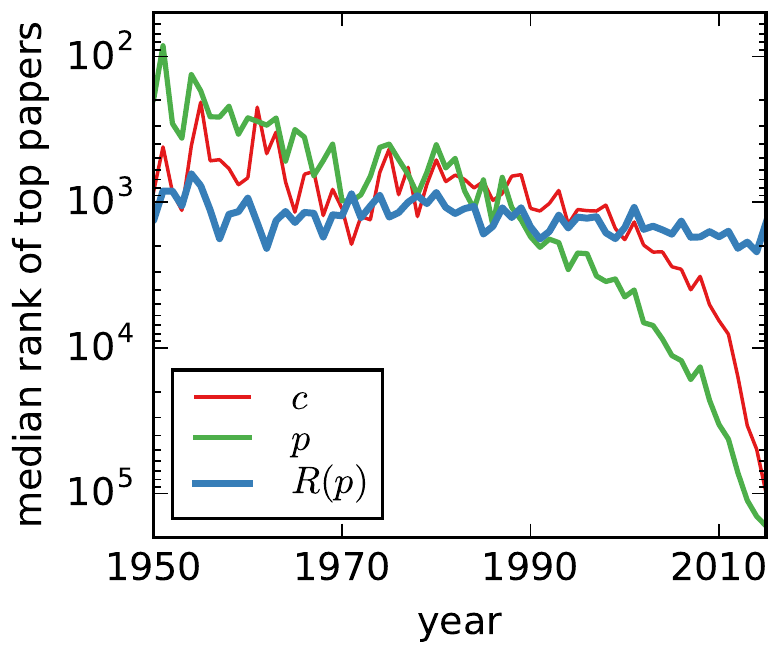}
\caption{For the APS citation data from the period 1893--2015 (560,000 papers in total), we compute the ranking of papers according to various metrics---citation count $c$, PageRank centrality $p$ (with the teleportation parameter $\alpha=0.5$), and rescaled PageRank $R(p)$. The figure shows the median ranking position of the top 1\% of papers from each year. The three curves show three distinct patterns. For $c$, the median rank is stable until approximately 1995; then it starts to grow because the best young papers have not yet reached sufficiently high citation counts. For $p$, the median rank grows during the whole displayed time period because PageRank applied on a time-ordered citation network results in advantage that grows with paper age. By contrast, the curve is approximately flat for $R(p)$ during the whole period which confirms that the metric is not biased by paper age and gives equal chances to all papers.}
\label{fig:APS_bias}
\end{figure}

Section~\ref{sec:rescaled} provides more details on this new metric, while its application to the identification of milestone papers in scholarly citation data is presented in paragraph~\ref{sec:milestones}. In addition to these results and the results presented in~\cite{mariani2016identification}, we illustrate here the metric's ability to actually remove the time bias in the model data that were already used to produce Figure~\ref{fig:PR_bias}. The results are presented in Figure~\ref{fig:PR_bias_removal}.
Panel A shows that the average age of the top 100 nodes as ranked by $R(p)$ is 
essentially independent of model parameters, especially when compared with the
broad range of $\tau_{\text{top 100}}(p)$ in Figure~\ref{fig:PR_bias}B. Panel
B compares the correlation between the rescaled PageRank and node fitness with that between PageRank and node fitness. In the region $\theta_R\approx\theta_A$, where we have seen before that PageRank is not biased toward nodes of specific age, there is no advantage to gain by using rescaled PageRank. There are nevertheless extended regions (the top left corner and the right side of the studied parameter range) where the original PageRank is heavily biased and the rescaled PageRank is thus able to better reflect the intrinsic node fitness. One can conclude that the proposed rescaling procedure removes time bias in model data and that this removal improves the ranking of nodes by their significance. The bias of the citation count and PageRank in real data as well as its removal by rescaled PageRank are exemplified in Figure~\ref{fig:APS_bias}. The benefits of removing the time bias in real data are presented in Section~\ref{sec:milestones}.

\subsection{Illusion of influence in social systems}
\label{sec:illusion}
We close with an example of an edge ranking problem where the task is to rank connections in a social network by the level of social influence that they represent.
In the modern information-driven era, the study of how information and opinions 
spread in the society is as important as it ever was. The research of
social influence is very active with networks playing an 
important role~\cite{marsden1993network,kempe2003maximizing,tang2009social,cha2010measuring}.
A common approach to assess the strength of social influence in a social network where every user is connected with their friends is based on measuring the probability that user $i$ collects item $\alpha$ if $f_{i\alpha}$ of $i$'s friends have already collected it~\cite{leskovec2007dynamics,cha2009measurement,steeg2011stops}. Here $f_{i\alpha}$ is usually called \emph{exposure} as it quantifies how much a user has been exposed to an item (the underlying assumption is that each friend who has collected item $\alpha$ can tell user $i$ about it). Note that the described setting can be effectively represented as a multilayer network~\cite{boccaletti2014structure} where users form a (possibly directed) social network and, at the same time, participate in bipartite user-item network which captures their past activities.

However, the described measurement is influenced by preferential attachment that is present in many social and information networks. Even when no social influence takes place and thus the social network has no influence on the items collected by individual users, the probability that a user has collected item  $\alpha$ is $k_{\alpha} / U$ where $U$ is the total number of users. Denoting the number of friends of user $i$ as $f_i$ (this is equal to the degree of user $i$ in the social network), there are on average $f_ik_{\alpha} /U$ friends who have collected item $\alpha$. At the same time, if preferential attachment is in effect in the system, the probability that user $i$ himself collects item $\alpha$ is also proportional to the item degree $k_{\alpha}$. Positive correlation between the likelihood of collecting an item and exposure is therefore bound to exist even when no real social influence takes place.

The normalized exposure
\begin{equation}
n_{i\alpha} = \max_{t<t_{i\alpha}}\bigg(\frac{f_{i\alpha}(t)}{f_i k_{\alpha}(t)}\,k_{\text{min}}\bigg)
\end{equation}
was proposed by Vidmer et al.~\cite{vidmer2015unbiased} to solve this problem. Here $k_{\alpha}(t)$ is the degree of item $\alpha$ at time $t$, $t_{i\alpha}$ is the time when user $i$ has collected item $\alpha$ (if user $i$ has not collected item $\alpha$ at all, the maximum is taken over all $t$ values), and $k_{\text{min}}$ is the minimum required item degree (without this constraint, items with low degree can reach high normalized exposure through normalization with $k_{\alpha}(t)$; such estimates based on a few events are noisy and thus little useful). Social influence measurements based on exposure and normalized exposure are compared on both real and model data by Vidmer et al.~\cite{vidmer2015unbiased}. When standard exposure is used, real data show social influence even when they are randomized---a clear indication that the bias introduced by the presence of preferential attachment is too strong to be ignored. By contrast, the positive correlation between item collection probability and normalized exposure disappears when the input data are normalized. In parallel, when no social influence is built in the model data, item collection probability is correctly independent of normalized exposure whereas it grows markedly with standard exposure. By taking time into account and explicitly factoring out the influence of preferential attachment, normalized exposure avoids the problems of the original exposure metric. In the future, this new metric could be used to rank items in order of likelihood to be actually consumed by a given user.

\section{Time-dependent node centrality metrics}
\label{sec:time}
In the previous section, we have presented important shortcomings of static metrics when applied to evolving systems, and score-rescaling methods to suppress these biases.
However, a direct rescaling of the final scores is not the only way to deal with the shortcomings of static metrics.
In fact, a substantial amount of research has been devoted to ranking algorithms that explicitly include time in their defining equation.
These time-dependent algorithms mostly aim at suppressing the advantage of old
nodes and thus improving the opportunity to recent nodes to acquire visibility.
Assuming that the network is unweighted, time-dependent ranking algorithms take as input the adjacency matrix $\mat{A}(t)$ of the network at a given time $t$ and information on node age and/or edge time-stamp. They can be classified into three (not all-inclusive) categories: (1) node-based rescaled scores (paragraph \ref{sec:rescaled}); (2) metrics that contain an explicit penalization
(often of an exponential or a power-law form) for older nodes or older edges (paragraph \ref{sec:explicit}); (3) metrics based on models of network growth that assume the existence of a latent fitness parameter, which represents a proxy for the node's success in the system (paragraph \ref{sec:model_based}).
Time-dependent generalizations of the static reputation systems introduced in paragraph \ref{sec:static_reputation} are also discussed in this section (paragraph \ref{sec:time_reputation}).


\subsection{Striving for time-balance: Node-based time-rescaled metrics}
\label{sec:rescaled}
As we have discussed in the previous section, network centrality metrics can be heavily biased by node age when applied to evolving networks. In other words, large part of the variation in node score is explained by node age, which may be an undesirable property of centrality score if we are interested in untangling the intrinsic fitness, or quality, of a node. In this paragraph we present metrics that explicitly require that node score is not biased by node age. These metrics simply take the original static centrality scores as input, and rescale them by comparing each node's score with only the scores of nodes of similar age.

\paragraph{Rescaling indegree}
In section \ref{sec:failure}, we have seen (Fig. \ref{fig:APS_bias}) that in real information networks, node indegree can be biased by age due to the preferential attachment mechanism.
This effect can be particularly strong for citation networks, as discussed in \cite{newman2009first, newman2014prediction} and in section \ref{sec:failure}. 
In~\cite{newman2009first}, the bias of paper citation count (node in-degree) in the scientific paper citation network is removed by 
evaluating the mean $\mu_i(c)$ and standard deviation $\sigma_i(c)$ of the citation count for papers published in a similar time as paper $i$. The systematic dependence of $c_i$ on paper appearance time is then removed by computing the rescaled citation count
\begin{equation}
R_{i}(c) = \frac{c_i-\mu_i(c)}{\sigma_i(c)}.
\label{rc}
\end{equation}
Note that $R_{c,i}$ represents the $z$-score of paper $i$ within its averaging window. Values of $R_c$ larger or smaller than zero indicate whether the paper is out- or under-performing, respectively, with respect to papers of similar age. There exists no principled criterion to choose the papers ``published in a similar time'' over which $\mu_i(c)$ and $\sigma_i(c)$ are computed. In the case of citation data, one can constrain this computation to a fixed number, $\Delta$, of papers published just before and just after paper $i$ \cite{mariani2016identification}, or weight papers score with a Gaussian function centered on the focal paper \cite{newman2009first}.
A statistical test in ref. \cite{mariani2016identification} shows that in the APS data, differently from the ranking by citation count, the ranking by rescaled citation count $R_c$ is not biased by paper age.


A different rescaling equation has been used in~\cite{radicchi2008universality, radicchi2011rescaling},
where the citation count of each scientific paper $i$ is divided by the average citation count of papers published in the same year and
belonging to the same scientific domain as paper $i$.
The resulting rescaled score, called $c_f$ in~\cite{radicchi2008universality}, 
has been found to follow a universal lognormal distibution -- independently of paper field and age -- in the Web of Science citation data.
This finding has been challenged by subsequent studies \cite{albarran2011skewness, waltman2012universality,vaccario2017quantifying}, which leaves it open whether an unbiased ranking of academic publications can be achieved through a simple rescaling equations. 
Since this review mostly focus on temporal effects, we refer the interested reader to the review article by Waltman \cite{waltman2016review} for details on the different field-normalization procedure introduced in the bibliometrics literature.
As there are various possibilities how to rescale indegree, one could also devise a reverse-engineering approach:
instead of using a rescaling equation and verifying whether it leads to an unbiased ranking of the nodes, one can
require that the scores of papers of different age follow the same distribution and then infer the correct rescaling equation \cite{radicchi2012reverse}.
However, this procedure has been only applied (see \cite{radicchi2012reverse} for details) to suppress the bias by field of papers published in the same year, and it remains unexplored for the time-bias suppression problem.

\paragraph{Rescaling PageRank score}
Differently from the indegree's bias towards old nodes, PageRank can develop a temporal bias towards either old or recent nodes (see paragraph~\ref{sec:pr_bias} and \cite{mariani2015ranking}) 
As shown in \cite{mariani2016identification} and illustrated here in Figure~\ref{fig:PR_bias_removal}, the algorithm's temporal bias can be suppressed with the equation
\begin{equation}
R_{i}(p) = \frac{p_i-\mu_i(p)}{\sigma_i(p)},
\label{rp}
\end{equation}
which is analogous to Eq.~\eqref{rc} above; by assuming that the nodes are labeled in order of decreasing age,
$\mu_i$ and $\sigma_i$ are computed over a "moving" temporal window $[i-\Delta/2, i+\Delta/2]$ centered on paper $i$.
Also in the APS data, differently from the ranking by PageRank, the ranking by rescaled PageRank $R_c$ is not biased by paper age \cite{mariani2016identification}. 
The performance of rescaled PageRank and rescaled indegree in identifying groundbreaking papers will be discussed in detail 
and compared with the performance of other ranking methods in paragraph \ref{sec:milestones}. Vaccario et al. \cite{vaccario2017quantifying} generalized this rescaling procedure based on the $z$-score to suppress both bias by age and bias by field of scientific papers' PageRank score. 

We emphasize that in principle the score by any structural centrality metric can be rescaled with an equation akin to Eq. \eqref{rp}.
The only parameter of the rescaled score is the size $\Delta$ of the temporal window used for the calculation of $\mu_i$ and $\sigma_i$; Mariani et al. \cite{mariani2016identification} point out that too large $\Delta$ values would result in a ranking biased towards old nodes in a similar way as the ranking by PageRank\footnote{The ranking by PageRank score and the ranking by $R(p)$ are the same for $\Delta=N$.}, whereas too small $\Delta$ values would result in a ranking heavily dependent on statistical fluctuations -- a node could score high just by happening to have only low-degree papers in its small temporal window. One should always avoid both extremes when using a rescaled metric; however, a statistically-grounded guideline to choose $\Delta$ is still lacking.


\subsection{Metrics with explicit penalization for older edges and/or nodes}
\label{sec:explicit}
We now consider variants of degree (paragraph \ref{sec:time_degree}) and of eigenvector-based centrality metrics (paragraphs \ref{sec:ecm}-\ref{sec:citerank}) where edges and/or nodes are weighted according to their age directly in the equation that defines node score.
Hence, differently from rescaled metrics that normalize \emph{a posteriori} the scores by static centrality metrics, the metrics presented in this paragraph directly include time information in the score calculation.

\subsubsection{Time-weighted degree and its use in predicting future trends}
\label{sec:time_degree}
Indegree (or degree) is the simplest method to rank nodes in a network.
As we have seen in section \ref{sec:first-mover}, node degree is biased by age in growing networks
that exhibit preferential attachment.
As a consequence, degree can fail in individuating the nodes that will become popular in the future \cite{zeng2013trend},
since node preferences may shift over time \cite{koren2010collaborative} and node attractiveness may decay over time \cite{medo2011temporal}.
One can consider a variant of degree where received links are weighted by a function of the edge age $\Delta t_{ij}:=t-t_{ij}$, where $t$ and $t_{ij}$ denote the time at which the scores are computed and the time at which the edge from $j$ to $i$ was created, respectively.
The weighted-degree score $s_{i}(t)$ of node $i$ at time $t$ can be defined as
\begin{equation}
s_{i}(t) = \sum_{j} A_{ij}(t)\,f(t-t_{ij}),
\label{weighted_degree}
\end{equation}
where $f(\Delta t)$ is a function of edge age $\Delta t_{ij}= t-t_{ij}$.
Zeng et al. \cite{zeng2013trend} set $f(\Delta t)=1-\lambda\, \Theta(\Delta t - T_r)$, where $\Theta(\cdot)$ denotes the Heaviside function -- $\Theta(x)$ is equal to one if $x\geq 0$, zero otherwise -- and $T_r$ is a parameter of the method
which implies that the resulting score $s^{rec}_{i}(t;T_r)$ (referred to as Popularity-Based Predictor PBP in \cite{zeng2013trend, zhou2015temporal})
is a convex combination of node degree $k_{i}(t)$ and node recent degree increase $\Delta k_{i}(t,T_r)=\sum_{j}A_{ij}\,\Theta(T_r-(t-t_{ij}))$,
\begin{equation}
s^{rec}_{i}(t;T_r)=k_{i}(t)-\lambda\,k_{i}(t-T_r)=(1-\lambda)\,k_{i}(t)+\lambda\,\Delta k_{i}(t,T_r).
\label{delta_k}
\end{equation}
The parameter $T_r$ specifies the length of the ``time window'' over which the recent degree increase is computed.
Note that $s^{rec}_{i}$ reduces to node degree or recent degree increase for $\lambda=0$ or $\lambda=1$, respectively.
Closely similar exponential penalization is considered by Zhou et al.~\cite{zhou2015temporal}, leading to
\begin{equation}
s_{i}^{exp}(t)=\sum_{j}A_{ij}(t)\,e^{-\gamma(t-t_{ij})}
\label{tbp}
\end{equation}
which is referred to as Temporal-Based Predictor in \cite{zeng2013trend, zhou2015temporal}; here $\gamma>0$ is a parameter of the method which is analogous to $T_r$ before.
Results obtained with time-weighted degree scores $s^{rec}$ and $s^{exp}$ on data from Netflix, Movielens and Facebook
show that: (1) $s^{rec}$ with $\lambda$ close to one performs better than degree
in identifying the most popular nodes in the future \cite{zeng2013trend}; (2) this performance is further improved by $s^{exp}$ 
(Fig. 3 in \cite{zhou2015temporal}).
Metric $s^{exp}$ (albeit with a different parameter calibration) is also used in \cite{ghosh2011time},
where it is referred to as the Retained Adjacency Matrix (RAM).
$s^{exp}$ is shown in \cite{ghosh2011time} to outperform indegree, PageRank and other static centrality metrics in ranking scientific papers according to their future
citation count increase and in identifying the top-cited papers in the future.
The performance of $s^{exp}$ is comparable to that of a time-dependent variant of Katz centrality, called Effective Contagion Matrix (ECM), that is discussed below (paragraph \ref{sec:ecm}).

Another type of weighted degree is the \emph{long-gap citation count} \cite{wasserman2015cross} which measures the number of edges received by a node
when it is at least $T_{min}$ years old.
Long-gap citation count has been applied to the network of citations between American movies and has been shown to be
a good predictor for the presence of movies in the list of milestone movies edited by the National Film Registry (see \cite{wasserman2015cross} for details).



\subsubsection{Penalizing old edges: Effective Contagion Matrix} 
\label{sec:ecm}
Effective Contagion Matrix is a modification of the classical Katz centrality metric where the paths that connect a node with another node 
are not only weighted according to their length, but also according to the ages of their links \cite{ghosh2011time}.
For a tree-like network (like citation networks), one defines a modified adjacency matrix $R(t,\gamma)$
(referred to as retained adjacency matrix by Ghosh et al.
\cite{ghosh2011time}) whose elements $R_{ij}(t,\gamma)=A_{ij}(t)\,\gamma^{\Delta t_{ij}}$, where $\Delta t_{ij}$ is the age of the edge $j\to i$ at the time $t$ when the scores are computed.
The aggregate score $s_{i}$ of node $i$ is defined as
\begin{equation}
 s_{i}(t)=\sum_{k=0}^{\infty} \alpha^{k}\, \sum_{j=1}^{N}[\mat{R}(t,\gamma)^{k}]_{ij}.
 \label{ecm}
\end{equation}
According to this definition, paths of length $k$ from thus have a weight which is given not only by $\alpha^{k}$ (as
in the Katz centrality), but also depends on the ages of the edges that compose the path.
By using the geometric series of matrices, and denoting by $\vek{e}$ the vector whose components are all equal to one, one can rewrite Eq. \eqref{ecm} as 
\begin{equation}
\vek{s}(t)=\sum_{k=0}^{\infty} \alpha^{k}\, \mat{R}(t,\gamma)^{k}\,\vek{e}=
(1-\alpha \,\mat{R}(t,\gamma))^{-1}\,\vek{e},
\end{equation}
which results in
\begin{equation}
s_{i}(t)=\alpha\,\sum_{j}[\mat{R}(t,\gamma)]_{ij}\,s_{j}(t)+1=\alpha\,\sum_{j}A_{ij}(t)\,\gamma^{\Delta t_{ij}}\,s_{j}(t)+1,
\end{equation}
which immediately shows that the score of a node is mostly determined by the scores of the nodes that recently pointed to it.
ECM score has been found \cite{ghosh2011time} to outperform other metrics (degree, weighted indegree, PageRank, age-based PageRank and CiteRank -- see below for CiteRank's 
definition) in ranking
scientific papers according to their future citation count increase and in identifying papers that will become popular in the future;
the second best-performing metric is weighted indegree as defined in Eq.~\eqref{tbp}.
Ghosh et al. \cite{ghosh2011time} also note that the agreement between the ranking and the future citation count increase can be improved by
using the various metrics as features in a support vector regression model.
The analysis in \cite{ghosh2011time} does not include TimedPageRank (described below); in future research, it will be interesting to 
compare the predictive power of all the existing metrics, and compare it with the predictive power of network growth models~\cite{wang2013quantifying}.

\subsubsection{Penalizing old edges: TimedPageRank}
\label{sec:tpr}
Similarly to the methods presented in the previous paragraph, one can 
introduce time-dependent weights of edges in PageRank's definition.
A natural way to enforce this idea 
was proposed by Yu et al. \cite{yu2005adding} in the form\footnote{Note that since $\sum_{j}A_{ij}\,f(\Delta t_{ij})/k_{j}^{out}\leq 1$ upon this definition, the vector of scores $\vek{s}$ is not normalized to one. Eq.~\eqref{timedpr} thus defines a \emph{pseudo-PageRank problem}, which can be put in a directed correspondence with a PageRank problem as defined in paragraph \ref{sec:pr_variants} (see paragraph \ref{sec:pr_variants} and \cite{gleich2015pagerank} for more details).}
\begin{equation}
s_{i}(t)=c\,\sum_{j}A_{ij}(t)\,f(\Delta t_{ij}) \,\frac{s_{j}(t)}{k^{out}_{j}}+1-c,
\label{timedpr}
\end{equation}
where $\Delta t_{ij}$ denotes the age of the link between nodes $i$ and $j$.
Yu et al. \cite{yu2005adding} use Eq.~\eqref{timedpr} with $f(\Delta t)=0.5^{\Delta t}$,
and define node $i$'s TimedPageRank score
as the product between its score $s_i$ determined by Eq. \eqref{timedpr}
and a factor $a(t-t_i)\in[0.5,1]$ which decays with node age $t-t_i$. 
A similar idea of an explicit aging factor to penalize older nodes was also implemented by Baeza-Yates~\cite{baeza2004web}.
Yu et al. \cite{yu2005adding} used scientific publication citation data to show that the top $30$ papers by TimedPageRank are more cited in the future years than the top papers by PageRank.

\subsubsection{Focusing on a temporal window: T-Rank, SARA}
\label{sec:window}
T-Rank \cite{berberich2004t, berberich2005time} is a variant of PageRank whose original aim is to rank websites. The 
algorithm assumes that we are mostly interested in the events that happened in a particular temporal window of interest $[t_1,t_2]$.
For this reason, the algorithm favors the pages that received many incoming links and were often updated within the temporal window of interest.
Time-stamps are thus weighted according to their
temporal distance from $[t_1,t_2]$, leading to a ``freshness'' function $f(t)$ which
is equal to or smaller than one if the link's creation time lies within the range $[t_1,t_2]$ or outside, respectively.
If a timestamp $t$ does not belong to the temporal window, 
its freshness $f(t)$ monotonously decreases with its distance from the temporal window of interest.
The freshness of timestamps are then used to compute the freshness of nodes, of links and of page updates, which are
the elements needed for the computation of the T-Rank scores.
As the temporal window of interest can be chosen at will, the algorithm is flexible and can provide interesting insights into the web users' reactions after an unexpected event, such as a terror attack \cite{berberich2005time}.

A simpler idea is to use only links that belong to a certain 
temporal window to compute the score of a node.
This idea has been applied to the citation network of scientific authors to compute an author-level score for a certain time window using a weighted variant of PageRank called Science Author Rank Algorithm (SARA) \cite{radicchi2009diffusion}. The main finding by Radicchi et al. \cite{radicchi2009diffusion} is that the authors that won a Nobel Prize are better identified by SARA than by indices based on indegree (citation count). The dynamic evolution of the ranking position of Physics' researchers by SARA can be explored in the interactive Web platform \url{http://www.physauthorsrank.org/authors/show}. 


\begin{figure*}
\centering
\includegraphics[scale=0.67]{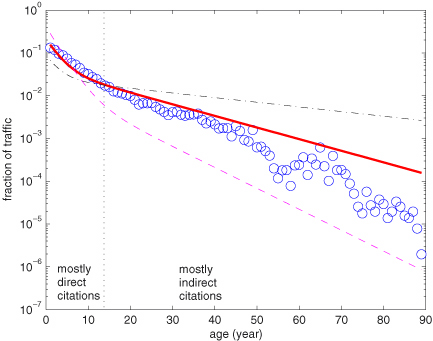}
\caption{A comparison of the age distribution of future incoming citations (red line) and of CiteRank score (blue dots). Both curves exhibit a two-step exponential decay. According to the analytic results by Walker et al. \cite{walker2007ranking}, this behavior can be interpreted in terms of two distinct citation mechanisms.
(Reprinted from~\cite{walker2007ranking}.)}
\label{fig:citerank}
\end{figure*}

\subsubsection{PageRank with time-dependent teleportation: CiteRank}
\label{sec:citerank}
In the previous paragraphs, we have presented variants of classical centrality metrics (degree, Katz centrality, and PageRank) where edges are penalized according to their age.
For the PageRank algorithm,
an alternative way to suppress the advantage of older nodes is to introduce an 
explicit time-dependent penalization of older nodes in the teleportation term.
This idea has led to the CiteRank algorithm which was introduced by Walker et al. \cite{walker2007ranking} to rank scientific publications.
The vector of CiteRank scores $\vek{s}$ can be 
found as the stationary solution of the following set of recursive linear equations\footnote{This definition is slightly different from that provided in \cite{walker2007ranking} -- the scores resulting from the definition adopted here are normalized to one and differ from those obtained with the definition in \cite{walker2007ranking} by a uniform normalization factor.}
\begin{equation}
s_{i}^{(n+1)}(t)=c\,\sum_{j}A_{ji}(t)\,\frac{s_{j}^{(n)}(t)}{k^{out}_{j}}+
(1-c)\,v(t,t_{i}) 
\label{cr}
\end{equation}
where, denoting by $t_{i}$ the publication date of paper $i$ and $t$ the time at which the scores are computed, we defined
\begin{equation}
v(t,t_{i})=\frac{\exp{(-(t-t_{i})/\tau)}}{\sum_{j=1}^{N}\exp{(-(t-t_{j})/\tau)}}.
\end{equation}
To choose the values of $c$ and $\tau$,
the authors
compare papers' CiteRank score with papers' future indegree increase $\Delta k^{in}$,
and find $c=0.5$, $\tau=2.6$ years as the optimal parameters. With this choice of parameters,
the age distribution of paper CiteRank score is in agreement with the age distribution of the papers' future indegree increase.
In particular, both distributions feature a two-step exponential decay (Fig. \ref{fig:citerank}).
A theoretical analysis based on CiteRank in~\cite{walker2007ranking} suggests that the two-step decay of $\Delta k^{in}$ can be explained by two distinct and co-existing citation mechanisms:
researchers cite a paper because they found it either directly or by following the references of a more recent paper. This example shows that well-designed time-dependent metrics are not only useful tools to rank the nodes, but can shed light into the behavior of the agents in the system.

\subsubsection{Time-dependent reputation algorithms}
\label{sec:time_reputation}
A reputation system ranks individuals by their reputation from the most to the least reputed one based on their past interactions or evaluations. Many reputation systems weight recent evaluations more than old ones and thus produce a time-dependent ranking. For example, the generic reputation formula presented by Sabater and Sierra~\cite{sabater2001regret} assigns evaluation $W_i\in[-1, 1]$ (here $-1$, $0$, and $+1$ represent absolutely negative, neutral, and absolutely positive evaluation, respectively) weight proportional to $f(t_i, t)$ where $t_i$ is the evaluation time and $t$ is the time when reputation is computed. The weight function $f(t_i, t)$ should be chosen to favor evaluations that were made short before $t$; a simple example of such a function is $f(t_i, t) = t_i / t$. Sabater and Sierra~\cite{sabater2001regret} proceed by analyzing an artificial scenario where a user first behaves reliable until reaching a high reputation value and then starts to commit fraud. Thanks to assigning high weight to recent actions, their reputation systems is able to quickly reflect the change in the user's behavior and correspondingly lower the user's reputation. Instead of weighting the evaluations on the basis of the time when they were made, it is also possible to order evaluations by their time stamps $t(W_1) < t(W_2) < \dots < t(W_N)$ and then assign weight $\lambda^{N-i}$ to evaluation $i$. This scheme is referred to as \emph{forgetting} as it corresponds to gradually lowering the weight of evaluations as more recent evaluations arrive~\cite{jsang2002beta}. When $\lambda=1$, no forgetting is effectively employed. When $\lambda=0$, the most recent evaluation alone determines the result. Note that the time information is also extensively used in otherwise time-unaware reputation systems to detect spammers and other anomalous and potentially malicious behavior~\cite{liu2010anomaly}.

\subsection{Model-based ranking of nodes}
\label{sec:model_based}
The ranking algorithms described up to now make assumptions that are regarded as plausible and rarely tested on real systems.
For instance, indegree assumes that a node is important, or central, if it is able to acquire many incoming links;
PageRank assumes that a node is important if it is pointed by other important nodes;
rescaled PageRank assumes that a node is important if its PageRank score exceeds PageRank scores of nodes of similar age.
To evaluate the degree to which these node centrality definitions are reliable, one can \emph{a posteriori} evaluate the performance of the ranking
algorithms in identifying high-quality nodes in the network or in predicting the future
evolution of the system (real-world examples will be discussed in section \ref{sec:applications}).

On the other hand, if one has a mathematical model which well describes how the network evolves, then one can assess node importance by simply fitting the model to the given data.
This section will review this approach in detail.
We will focus on preferential-attachment models where each node is endowed with a fitness parameter \cite{bianconi2001competition}, which represents
the perceived importance of the node by other nodes in the network.

\paragraph{Ranking based on the fitness model}
Bianconi and Barabasi~\cite{bianconi2001competition}, extended the original Barabasi-Albert model  by assigning to each node an intrinsic quality parameter, called \emph{fitness}.
Node fitness represents the nodes' inherent ability to acquire new incoming links.
In the model (hereafter \emph{fitness model}), 
the probability $\Pi_{i}(t)$ that a new link will be attached to node $i$ at time $t$ is given by 
\begin{equation}
\Pi_{i}(t)\sim k^{in}_{i}(t)\,\eta_i,
\end{equation}
where $\eta_i$ represents node $i$'s fitness.
This attachment rule leads to the following equation for the dependence of the expected node degree on time
\begin{equation}
\braket{ k^{in}_{i}(t) }=m\biggl(\frac{t}{t_{i}} \biggr)^{\beta(\eta_{i})},
\label{FM}
 \end{equation}
where if the support of the fitness distribution is bounded, $\beta(\eta)$ is solution of the self-consistent equation of the form $\beta(\eta)=\eta/C[\beta(\eta)]$ and $C[\beta]$ only depends on the chosen fitness distribution (see \cite{bianconi2001competition}
for the details). Remarkably, the model is flexible and can reproduce several degree distributions by appropriately choosing the functional form of the fitness distribution \cite{bianconi2001competition}.

Using Eq.~\eqref{FM}, we can estimate node fitness in real data. To do this, it is sufficient to follow the degree evolution $k^{in}(t)$ of the nodes
and properly fit $k^{in}(t)$ to a power-law to obtain the exponent $\beta$, which is proportional to node fitness.
This procedure has been applied by Kong et al. \cite{kong2008experience} to data from the WWW, leading to two interesting conclusions: (1) 
websites' fitness distribution is narrow; (2) fitness distribution is not dependent on time, 
and can be well approximated by an exponential distribution.

\begin{figure*}
\centering
\includegraphics[width=\textwidth]{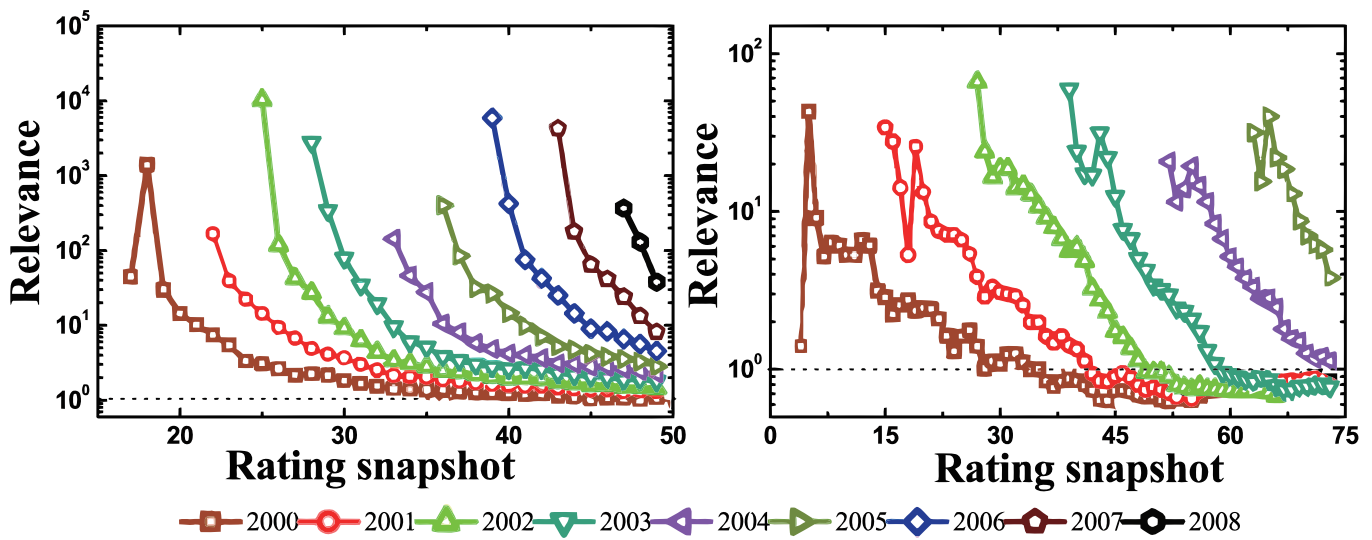}
\caption{An illustration of the relevance decay for movies posted in different
years (reprinted from~\cite{ren2016characterizing}). Time is divided into equal-duration 
time windows (``rating snapshots'' in the horizontal axis), and the average 
relevance $r(t)$ within each snapshot for movies published in a given year is 
calculated (vertical axis). The left panel shows the average relevance decay for 
MovieLens movies posted from $2000$ to $2008$; the right panel shows the same for 
Netflix movies posted between $2000$ to $2005$.} \label{fig:relevance_decay}
\end{figure*}

\paragraph{Ranking based on the relevance model}
While the fitness model describes the growth of information networks better than the original Barabasi-Albert model \cite{bianconi2001competition, kong2008experience},
it is still an incomplete description of degree evolution for many evolving systems. 
The missing element in the fitness model is node aging \cite{dorogovtsev2000evolution}: the ideas contained in scientific papers are incorporated in subsequent papers, reviews and
books, which results in a gradual decay of papers' attractiveness to new citations.
The \emph{relevance model} introduced by Medo et al. \cite{medo2011temporal} incorporates node aging by assuming 
\begin{equation} 
\Pi_{i}(t)\sim k^{in}_{i}(t)\,\eta_i\,f(t-t_i),
\label{RM}
\end{equation}
where $f(t-t_i)$ is a function of node age which tends to zero \cite{medo2011temporal, wang2013quantifying} or to a small constant value~\cite{medo2014statistical} for large age. For an illustration of node relevance decay in real data from Movielens and Netflix, see Fig.~\ref{fig:relevance_decay} and \cite{ren2016characterizing}. Assuming that the normalization factor of $\Pi_{i}(t)$ converges to a finite value $\Omega^*$, one can use Eq.~\eqref{RM} to show that
\begin{equation}
\braket{k^{in}(t)}\simeq \exp\Big[\eta_{i}\int_0^t f(t')\,\dd t' / \Omega^*\Big].
\label{relevance}
\end{equation}
The value of $\Omega^*$ depends on the distribution of node fitness and the aging function (see~\cite{medo2011temporal} for details).

Eq.~\eqref{RM} can be fitted to real citation data in different ways.
Medo et al.~\cite{medo2011temporal} computed the relevance $r_i(t)$ of node $i$ at time $t$ as the ratio between the fraction $l_{i}(t)$ of links obtained by node $i$ in a suitable time window and the fraction $l_{i}^{PA}(t)$ expected from pure preferential attachment. That is, one writes
\begin{equation}
r_{i}(t)=\frac{l_{i}(t)}{l_{i}^{PA}(t)}
\label{empirical_relevance}
\end{equation}
where $l_{i}^{PA}(t)=k_{i}^{(in)}(t)/\sum_{i}k^{(in)}_{j}(t)$.
A fitness estimate for node $i$ is then provided by the node's total relevance $T_{i}=\sum_{t}r_{i}(t)$;
similarly to the fitness distribution found in the WWW \cite{kong2008experience}, 
the total relevance distribution has an exponential decay in the APS citation data \cite{medo2011temporal}.
Alternatively, one can estimate node fitness through standard maximum likelihood estimation (MLE) \cite{medo2014statistical}.
Wang et al.~\cite{wang2013quantifying} used MLE to
analyze the citation data from the APS and Web of Science,
and found that the time-dependent factor $f(t)$ follows a universal lognormal form
\begin{equation}
f(t)=\frac{1}{\sqrt{2\pi t}\sigma_{i}}\exp\Biggl[-\frac{(\log{(t)}-\mu_i)^2}{2\,\sigma_i^2}\Biggr]
\label{fit_barabasi}
\end{equation}
where $\mu_i$ and $\sigma_i$ are parameters specific to the temporal citation pattern of paper $i$.
By fitting the relevance model with Eq. \eqref{fit_barabasi} to the data,
the authors estimate
paper fitness and use this estimate to predict the future
impact of papers (we refer to \cite{wang2013quantifying} and paragraph \ref{sec:papers} for more details).
Medo \cite{medo2014statistical} uses MLE to compare various network growth models with respect to their ability to fit data
from the Econophysics forum by Yi-Cheng Zhang [\url{www.unifr.ch/econophysics}], and the Relevance Model turns out to better fit the data with respect to other existing models.

\paragraph{Ranking based on a model of PageRank score evolution}
In the previous paragraph, we discussed various procedures to estimate node fitness based on the same general idea:
if we know the relation between node degree and node fitness, we can properly fit the relation to the data to infer the fitness values.
In principle, the same idea can find application for network centrality metrics other than degree.
For PageRank score $s$, for example, one can assume a growth model in the form \cite{berberich2006buzzrank} 
\begin{equation}
s_{i}(t)=\exp\biggl[\int_{0}^{t}dt\,\alpha_{i}(t)\biggr],
\label{buzz}
\end{equation}
where the behavior of the function $\alpha_{i}(t)$ has to be inferred from the data. 
Berberich et al. \cite{berberich2006buzzrank} assume that $\alpha_{i}(t)$ is independent of time within the system time span $[t_{min},t_{max}]$ and show that this assumption is reasonable in the academic citation network they analyzed.
With this assumption, Eq.~\eqref{buzz} becomes
\begin{equation}
s_{i}(t)=s_{i}(t_0)\,\exp\bigl[\alpha_{i}(t-t_{min})\bigr],
\label{buzz1}
\end{equation}
where $\alpha_{i}(t)=\alpha_i$ if $t\in[t_{min},t_{max}]$.
A fit of Eq.~\eqref{buzz} to the data allows us to estimate $\alpha_i$, which is referred to as the BuzzRank score of node $i$ Eq.~\eqref{buzz} is conceptually similar to Eq.~\eqref{relevance} for the RM: the growth of nodes' centrality score depends exponentially on a quantity -- node fitness $\eta$ in the case of the RM, node growth rate $\alpha$ in the case of BuzzRank -- that can be interpreted as a proxy for node's success in the system.

\section{Ranking nodes in temporal networks}
\label{sec:temporal}
In the previous sections, we have illustrated several real-world and model-based examples that show that the temporal ordering of interactions 
is a key element that cannot be neglected when analyzing an evolving network.
For instance, we have seen that some networks exhibit a first-mover advantage (paragraph \ref{sec:first-mover}), a diffusion process on a network can be strongly biased
by the network's temporal linking patterns (paragraph \ref{sec:pr_bias}), and neglecting the ordering of interaction may lead to
wrong conclusions when evaluating the social influence between two individuals (paragraph \ref{sec:illusion}).
Various approaches to solve the consequent drawbacks of static centrality metrics (such as indegree and PageRank) have been presented in section \ref{sec:time}.
In particular, we have introduced  metrics (such as the rescaled metrics and CiteRank) that take as input not only the network's adjacency matrix $\mat{A}$, but also the time stamps of the edges and/or the age of the nodes.
These approaches take into account the temporal dimension of the data.
They are useful for their ability to suppress the bias of static metrics (see section \ref{sec:failure}) and to highlight recent contents in growing networks (see section \ref{sec:time}).

However, in the previous sections, we have focused on \emph{unweighted} networks, 
where at most one interaction occurs between two nodes $i$ and $j$.
This assumption is limiting in many real scenarios: for example,
individuals in social networks typically interact repeatedly (e.g. through emails, mobile phone calls, or text messages), and the number and order of interactions can bring important information on the strength of a social tie.
In this section, we focus on systems where repeated node-node interactions can occur and the timing of interactions plays a critical role.
To study this kind of systems, we adopt now the temporal networks' framework \cite{holme2012temporal}.
A \emph{temporal network} $\mathcal{G}^{(temp)}=(N,E^{(temp)})$ is completely specified by the number $N$ of nodes in the system and the list $E^{(temp)}$ of time-stamped edges (or contacts) ${(i,j,t)}$ between the nodes\footnote{For some systems, each edge is also endowed with a temporal duration. For our purposes, by assuming that there exist a basic time unit $u$ in the system, edges endowed with a temporal duration can be still represented through time-stamped edges. We simply represent an edge that spans the temporal interval $[t_1,t_2]$ as a sequence of $(t_1-t_2)/u$ subsequent time-stamped edges.}.
In the literature, temporal networks are also referred to as temporal graphs \cite{kempe2000connectivity, kostakos2009temporal}, time-varying networks/graphs \cite{casteigts2012time}, dynamic 
networks/graphs \cite{casteigts2012time}, scheduled networks \cite{berman1996vulnerability}, among others.

\subsection{Getting started: how to represent a temporal network?} 
Recent studies \cite{rosvall2014memory,scholtes2014causality} have pointed out that projecting a temporal network $\mathcal{G}^{(temp)}$ to a static weighted
(time-aggregated) network discards essential information and, as a result, often yields misleading results. 
To properly deal with the temporal nature of temporal networks, there are (at least) two possible strategies.
The first is to revisit the way the network representation of the data is constructed.
While in principle several representation of temporal networks can be adopted (see \cite{holme2015modern} for an overview), a simple and useful representation is the \emph{higher-order network representation} of the temporal network, (also referred to as \emph{memory network representation} in the literature). 
Differently from the time-aggregated representation which only preserves the number of contacts between pairs of nodes for a given temporal network, a higher-order network representation \cite{rosvall2014memory,scholtes2014causality,xu2016representing} transforms the edge list into a set of \emph{time-respecting paths} (paragraph \ref{sec:time_respecting_paths}) on the network's nodes, and it preserves the statistics of time-respecting paths up to a certain topological length (details are provided below).
Standard ranking algorithms (such as PageRank or closeness centrality) can be thus run on higher-order networks (paragraph \ref{sec:higher_order_centrality}). Higher-order networks are particularly powerful in highlighting the effects of memory on the speed of network diffusion processes (paragraph \ref{sec:effects}) and on the ranking of the nodes (paragraphs \ref{sec:higher_order} and \ref{sec:journals}).

The second possible strategy is to divide the system's time span into temporal slices (or layers) of equal duration, and consider a set $\{\mat{w}(t)\}$ of adjacency matrices, one for each layer -- we refer to this representation as the \emph{multilayer representation} of the original temporal network \cite{kivela2014multilayer}. The original temporal network is then represented by a number of static networks where usual network analysis methods can be employed. Note that the information encoded in the multilayer representation is exactly the same as that present in the original temporal network only if the temporal span of each layer is equal to the basic time unit of the original dataset. The problem of choosing a suitable time span for the temporal layers is a non-trivial and dataset-dependent problem. In this review, we do not address this issue; the reader can find interesting insights into this problem in \cite{holme2015modern,krings2012effects,sekara2016fundamental}, among others. In the following, when presenting methods based on the multilayer representation of temporal networks, we simply assume that each layer's duration has been chosen appropriately and, as a result, we have a meaningful partition of the original data into temporal layers. Diffusion-based ranking methods based on this construction are described in paragraph \ref{sec:multilayer}. We also present generalizations to temporal networks of centrality metrics based on shortest paths (paragraph \ref{sec:shortest}).







\begin{figure*}[t]
\centering
\includegraphics[scale=0.5]{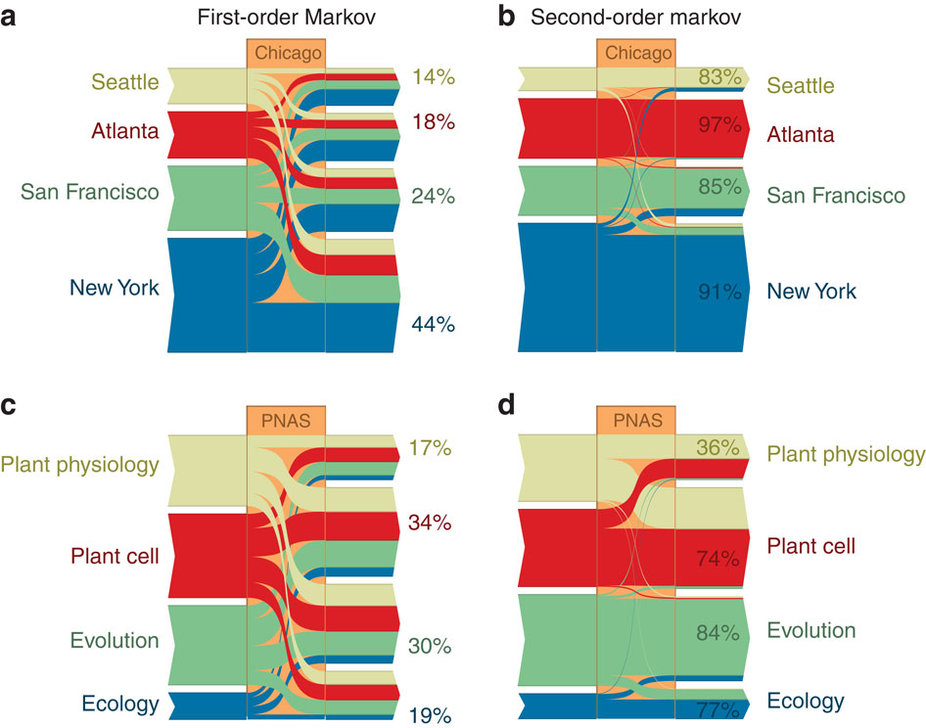}
\caption{An illustration of the difference between the first-order Markovian (time-aggregated) and second-order network representation of the same data (reprinted from~\cite{rosvall2014memory}). Panels a-b represent the destination cities (the right-most column) of flows of passengers from Chicago to other cities, given the previous location (the left--most column). When including memory effects (panel b), the fraction of passengers coming back to the original destination is large, in agreement with our intuition. A similar effect is found for the network of academic journals (panels c-d) -- see \cite{rosvall2014memory} and paragraph \ref{sec:journals} for details.}
\label{fig:rosvall_flows}
\end{figure*}

\subsection{Memory effects and time-preserving paths in temporal networks}
\label{sec:time_respecting_paths}
Considering the order of events in a temporal network has been shown to be fundamental to properly describe how different kinds of flow processes (e.g., flow of information, flow of traveling passengers) unfold on a temporal network.
In this section, we present a paradigmatic example that has been provided by Rosvall et al. \cite{rosvall2014memory} and highlights the importance of memory effects
in a temporal network. The example is followed by a general definition of time-preserving paths for temporal networks, which
constitute a building block for the causality-preserving diffusive centrality metrics that are discussed in section \ref{sec:higher_order_centrality}.

\paragraph{An example}
The example shown in Fig.~\ref{fig:rosvall_flows} is taken from \cite{rosvall2014memory} and well exemplifies the role of memory in temporal networks.
The figure considers two datasets: (1)
the network of cities connected by airplane passengers' itineraries and (2) the JSTOR
[\url{www.jstor.org}] network of scientific journals connected by citations between journals' articles.
We consider first the network of cities. The Markovian assumption \cite{rosvall2014memory} is that a flow's destination depends only on the current location, and not on its more distant history.
According to this assumption, if we consider the passengers who started from Seattle and passed through Chicago, for example, the fraction of them coming back is only $14\%$ (Fig.~\ref{fig:rosvall_flows}a).
One might conjecture that this is a poor description of real passenger traffic as travelers are usually based in a city and are likely to come back to
the same city at the end of their itineraries.
To capture this effect, one needs to consider a ``second-order'' description of the flow.
This simply means that instead of counting how many passengers move from Chicago to Seattle, we count how many
move to Seattle, given that they started in a certain city. We observe that the fraction of passengers who travels from Chicago to Seattle,
given that they reached Chicago from Seattle, is $83\%$, which is much larger than what was expected under the Markovian assumption ($14\%$).
Similar considerations can be made for the other cities (Figs.~\ref{fig:rosvall_flows}a-b).
To capture memory effects in the citation network of scientific journals, Rosvall et al. \cite{rosvall2014memory} analyze the statistics of triples of journals $j_1\to j_2\to j_3$ resulting from triples of articles $a_1\to a_2\to a_3$ such that $a_1$ cites $a_2$ and $a_2$ cites $a_3$, and paper $a_i$ is published in journal $j_i$.  Based on this construction, Rosvall et al. \cite{rosvall2014memory} show that memory effects heavily impact flows of citations between scientific journals (Figs.~\ref{fig:rosvall_flows}c-d).
We can conclude that including memory is critical for the prediction of flows in real world (see~\cite{xu2016representing}  and paragraph \ref{sec:higher_order}).

\paragraph{Time-respecting paths}
The example above shows that when building a network representation from a certain dataset, memory effects need to be taken into account to properly describe the system.
To account for memory, in the example of traveling passengers, we are no more interested in how many passengers travel between two locations, but in how many passengers follow a certain (time-ordered) itinerary across different locations. To enforce this paradigm shift, let us introduce a general formalism to describe time-preserving 
(or time-respecting, or time-ordered)
paths \cite{kempe2000connectivity,moody2002importance,scholtes2014causality}.
Given a static network and three nodes $a,b,c$, the existence of the links $(a,b)$ and $(b,c)$ directly implies that $a$ is connected
to $c$ through at least one path $a\to b\to c$ (\emph{transitivity} property).
The same does not hold in a temporal network: for a node $a$ to be connected with node $c$ through a path $a\to b\to c$,
the link $(a,b,t_1)$ must happen \emph{before} link $(b,c,t_2)$, \ie, $t_1\leq t_2$.
Time-aggregated networks thus overestimate the number of paths between pairs of nodes -- Lentz et al. \cite{lentz2013unfolding} present an elegant framework to quantify this property.
The first general definition of time respecting path of length $n$ is the following: ${(i_1,i_2,t_1), (i_2,i_3,t_2), \dots, (i_{n}, i_{n+1},t_n)}$
is a time-respecting path if and only 
if $t_1< t_2 < \dots<t_n$ \cite{kempe2000connectivity,tang2010analysing,nicosia2013graph, scholtes2014causality}.

In many practical applications, we are only interested in dynamical processes that occur at timescales much smaller than
the total time over which we observe the network.
For this reason, the existence of a time-respecting 
path $\{(i_1,i_2,t_1), (i_2,i_3,t_2)\}$ with a long waiting time between $t_1$ and $t_2$ may not be
informative at all about the influence of node $i_1$ on $i_3$.
A practical solution is to set a threshold $\delta$ for the maximum acceptable inter-event time \cite{pan2011path, scholtes2014causality, scholtes2016higher}.
Accordingly, one says that $\{(i_1,i_2,t_1), (i_2,i_3,t_2), \dots, (i_{n}, i_{n+1},t_n)\}$
is a time-respecting path (with limited waiting time $\delta$) if and only if $0<t_{n+1}- t_n\leq \delta$.
The suitability of a particular choice of $\delta$ depends on the system being examined. While too small $\delta$ values can result in a too sparse network from which it is hard to extract useful information, too large $\delta$ values can include in the network also paths that are not consistent with the real flows of information.
In practice, values as different as six seconds (used for pairwise interactions in an ant colony \cite{scholtes2014causality}) and one hour 
(used for e-mail exchanges between employees in a company \cite{scholtes2014causality}) can be used for $\delta$, depending on the context.

\subsection{Ranking based on higher-order network representations}
\label{sec:higher_order_centrality}
In this paragraph, we use the notion of a time-preserving path presented in the previous paragraph to introduce centrality metrics built on higher-order representation of temporal networks. Before introducing the higher-order representations (paragraph \ref{sec:second_order}), let us briefly revise the Markovian assumption.
Most of the methods presented in this paragraph are implemented in the Python 
library \url{pyTempNets} [available at \url{https://github.com/IngoScholtes/pyTempNets}].

\subsubsection{Revisiting the Markovian assumption}
\label{sec:first_order}
When studying a diffusion process on a (weighted) time-aggregated network,
standard network analysis 
implicitly assumes the \emph{Markovian property}:
the next move of a random walker does not depend on the walker's previous steps, but only on the node $i$ where he is currently located and on the weights of the links attached to node $i$.
Despite being widely used in network centrality metrics, community detection algorithms and spreading models,
this assumption can be unrealistic for a wide range of real systems (see the two examples provided in Fig.~\ref{fig:rosvall_flows}, for example).

Let us revisit the Markovian assumption in general terms.
In the most general formulation of a walk on a time-aggregated network, 
the probability $P^{(n+1)}_i$ that a random walker is located node $i$ at (diffusion-)time $n+1$ depends on the whole past.
Using the same notation as \cite{rosvall2014memory}, if we denote by $X_n$ the random variable which marks the position of the walker at time $n$, we write
\begin{equation}
\begin{split}
P_i^{(n+1)}=P(&X_{n+1}=i | X_n=i_n,X_{n-1}=i_{n-1}, \\ 
&\dots, X_1=i_{1},X_0=i_{0}),
\end{split}
\end{equation}
where $X_0=i_0$ is the initial position of the walker and $i_1,\dots,i_n$ are all his consequent past positions. 
The Markovian assumption 
\begin{equation}
P_i^{(n+1)}=P(X_{n+1}=i|X_n=i_n)=P(X_{1}=i|X_0=i_n),
\end{equation}
which implies that the probability of finding a random walker at node $i$ at time $t+1$ is given by 
\begin{equation}
P^{(n+1)}_{i}=\sum_j P(j\to i)\,P^{(n)}_j=\sum_j P_{ij}\,P^{(n)}_j,
\label{markovian_rw}
\end{equation}
where
\begin{equation}
P_{ij}:=P(j\to i)= P(X_{t+1}=i|X_t=j)=\frac{W_{ij}}{\sum_k W_{kj}}
\end{equation}
are the elements of the (column-stochastic) transition matrix $\mat{P}$ of the network ($P_{ij}=W_{ij}/\sum_k W_{kj}$, where $W_{ij}$ is the observed number of times a passenger traveled from node $j$ to node $i$) \footnote{Note that we are also assuming that the random walk process is \emph{stationary}, i.e., the transition probabilities do not depend on diffusion time $n$. This assumption still holds for the random walk process based on higher-order network representations presented below (paragraphs \ref{sec:second_order}-\ref{sec:higher_order}), whereas it does not hold for random walks based on multilayer representations of the underlying time-stamped data (paragraph \ref{sec:multilayer}).}.

As we have seen in section \ref{sec:static}, PageRank and its variants
build on the random walk defined by Eq.~\eqref{markovian_rw} and add an additional element, the teleportation term.
While the time-dependent PageRank variants defined in section \ref{sec:time} are based on a different transition matrix than the original PageRank algorithm, they also implicitly assume the Markovian property.
We now shall discuss how to relax the Markovian assumption and construct higher-order Markov models where also the previous trajectory of the walker
influences where the walker will jump next.

\subsubsection{A second-order Markov model}
\label{sec:second_order}
Following \cite{rosvall2014memory}, we explain in this section how to build a second-order Markov process on a given network.
Consider a random walker located at a certain node $j$. According to the second-order Markov model,
the probability that the walker will jump from node $j$ to node $i$ at time $n+1$
depends also on the node $k$ where the walker was located at the previous step of the dynamics.
In other words, the move $\vec{ji}$ from node $j$ to $i$ actually depends on the edge $\vec{kj}$. It follows that the second-order Markov model can be represented as a first-order Markov model on edges in the network. That is to say, the random walker jumps and locates not on nodes, but on edges. 

The probability $P(j\to i,n+1|k\to j, n)$
that the walker performs a jump from node $j$ to node $i$ at time $n+1$ after having performed a jump from node $k$ to 
node $j$ at time $n$ can be written as
\begin{equation}
P(j\to i,n+1|k\to j, n)=P(\vec{kj} \to \vec{ji})=
\frac{W(\vec{kj} \to \vec{ji})}{\sum_{\vec{jl} }W(\vec{kj} \to \vec{jl})},
\label{second_order}
\end{equation}
where $W(\vec{kj} \to \vec{ji})$ denotes the observed number of occurrences of the directed time-respecting path $\vec{kj} \to \vec{ji}$ in the data -- with the notation introduced in paragraph \ref{sec:time_respecting_paths}, $\vec{kj} \to \vec{ji}$ denotes here any length-two path $(k,j,t_1)\to (j,i,t_2)$ such that $0<t_2-t_1\leq\delta$.
Eq.~\eqref{second_order} allows us to study a non-markovian diffusion process that preserves the observed frequencies of time-respecting paths of length two.
This non-markovian diffusion process can be interpreted as a markovian diffusion process on the \emph{second-order representation} of the temporal network
(also referred to as the \emph{second-order network} or \emph{memory network})
whose nodes (also referred to as \emph{memory nodes} or \emph{second-order nodes}) are the edges of the original network.
For this reason, one can use standard static network tools to simulate diffusion and compute the corresponding node centrality
(see below) on second (and higher) order networks.
Importantly, in agreement with Eq.~\eqref{second_order}, the edge weights in the second-order network are the relative frequencies of the time-respecting paths of length two \cite{scholtes2014causality}.

The probability of finding the random walker at memory node 
$\vec{ji}$ at time $n+1$ is thus given by
\begin{equation}
P^{(n+1)}(\vec{ji})=\sum_{k}P(\vec{kj}\to\vec{ji})\, P^{(n)}(\vec{kj}).
\end{equation}
The probability $P_i^{(n+1)}$ of finding the random walker at node $i$ at time $n+1$ is thus
\begin{equation}
P_i^{(n+1)}=\sum_j P^{(n+1)}(\vec{ji})=\sum_{jk}P(\vec{kj}\to\vec{ji})\,P^{(n)}(\vec{kj}).
\end{equation}
The random walk based on the second-order model presented in this paragraph
is much more predictable than a random walk based on a network without memory, in the sense that the next step of the walker, given its current position, has much smaller uncertainty as measured
by conditional entropy (this holds for all the temporal-network datasets studied by \cite{rosvall2014memory}). 
An attentive reader may have noticed that while constructing the higher-order temporal network has the advantage to allow us to use standard network techniques, it comes at the price of an increased computational complexity. Indeed, in order to describe a non-Markovian process with a Markovian process, we had to increase the dimension of the system: the second-order network is comprised of $E>N$ memory nodes, where $E$ is the number of edges that have been active at least once during the whole observation period. This also points out that the problem of increased complexity requires more attention for dense networks than for sparse networks.

\subsubsection{Second-order PageRank}
\label{sec:second_pr}
In this paragraph, we briefly discuss how the PageRank algorithm can be generalized to higher-order networks.
The PageRank score $s(\vec{ji})$ of a memory node $\vec{ji}$ can be defined as the corresponding component of the stationary state of the following process \cite{rosvall2014memory}
\begin{equation}
s(\vec{ji},n+1)=c\,\sum_{k}s(\vec{kj},n)\,P(\vec{kj}\to\vec{ji})+
(1-c)\,\frac{\sum_{l}W_{il}}{\sum_{lm}W_{lm}},
\end{equation}
where the term multiplied by $1-c$ is the component $i$ of a ``smart'' teleportation vector (see paragraph \ref{sec:pr_variants}).
The second-order PageRank score of node $i$, $s_{i}$, is defined as
\begin{equation}
s_{i}=\sum_{j}s(\vec{ji}).
\label{pr_second_order}
\end{equation}
The effects of second-order flow constraints of the ranking of scientific journals by PageRank 
will be presented in paragraph \ref{sec:journals}.

\subsubsection{The effects of memory on diffusion speed}
\label{sec:effects}
From a physicist's point of view, it is interesting to study how the temporal order of interactions affects the speed of diffusion on the network.
Recent literature used standard models for spreading processes
\cite{karsai2011small,rosvall2014memory}, simulated random walks \cite{starnini2012random}, and graph Laplacian spectral properties \cite{masuda2013temporal},
to show that considering the timing of interactions can slow down spreading.

However, diffusion can also speed up when memory effects are included. A general analytic framework introduced by Scholtes et al.~\cite{scholtes2014causality} predicts whether including second-order memory effects will slow-down or speed-up diffusion. 
The main idea is to exploit the property that the convergence time of a random walk process is related to the second largest eigenvalue $\lambda_2$ of its
transition matrix. 
First, one constructs the transition matrix $\mat{T}^{(2)}$ associated to the second-order Markov model of the network.
It can be proven \cite{scholtes2014causality}, that for any stochastic matrix with eigenvalues $1=\lambda_1, \lambda_2, \dots, \lambda_{m}$, and for any $\epsilon>0$, the number of diffusion steps after which the distance between the vector $\vek{p}^{(n)}$ of visitation frequencies and the random-walk stationary vector $\vek{p}$ becomes smaller than $\epsilon$ is proportional to $1/\log(\abs{\lambda_2})$. 
To establish whether second-order memory effects will slow down or speed up diffusion, one thus constructs a randomized second-order transition matrix
$\tilde{\mat{T}}^{(2)}$ where the expected frequency of physical paths of length two is consistent with the time-aggregated network.
The second largest eigenvalue of the resulting matrix is denoted as $\tilde{\lambda}_2$.
The analytic prediction for the change of diffusion speed is thus given by
\begin{equation}
S(\tilde{\mat{T}}^{(2)}):=\frac{\log\abs{\tilde{\lambda}_2}}{\log\abs{\lambda_2}}.
\end{equation}
Values of $S(\tilde{\mat{T}}^{(2)})$ larger (smaller) than one predict slow-down (speed-up) of diffusion. The resulting predictions
accurately match the results of diffusion simulations on real data (see Fig.~2 in~\cite{scholtes2014causality}).

\begin{figure}[t]
\centering
\includegraphics[scale=0.3]{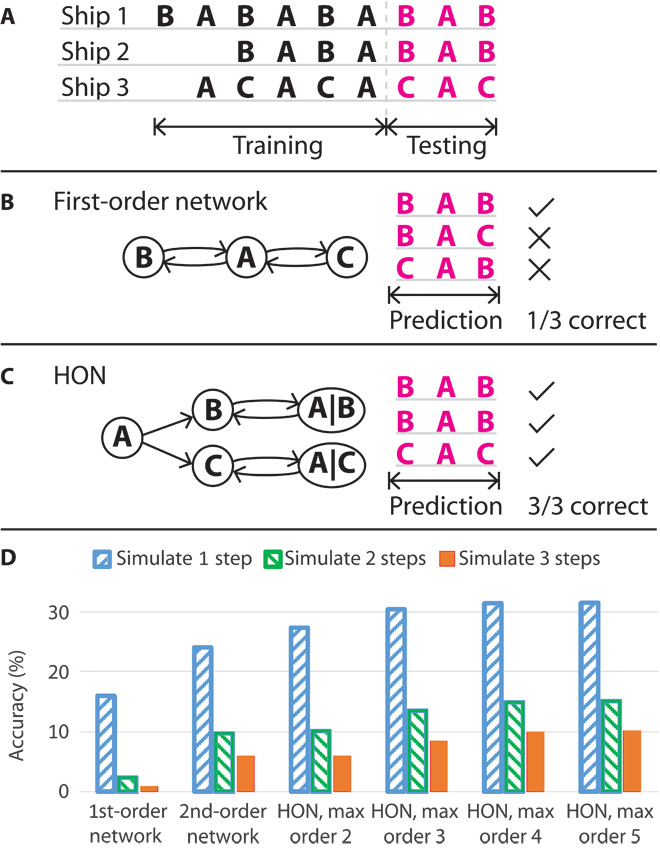}
\caption{An illustration of prediction of real-world ship trajectories between three locations $A$, $B$, and $C$ (reprinted  from~\cite{xu2016representing}). A part of the data is used to build the higher-order network (training); the remaining part is used to test the predictive power of different network representations (testing, panel A). In the first-order (time-aggregated) network (panel B), the nodes represent the three locations and the directed edge from location $A$ and location $B$ is weighted with the number of occurrences of the displacement $A\to B$ (see paragraph \ref{sec:first_order}). Differently from the higher-order networks used in \cite{rosvall2014memory, scholtes2014causality}, in the higher-order network introduced by Xu et al. \cite{xu2016representing}, nodes of different order co-exist (panel C). Accounting for memory effects dramatically improves prediction accuracy (panel D). More details can be found in \cite{xu2016representing}.}
\label{fig:hon-prediction}
\end{figure}

\begin{figure}[t]
\centering
\includegraphics[scale=0.3]{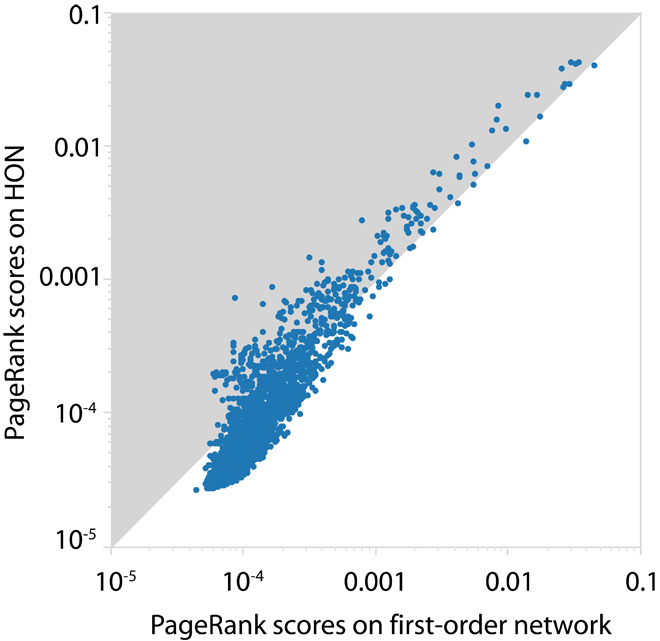}
\caption{The relation between higher-order PageRank and first-order PageRank score in WWW data (reprinted from~\cite{xu2016representing}). As reported in \cite{xu2016representing}, when higher-order effects are included, more than $90\%$ of the webpages descrease their PageRank score. This may indicate that standard PageRank score overestimate the traffic on the majority of websites, which may mean that the distribution of the attention received by websites is even more uneven than previously thought.}
\label{fig:hon-pr}
\end{figure}

\subsubsection{Generalization to higher-order Markov models and prediction of real-world flows}
\label{sec:higher_order}
In paragraph \ref{sec:second_order}, we have considered random walks based on a second order Markov model.
It is natural to generalize this model to take into account higher order dependencies.
If we are interested, for example, in a third-order Markov model, we could simply build a higher-order network whose memory nodes
are triples of physical nodes \cite{scholtes2014causality}. The memory node $\overrightarrow{ijk}$ would then represent a time-respecting path $i\to j\to k$.
Xu et al. \cite{xu2016representing} consider a more general higher-order model where, after a maximum order of dependency is fixed, all orders of dependencies can coexists (see Fig.~\ref{fig:hon-prediction}C).
The authors use global shipping transportation data to show that by considering more than two steps of memory, the random walk is even more predictable, which results in an enhanced accuracy in out-of-sample prediction of real-world flows (Fig.~\ref{fig:hon-prediction}D).

Interestingly, Xu et al. \cite{xu2016representing} also discuss the effects of including memory effects on ranking websites in the World Wide Web. Similarly to Eq.~\eqref{pr_second_order}, the authors define the higher-order PageRank score of a given node $i$ as the sum of the PageRank scores of higher-order nodes that represent node $i$.
Fig.~\ref{fig:hon-pr} shows that despite being positively correlated, standard and higher-order PageRank bring substantially different information on node importance.
Since higher-order PageRank describes user traffic better than standard PageRank, one could conjecture that it provides a better estimate of website relevance; however, further research is needed to test this hypothesis and investigate possible limitations of the higher-order network approach.

The possibility of including all order of dependencies up to a certain order leads to a natural question: is there an optimal order to reliably describe a temporal-network dataset? Ideally, we would like our model to describe the system accurately enough with the smallest possible memory order; in other words, an ideal model would have good predictive power and, at the same time, low complexity. The problem of determining the optimal order is non-trivial and it is addressed by Scholtes \cite{scholtes2017network} with maximum likelihood estimation techniques. The statistical framework introduced by Scholtes \cite{scholtes2017network} allows us to determine the optimal order $k_{opt}$ to describe a temporal network, thereby distinguishing among systems for which the traditional time-aggregated network representation is satisfactory ($k_{opt}=1$) and systems for which temporal effects play an essential role ($k_{opt}>1$).

\subsection{Ranking based on a multilayer representation of temporal networks}
\label{sec:multilayer}
The higher-order Markov models presented above are particularly convenient because: (1) they preserve the temporal order of interactions up to a certain order;
(2) centrality scores on the respective networks can be simulated by means of standard network techniques.
However, one can instead consider the temporal network as a sequence of temporal snapshots, each of which is composed of the set
of links that occurred within a temporal window of given length \cite{masuda2013temporal,rocha2014random, taylor2015eigenvector}.
Within this framework, a temporal network composed of $N$ nodes and of total temporal duration $T$ is represented by $L$ adjacency matrices $\{w(t)\}$ ($t=1,\dots, L$) with dimension $N\times N$, where $L$ is the number of temporal snapshots the total observation time is divided into.
The element $w_{ij}(t)$ of the adjacency matrix for the snapshot $t$ represents the number of interactions between $i$ and $j$
within the time period $[t,t+T/L)$.
The set of adjacency matrices is typically referred to as an adjacency tensor \cite{holme2015modern,taylor2015eigenvector}.

\subsubsection{TempoRank}
Building on the random walk process proposed by Starnini et al.~\cite{starnini2012random}, Rocha and Masuda \cite{rocha2014random} design a random walk-based ranking method for undirected networks, called TempoRank. 
The method assumes that at each diffusion time $t$, the random walker only ``sees'' the network edges active within the temporal layer $t$ ($w_{ij}(t)=1$ for an edge $(i,j)$ active at layer $t$), which means that there is an exact correspondence between the diffusion time and the real time -- diffusion and network evolution unfold with the same time-scale. This is very different from diffusion-based static ranking algorithms which implicitly assume that diffusion takes place at a much faster timescale than the time scale at which the network changes.

With respect to the random walk processes studied in \cite{starnini2012random}, TempoRank features a ``sojourn'' (a temporary stay) probability according to which a random walker located at node $i$
has the probability $q^{k_{i}(t)}$ of remaining located at node $i$, where $k_{i}(t)=\sum_{j}w_{ij}(t)$ is the number of edges of node $i$ in layer $t$.
The (row-stochastic) transition matrix $P_{ij}(t)$ of the random walk at time $t$ is given by
\begin{equation}
P_{ij}(t)= \begin{cases}
\delta_{ij} & \text{if } k_{i}(t)=0,\\ 
q^{k_{i}(t)} & \text{if } k_{i}(t)>0,\ i=j, \\
(1-q^{k_{i}(t)})\,w_{ij}(t)/k_{i}(t) & \text{if } k_{i}(t)>0,\ i\neq j.
\end{cases}
\label{walk}
\end{equation}
It can be shown that a non-zero sojourn probability ($q>0$) is sufficient for the convergence of the random walk process if and only if the time-aggregated network is connected; by contrast, if $q=0$, the convergence of the random walk process to a stationary vector is not guaranteed even if the time-aggregated network is connected.

The need for a non-zero sojourn probability is motivated in paragraph 2.3 of \cite{rocha2014random} and well illustrated by the following example. Consider a network of three nodes $1,2,3$ whose time-aggregated representation is triangular ($w_{12}=w_{23}=w_{13}=1$). Say that only one edge is active at each temporal layer: $w_{12}(1)=w_{21}(1)=w_{23}(2)=w_{32}(2)=w_{13}(3)=w_{31}(3)=1$, while all the other $w_{ij}(t)$ are equal to zero. For this simple temporal network, if $q=0$, a random walker starting at node $1$ will always end at node $1$ after three steps, which implies that the random walk is not mixing. This issue is especially troublesome for temporal networks with fine temporal resolution for which there exist only a limited number of possible pathways for the random walker, and it is solved by imposing $q>0$. We remark that by defining the sojourn probability as $q^{k_{i}(t)}$, we assume that the random decision whether the walker located at node $i$ will move (probability $q$) or not (probability $1-q$) is made independently for each edge.

It is important to study the asymptotic properties of the walk defined by Eq.~\eqref{walk}. However, since TempoRank's diffusion time is equal to the real-system time $t$, it is not clear a priori how the random walk should behave after $L$ diffusion steps, \ie, after the temporal end of the dataset is reached. To deal with this ambiguity, Rocha and Masuda \cite{rocha2014random} impose periodic boundary conditions on the transition matrix: $\mat{P}(t+L)=\mat{P}(t)$.
With this choice, when the last temporal layer of the dataset $t=L$ is reached, the random walker continues with the oldest temporal layer $t=1$.
The random walk process that defines the TempoRank score thus runs cyclically multiple times on the dataset. Below, we denote by $m$ the number of cycles performed by the random walk on the dataset; we are interested in the properties of the random walk in the limit of large $m$.
The one-cycle transition matrix $\mat{P}^{(temp, 1c)}$ is defined as
\begin{equation}
\mat{P}^{(temp, 1c)} = \prod_{t=1}^{L}\mat{P}(t)=\mat{P}(1)\dots \mat{P}(L).
\end{equation}
The TempoRank vector after $n$ time steps is thus given by the leading left eigenvector\footnote{We choose the left-eigenvector here because the transition matrix defined by Eq. \eqref{walk} is row-stochastic.} $\vek{s}(n)$ of matrix $\mat{P}^{(temp)}(n)$, where
$\mat{P}^{(temp)}(n)$ is given by the (periodic) time-ordered product of $n=m\,L+t$ transition matrices $\mat{P}(t)$. 
This eigenvector corresponds to the stationary density of the random walk after $m$ cycles over the dataset and $t$ additional temporal layers. Even when $m$ is large, the vector $\vek{s}(n)$ fluctuates within a cycle due to the displacements of the walker along the network temporal links.
The TempoRank vector of scores $\vek{s}$ is thus defined as the average within one cycle of the leading eigenvector of the matrix $\mat{P}^{(temp)}(n)$ for large $m$
\begin{equation}
\vek{s}=\frac{1}{L}\sum_{t=1}^{L}\vek{s}(m L+t)
\end{equation}
As shown in~\cite{rocha2014random}, the TempoRank score is well approximated by a local time-aware variable, referred to as the in-strenght approximator, yet how this result depends on network's structural and temporal patterns remains unexplored. 
Importantly, TempoRank scores are little correlated with the scores obtained with the
time-aggregated network representation. This finding constitutes yet another evidence that node centrality computed with a temporal network representation can substantially differ from node centrality computed with the time-aggregated representation, which  is in a qualitative agreement with the findings obtained with higher-order Markov models \cite{scholtes2014causality, rosvall2014memory}.


\subsubsection{Coupling between temporal layers}
Taylor et al. \cite{taylor2015eigenvector} generalizes the multilayer approach to node centrality by including the coupling between consecutive temporal network layers.
The aim of this coupling is to connect each node with itself between two consecutive layers and to 
rule the magnitude of variations of node centrality over time.
Each node's centrality for a certain temporal layer is thus dependent not only on the centrality of its neighbors in that layer,
but also on its centrality in the precedent and subsequent temporal layer.
The method can be in principle applied to any structural centrality metric.
A full description of the mathematics of the method involves tensorial calculations and can be found in~\cite{taylor2015eigenvector}.
The method has been applied to rank mathematics departments and the Supreme Court decisions. 
In the case of mathematics departments, the regime of strong coupling between layers represents the basic idea the prestige of a mathematics department should not widely fluctuate over time.
This method provides us with an analytic framework to identify not only the most central nodes in the network, but also the ``top-movers'', i.e., the nodes whose centrality is rising most quickly. 

\subsubsection{Dynamic network-based scoring systems for ranking in sport}
\label{sec:dynwinlose}
As shown by Motegi and Masuda~\cite{motegi2012network}, both Park and Newman's win-lose score \cite{park2005network} and Radicchi's prestige score \cite{radicchi2011best} can be generalized to include the temporal dimension.
There are at least two main reasons to include time in sport ranking:
(1) old performances are less indicative of the players' current strength \cite{junior2012time}; 
(2) it is more rewarding to beat a certain player when he/she is at the peak of his/her performance than doing so when he/she is a novice or short before retirement \cite{motegi2012network}.
To account for the former effect, some time-aware variants of structural centrality metrics simply introduce time-dependent penalization
for older interactions \cite{junior2012time}, in the same spirit as the time-dependent algorithms presented in paragraph \ref{sec:explicit}.

In the dynamic win-lose score introduced by Motegi and Masuda \cite{motegi2012network}, the latter effect is taken into account by assuming that the contribution to player $i$'s score coming from a win against player $j$ at time $t$ is proportional to the score of player $j$ at that time $t$.
The equations that define the vector $\vek{w}$ of win scores and the vector $\vek{l}$ of lose scores are provided in \cite{motegi2012network}.
By assuming that we have a reasonable partition of time into temporal layers $1,\dots, t$
-- Motegi and Masuda \cite{motegi2012network} set the temporal duration of each layer to one day --, the vector 
of dynamic win scores $\vek{w}(t)$ at time $t$ is defined
as
\begin{equation}
\vek{w}(t)=\mat{W}\TT(t)\,\vek{e},
\end{equation}
where
\begin{equation}
\begin{split}
&\mat{W}\TT(t)=\,\,\mat{A}(t)+\ee^{-\beta}\sum_{m_t\in\{0,1\}}\alpha^{m_t}\,\mat{A}(t-1)\,\mat{A}(t)^{m_t}+\\
 &+ \ee^{-2\,\beta}\sum_{m_t, m_{t-1}\in\{0,1\}}\alpha^{m_t+m_{t-1}} \mat{A}(t-2)\,\mat{A}(t-1)^{m_{t-1}}\,\mat{A}(t)^{m_t}+\\
 &+\dots+\\
 &+ \ee^{-\beta\,(t-1)}\sum_{m_{2},\dots, m_{t}\in\{0,1\}}\alpha^{\sum_{j=2}^{t}m_{j}}\mat{A}(1)\,\mat{A}(2)^{m_{2}}\dots \mat{A}(t)^{m_t}.
\end{split}
\label{long_definition}
\end{equation}
The vector of loss scores $\vek{l}(t)$ is defined analogously by replacing $\mat{A}(t)$ by $\mat{A}\TT(t)$. As in paragraph~\ref{sec:winlose}, the dynamic win-lose score $\vek{s}(t)$ is defined as the difference $\vek{s}(t):=\vek{w}(t)-\vek{l}(t)$.

We can understand the rationale behind this definition by considering the first terms of Eq.~\eqref{long_definition}.
The contributions to $i$'s score that come from wins at a past time $t'$, with $t'\leq t$, are modulated by a factor $\exp[-\beta(t-t')]$, where $\beta$ is a parameter of the method. Similarly to Park and Newman's static win-lose score, also indirect wins determine player $i$'s score; differently from the static win-lose score, only indirect wins that happened \emph{after} player's $i$ respective direct win are included.
With these assumptions, the contributions to player $i$'s score at time $t$ that come from the two past time layers $t-1$ and $t-2$ are
\begin{equation}
\begin{split}
&w_i(t)=\,\sum_{j}A_{ij}(t)+\\
&+e^{-\beta}\,\sum_{j}A_{ij}(t-1)+
\alpha\,e^{-\beta}\,\sum_{j,k}A_{ij}(t-1)A_{jk}(t)+\\
&+e^{-2\,\beta}\,\sum_{j}A_{ij}(t-2)
+ \alpha e^{-2\,\beta}\,\sum_{j,k}A_{ij}(t-2)A_{jk}(t)+\\
&+ \alpha e^{-2\,\beta}
\,\sum_{j,k}A_{ij}(t-2)A_{jk}(t-1)+\\
&+ \alpha^2 e^{-2,\beta}
\,\sum_{j,k,l}A_{ij}(t-2)A_{jk}(t-1)A_{kl}(t)+\mathcal{O}(e^{-3\,\beta})
\end{split}
\end{equation}
where $\mathcal{O}(e^{-3\,\beta})$ terms correspond to wins that happened before time $t-2$.
Motegi and Masuda \cite{motegi2012network} applied different player-level metrics to the network of tennis players, and found that the dynamic win-lose score has a better predictive power than its static counterpart by Park and Newman.

\subsection{Extending shortest-path-based metrics to temporal networks}
\label{sec:shortest}
Up to now, we have only discussed diffusion-based metrics for temporal networks. It is also possible to generalize shortest-path-based centrality metrics to temporal networks, as it will be outlined in this paragraph.

\subsubsection{Shortest and fastest time-respecting paths}
\label{sec:fastest}
It is natural to define the length of a static path as the number of edges traveled through the path.
As we have seen in paragraph~\ref{sec:time_respecting_paths}, the natural way to enforce the causality of interactions is to consider time-respecting paths instead of static paths. There are two possible ways to extend the notion of shortest paths to time-respecting paths.
In analogy with the definition of length for static paths,
one can simply define the length of a time-respecting path as the number of edges traversed along the path.
This is called the \emph{topological length} of path $\{(i_1,i_2,t_1), (i_2,i_3,t_2), \dots, (i_{n}, i_{n+1},t_n)\}$ which is simply equal to $n$~\cite{nicosia2013graph, scholtes2016higher}.
Alternatively, one can also define the \emph{temporal length} of path $\{(i_1,i_2,t_1), (i_2,i_3,t_2), \dots, (i_{n}, i_{n+1},t_n)\}$ which is the time distance between the first and the last path's edge, i.e., $t_n-t_1+1$~\cite{pan2011path}.

Hence, the two alternatives to generalize the notion of shortest path to temporal networks correspond to choosing the time-respecting paths with the smallest topological length \cite{scholtes2016higher}
or those with the smallest temporal duration \cite{tang2010analysing, holme2012temporal}, respectively.
The former and latter choice lead to the definition of \emph{shortest} and \emph{fastest time-respecting paths}, respectively. More details and subtleties associated to the definition of a distance in temporal networks can be found in \cite{holme2015modern} in paragraph 4.3.

\subsubsection{Temporal betweenness centrality}
Static betweenness score of a given node is defined as the fraction of geodesic paths that pass through the node (see section~\ref{sec:shortestpath}).
It is natural to extend this definition to temporal networks by considering the fraction of shortest \cite{scholtes2016higher} or fastest \cite{tang2010analysing} time-respecting paths that pass
through the node.
Denoting by $\sigma^{(i)}(j,k)$ the number of shortest (or fastest) time-respecting paths between $j$ and $k$ that pass through node $i$,
we can define the temporal betweenness centrality $B^{(temp)}_i$ of node $i$ as 
\cite{scholtes2016higher}
\begin{equation}
B^{(temp)}_i=\sum_{j\neq i \neq k} \sigma^{(i)}(j,k).
\label{temp_betw}
\end{equation}
Temporal betweenness centrality can be also normalized by the total number of shortest paths that have not $i$ as one of the two extreme nodes \cite{kim2012temporal}.
A different definition of temporal betweenness centrality, based on fastest time-respecting paths, has been introduced by Tang et al. \cite{tang2010analysing}.

\subsubsection{Temporal closeness centrality}
In section \ref{sec:shortestpath}, we defined static closeness score of a given node as the inverse of the mean geodesic distance of the node
from the other nodes in the network. In analogy with static closeness, one can define the temporal closeness score of a given node
as the inverse of the temporal distance (see section \ref{sec:fastest}) of the node
from the other nodes in the network.
Let us denote the temporal distance between two nodes $i$ and $j$ at time $t$ as $d^{(temp)}_{ij}$ -- here temporal distance $d^{(temp)}_{ij}$ is a placeholder variable to represent either the topological length of the shortest paths\footnote{In this case, the use of the adjective "temporal" might feel inappropriate since we are using it to label a topology-related quantity. However, the adjective "temporal" reflects the fact that we are only considering time-respecting paths.} or the duration of the fastest paths between $i$ and $j$,
the temporal closeness score $C^{(temp)}_i$
of node $i$ thus reads \cite{holme2012temporal}
\begin{equation}
C^{(temp)}_i=\frac{N-1}{\sum_{j\neq i}d^{(temp)}_{ij}}.
\label{temp_clos1}
\end{equation}
This definition suffers from the same problem encountered for Eq. \eqref{closeness1}: if two nodes are not connected by any temporal path
within the observation period, their temporal distance
is infinite which results in zero closeness score.
In analogy with Eq. \eqref{closeness2},
to overcome this problem, closeness score can be redefined in terms of the harmonic average of the temporal distance between the nodes \cite{kim2012temporal,scholtes2016higher}
\begin{equation}
C^{(temp)}_i=\frac{1}{N-1}\sum_{j\neq i}\frac{1}{d^{(temp)}_{ij}}.
\label{temp_clos2}
\end{equation}
In general, the correlation between temporal and time-aggregated closeness centrality
is larger than the correlation between temporal and time-aggregated betweeness centrality \cite{scholtes2016higher}.
This is arguably due to the fact that 
node closeness depends only on paths' length, whereas betweeness centrality takes also into account the paths' actual structure,
which is deeply affected when considering the temporal ordering of interactions \cite{scholtes2016higher}.
A different definition of temporal closeness based on the average of the temporal duration of fastest time-respecting paths is provided by Tang et al. \cite{tang2010analysing}.

Computing centrality metrics defined by Eqs. \eqref{temp_betw} and \eqref{temp_clos2} preserves the causal order of interaction,
but requires a detailed analysis of all the time-respecting paths in the network.
On the other hand, centrality metrics based on higher-order aggregated networks (section \ref{sec:higher_order_centrality}) can be 
computed through static methods which are computationally faster.
A natural question arises: can we use metrics computed on higher-order time-aggregated networks to approximate temporal centrality metrics?
A first investigation has been carried out by Scholtes et al. \cite{scholtes2016higher}, who show 
that static betweenness and closeness centrality computed on second-order time-aggregated networks tend to better approximate the corresponding temporal
metrics than the respective static metrics computed on the first-order time-aggregated network.
These findings indicate that taking into account the first few markovian orders of memory could be sufficient to
approximate reasonably well the actual temporal centrality metrics.

\section{Temporal aspects of recommender systems}
\label{sec:recommender}
While most of this review concerns with constructing a general ranking of nodes in a network, there are situations where such a general ranking is of a limited applicability. This is particularly relevant in bipartite user-item networks where users' personal tastes and preferences play an important role. For example, a reader is more likely to be interested in a new book by their favorite author than in a book of a recently announced laureate of the Nobel Prize in literature. For this reason, one seeks not to establish a general ranking of all items in the network but rather multiple \emph{personalized} item rankings---ideally as many of them as there are different users in the system. This is precisely the goal of recommender systems that aim to use data on users' past preferences to identify the items that are likely to be appreciated by a given individual user in the future~\cite{schafer2007collaborative,koren2011advances}.

The common approach to recommendation is static in nature: the input data are considered without taking time into account. While this typically produces satisfactory results, there is now increasing evidence that the best results are obtained by considering the time information~\cite{bell2007lessons}. Winners of several recent high-profile machine-learning competitions, such as the NetflixPrize~\cite{bennett2007netflix}, have emphasized the importance of taking temporal dynamics of the input data into account~\cite{gama2014survey}. In this section, we present the role played by time in recommendation.

\subsection{Temporal dynamics in matrix factorization}
Recommendation methods have been traditionally classified as memory and model based~\cite{breese1998empirical}. \emph{Memory-based methods} rely on selecting a set of similar users (``neighbors'') for every user and recommend items on the basis of what the neighbors have favored in the past. \emph{Model-based methods} rely on formulating a statistical model for user behavior and than use this model to choose which items to recommend. However, this classification now seems outdated as modern matrix factorization methods combine the two approaches: they are model based (Equations~\eqref{basic_model} and \eqref{model_with_baseline} below are examples of how the ratings can be modeled), yet they also have memory-based features as they sum over all memory-stored ratings~\cite{koren2011advances}.

Motivated by its good performance in many practical 
contexts~\cite{takacs2007major,koren2009matrix}, we describe here a recommendation method 
based on matrix factorization. The input data of this method take form of ratings given by
users to items; there are $U$ users and $I$ items in total. The method assumes that
the rating of user $i$ to item $\alpha$ as $r_{i\alpha}$ is an outcome of a
matching between the user's preferences and the item's properties. 
In particular, both users and items are assumed to be representable 
in a latent (\ie, hidden) space of dimension $D$; the vectors corresponding to
user $i$ and item $\alpha$ are $\vek{p}_i$ and $\vek{q}_{\alpha}$, respectively. Assuming that 
the vectors for all users and items are known, a yet unexpressed rating is estimated as
\begin{equation}
\label{basic_model}
\hat r_{i\alpha} = \vek{p}_i \cdot \vek{q}_{\alpha}.
\end{equation}
In other words, we factorize the sparse rating matrix $\mat{R}$ into a product of two matrices---the $U\times D$ matrix $\mat{P}$ with user taste vectors and the $I\times D$ matrix $\mat{Q}$ with item property vectors as $\mat{R} = \mat{P}\mat{Q}\TT$. This simple recipe can be further extended to include other factor such as, for example, which items have been actually rated by the users~\cite{koren2008factorization}. The increased model complexity is then justified by an improved prediction accuracy.

Of course, the vectors are initially unknown and need to be learned (estimated) from the available data. The learning procedure is formally represented as an optimization problem where the task is to minimize the difference between the estimated and actually expressed ratings
\begin{equation}
\sum_{r_{i\alpha}\in\mathcal{R}} (r_{i\alpha} - \vek{p}_i\cdot\vek{q}_{\alpha})^2
\end{equation}
where the sum is over ratings $r_{i\alpha}$ in the set of all ratings $\mathcal{R}$. Since this is a high-dimensional optimization problem, it is useful to employ some regularization technique to prevent over-fitting. This is usually done by adding regularization terms that penalize user and item vectors with large elements
\begin{equation}
\label{to_optimize}
\sum_{r_{i\alpha}\in\mathcal{R}} (r_{i\alpha} - \vek{p}_i\cdot\vek{q}_{\alpha})^2 + \lambda\big(\|\vek{p}\|^2+\|\vek{q}\|^2\big)
\end{equation}
where parameter $\lambda$ is determined by cross-validation (\emph{i.e}, by comparing the predictions with a hidden part of data)~\cite{picard1984cross,arlot2010survey}. While Eq.~(\ref{to_optimize}) presents a seemingly difficult optimization problem with many parameters, simple optimization by gradient descent works well and quickly converges to a good solution. This optimization method can be implemented efficiently because partial derivatives of Eq.~(\ref{to_optimize}) with respect to elements of vectors $\vek{p}_i$ and $\vek{q}_{\alpha}$ can be written down analytically (see~\cite{takacs2007major} for a particular example).

When modeling ratings in a bipartite user-item system, it is useful to be more specific than the simple model provided by Eq.~(\ref{basic_model}). A common form~\cite{koren2010collaborative} of estimated rating is
\begin{equation}
\label{model_with_baseline}
\hat r_{i\alpha} = \mu + d_i + d_{\alpha} + \vek{p}_i \cdot \vek{q}_{\alpha}
\end{equation}
where $\mu$ is the overall average rating, and $d_i$ and $d_{\alpha}$ are the average observed deviations of ratings given by user $i$ and to item $\alpha$, respectively. The interpretation of this ``separation of effects'' is straightforward: the rating of a strict user to a generally badly rated movie is likely to be low unless the match between the user's taste and the movie's features turns out to be particularly favorable. A more detailed discussion of the baseline estimates can be found in~\cite{koren2010factor}. It is useful to separate the baseline part of the prediction
\begin{equation}
\label{baseline}
\hat r_{i\alpha} = \mu + d_i + d_{\alpha}
\end{equation}
for a comparison in recommendation evaluation which can tell us how much the present matrix factorization method can improve over this elementary approach.

While generally well-performing, the above-described matrix factorization framework is \emph{static} in nature: all ratings regardless of their age enter the estimation process with the same weight and all parameters are assumed constant. In particular, even when estimated parameter values can change when the input data changes (because of, for example, accumulating more data), a given set of parameter values is assumed to model all available input data---the oldest and the most recent alike. An analysis of the data shows that such a stationarity assumption is generally not satisfied. In the classical NetflixPrize dataset, for example, two non-stationary features become quickly apparent. First, the average item rating has increased suddenly by approximately 0.3 (in the $1--5$ scale) in 2004. Second, item rating generally increases with item age. Koren~\cite{koren2010collaborative} connected the former effect with improving the function of the Netflix's internal recommendation system and, also, the company's change of word explanations given to different ratings (from ``superb'' to ``loved it'' for rating 5, for example). Similarly, the latter effect is linked with young items being chosen at random at a higher proportion than old items for which the users apparently have more information and only decide to consume and rate them when the chances are good that they might like the item.

To include time effects, the baseline prediction provided by Eq.~(\ref{baseline}) can be readily generalized to the form
\begin{equation}
\label{baseline_with_time}
\hat r_{i\alpha} = \mu + d_i(t) + d_{\alpha}(t)
\end{equation}
where the difference terms $d_i(t)$ and $d_{\alpha}(t)$ are assumed to change with time. How to model the time effects that influence $d_i(t)$ and $d_{\alpha}(t)$ is an open and context-dependent problem. In any given setting, there are multiple sources of variation with time that can be considered (such as a slowly-changing appreciation of a movie, rapidly-changing mood of a user, and so forth). Which of them have significant impact on the resulting performance is best decided by formulating a complex model and measuring the prediction precision when components of this model are ``turned on'' one after another. The evolution of slowly-changing features can be modeled by, for example, dividing the data into bins. If time $t$ falls in bin $B(t)$, we can write $d_{\alpha}(t) = d_{\alpha} + d_{\alpha, B(t)}$ where the overall difference $d_{\alpha}$ is complemented with the difference in a given bin. By contrast, the bins needed to model fast-changing features need to be very short and therefore suffer from poor statistic for each of them; Fast-changing features are thus better modeled with smooth functions. For example, the gradual shift in a user's bias can be modeled by writing $d_i(t) = d_i + \delta_i \text{sign}(t - t_i)\,\lvert t - t_i\rvert^{\beta}$ where $t_i$ is the average time stamp of $i$'s ratings. As before, parameter $\beta$ is to be set by cross validation. For further examples on which temporal effects to model and how, see~\cite{koren2010collaborative} for an example.

We can now finally come back to the matrix factorization part and introduce time effects there. Out of the two set of computed vectors---user preferences $\vek{p}_i$ and item features $\vek{q}_{\alpha}$, it is reasonable to assume that only the user preferences change. To keep things simple, we follow here the same approach as above to model user differences $d_i(t)$. Element $k$ of $i$'s preference vector thus becomes
\begin{equation}
p_{ik} = p_{ik} + \delta_{ik}\,\text{sign}(t - t_i)\,\lvert t - t_i\rvert^{\beta}.
\end{equation}
The complete prediction thus reads
\begin{equation}
\label{model_with_time}
\hat r_{i\alpha}(t) = \mu + d_i(t) + d_{\alpha}(t) + \vek{p}_i(t) \cdot \vek{q}_{\alpha}.
\end{equation}
Note that there are many variants of matrix factorization and also many ways how to incorporate time in them. For the sake of simplicity, we have omitted here the day-specific effect which models the fact that a user's mood on a given single day can be anything from pessimistic through neutral to optimistic and the predictions should include this. When this effect is included in the model, the accuracy of results can improve considerably (see~\cite{koren2010collaborative} for details). The matrix model can be further modified to include, for example, the very fact which items have been chosen and rated by the user. Koren~\cite{koren2008factorization,koren2010factor} has shown that this modification also improves the prediction accuracy.

To achieve further improvements, several well-performing methods which can be generally very diverse and unrelated can be combined in a unique predictor~\cite{breiman1996bagging,jahrer2010combining}; in the literature, this is referred to as ``blending'' or ``bagging''. Notably, some sort of method combination have been employed by all successful teams in the NetflixPrize~\cite{bell2007lessons}.
This is a sign that to obtain ultimate achievements, there is no single method that outperforms all the others but rather that all methods capture some part of the true user behavior.

\subsection{Temporal dynamics in network-based recommendation}
\label{sec:rec-time}
A popular approach to recommendation is based on representing the input data with bipartite networks and studying various diffusive processes on them~\cite{zhou2007bipartite, zhang2007heat}. {Given a set of ratings $\{r_{i\alpha}\}$ given by users to items (the same ratings that served as input for the matrix factorization methods in the previous section), all ratings above a chosen threshold are represented as links between the respective users and items. When computing the recommendation scores of items for a given user $i$ in the original ``probabilistic spreading'' method~\cite{zhou2007bipartite}, one assigns one unit of ``resource'' to all items connected with the given user and zero resource to the others. The 
resource is then propagated to the user side and back in a random walk-like process and the resulting amounts of resource on the item side are interpreted as item recommendation scores; items with high score are assumed to be likely to be appreciated by the given user $i$. The main difference from matrix factorization methods is that network-based methods assume binary user-item data as input\footnote{This makes the methods applicable also in systems where the users give no explicit ratings but only attach themselves with items that they have bought, watched, or otherwise interacted with.} and use physics-motivated processes on them.


Despite high activity in this field~\cite{yu2016network}, little attention has been devoted to the role that time plays in the recommendation process. While it has been noticed, for example, that the most successful network-based methods tend to recommend popular items~\cite{liu2011information,qiu2011item}, this has not been explicitly connected with the fact that to gain popularity, items need time and therefore the recommended items are likely to be old. The role of time in network-based recommendation has been directly discussed by Vidmer and Medo \cite{vidmer2016role}. The main contribution of that paper is two-fold. First, it points out a deficiency of the usual evaluation procedure that is based on removing links from the network at random and examining how they are reproduced by a recommendation method. Random removal of links targets preferentially popular items---the evaluation procedure thus shares the bias that is innate to network-based evaluation methods which then leads to an overestimation of the methods' performance. To make the evaluation more fair and also more relevant from the practical point of view, the latest links need to be removed instead. All the remaining (preceding) links are then used to reconstruct ``near future'' of the network. Note that this is fundamentally different from the evaluation based on random link removal where both past and future links are used to reconstruct the missing part.

\begin{figure*}
\centering
\includegraphics[scale=0.7]{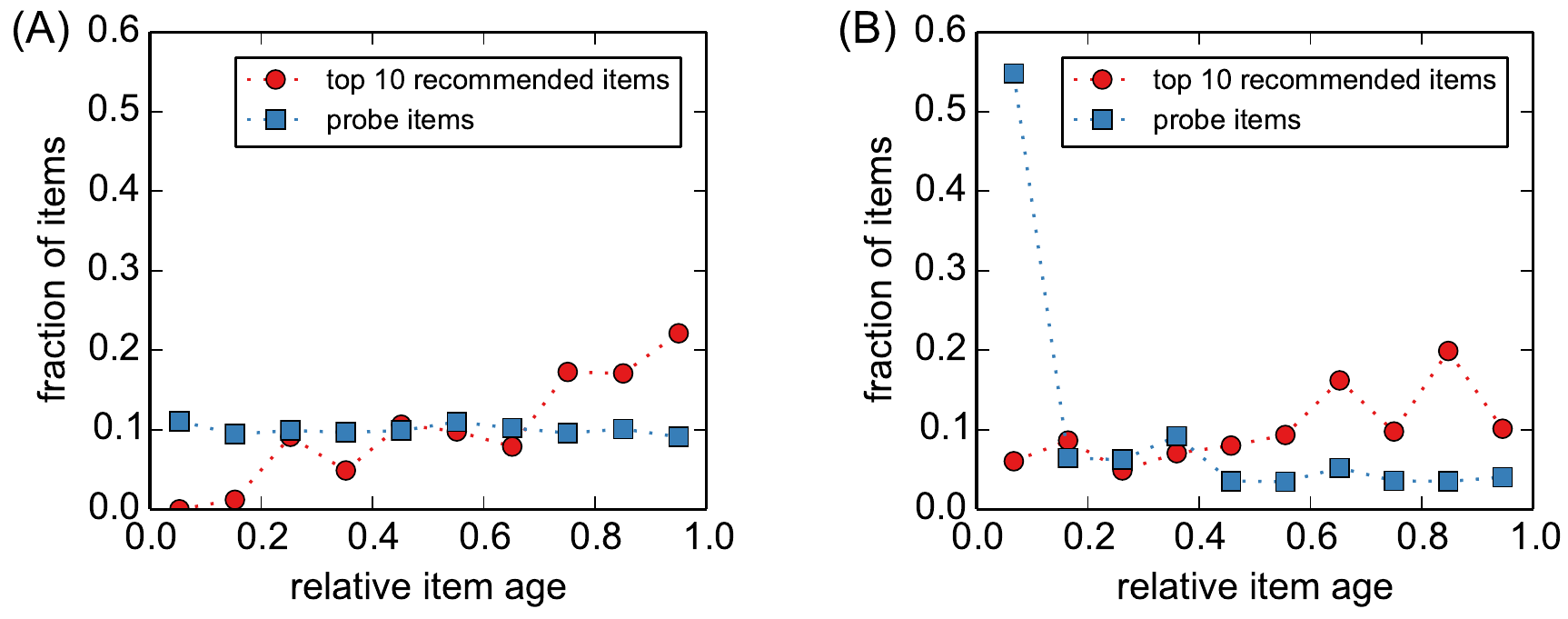}
\caption{For two distinct datasets (the popular Netflix dataset~\cite{bennett2007netflix} in panel A and the Econophysics Forum dataset~\cite{liao2014network} in panel B), we remove the most recent 1\% of links and use the remaining 99\% to compute recommendations with the classical probability spreading method~\cite{zhou2007bipartite}. The curves show the age distribution among the items at top 10 places of users' recommendation lists and among the items in the removed 1\% of links, respectively (item age is computed relatively with respect to the whole dataset's duration). Recent items are underrepresented in recommendations (and, correspondingly, old items are overrepresented) with both datasets. The disparity is greater for the Econophysics Forum data because aging is faster there than it is in the Netflix data.}
\label{fig:time_probe}
\end{figure*}

As shown in \cite{vidmer2016role}, measured performance of recommendation indeed deteriorates once time-based evaluation is employed. The main reason why network-based methods fail in reproducing the most recent links in a network is that while the methods generally favor popular items, the growth of real networks is typically determined not only by node popularity but also by node age that helps recent nodes to gain popularity despite the presence of old and already popular nodes~\cite{medo2011temporal}. In other words, the omnipresent preferential attachment mechanism~\cite{jeong2003measuring} is modulated with other influences---node age and node fitness, above all. The difference between the age of items attached with the latest links and the age of items scoring highly in a classical network-based recommendation method is illustrated in Figure~\ref{fig:time_probe}.

This suggests that the performance of network-based methods could be improved by enhancing the standing of recent nodes in the recommendation lists. While there are multiple ways how to achieve that, Vidmer and Medo \cite{vidmer2016role} present a straightforward combination of a traditional method's recommendation scores with item degree increase in a recent time period (due to the aging of nodes, high degree increase values tend to be achieved by recent nodes; this is also illustrated in Figure~\ref{fig:time_probe}). Denoting the recommendation score of item $\alpha$ for user $i$ at time $t$ as $f_{i\alpha}(t)$ and the degree increase of item $\alpha$ over the last $\tau$ time steps as $\Delta k_{\alpha}(t, \tau)$, the best-performing modified score was
\begin{equation}
f_{i\alpha}(t)' = f_{i\alpha}(t)\,\frac{\Delta k_{\alpha}(t,\tau)}{k_{\alpha}(t)}.
\end{equation}
Besides multiplying the original score with the recent degree increase, we divide here with the current item degree $k_{\alpha}(t)$ which aims to correct the original method's bias towards popular items (both baseline methods utilized in \cite{vidmer2016role}, preferential spreading and similarity-preferential diffusion, have this bias). The above-described modification significantly improves recommendation accuracy (as measured by, for example, recommendation recall) and, at the same time, it tends to recommend less popular items than the original baseline methods (and thus follows the long-lasting pursuit for diversity-favoring recommendation methods~\cite{ziegler2005improving,zhang2008avoiding,adomavicius2012improving}). The described modified method is very elementary; How best to include time in network-based recommendation methods is thus still an open question.

\subsection{The impact of recommendation ranking on network growth}
Once we succeed in finding recommendation methods with satisfactory accuracy, the natural question to ask is what would happen if the users would actually follow the recommendations. In other words, once input data are used to compute recommendations, what is the potential effect of the recommender system on the data? This can be answered by coupling a recommendation method with a simple agent-based model where a user can become active, choose an item (based on the received recommendations or at random), and thus establish a new link in the network that affects the future recommendations for this user as well as, indirectly, for all the others.

As first shown by Zeng et al.~\cite{zeng2012reinforcing}, iterating recommendation results tends to magnify popularity differences among the items with popular items benefiting most from recommendation. To be able to study long-term effects of iterating recommendation, Zeng et al. \cite{zeng2015modeling} proposes to simulate a closed system where new links are added and, to prevent the bipartite user-item network from eventually becoming complete, the oldest links are removed. The authors show that popularity-favoring recommendation methods generally lead to a pathological stationary state where only a handful of items occupy the attention of all users. This concentration of the users' attention is measured by the Gini index which is the most commonly used measure of inequality~\cite{ceriani2012origins}; its values of zero and one correspond to the perfectly equal setting (where in our case, the degree of all items is the same) and the maximal inequality (where one item monopolizes all the links), respectively. Figure~\ref{fig:gini} shows the stationary value of the item degree Gini coefficient that emerges by iterating the above-described network rewiring process. The authors use the popular ProbS-HeatS hybrid method~\cite{zhou2010solving} which has one parameter, $\theta$, that can be used to tune from accurate and popularity-favoring recommendations for $\theta=0$ to little accurate and diversity-favoring recommendations for $\theta=1$. As can be seen in the figure, $\theta$ values of below approximately 0.6 produces recommendations of high precision but the stationary state is very unequal. By contrast, $\theta$ close to one produces recommendations of low precision and a low Gini coefficient in the stationary state. Figure~\ref{fig:gini} shows a hopeful intermediate regime where the sacrifice of some little fraction of the recommendation precision can lead to a stationary state with substantially lower Gini coefficient. This kind of analysis is important as it questions an important limiting factor of information filtering algorithms and suggests possible ways how to overcome it.

\begin{figure*}
\centering
\includegraphics[width=\textwidth]{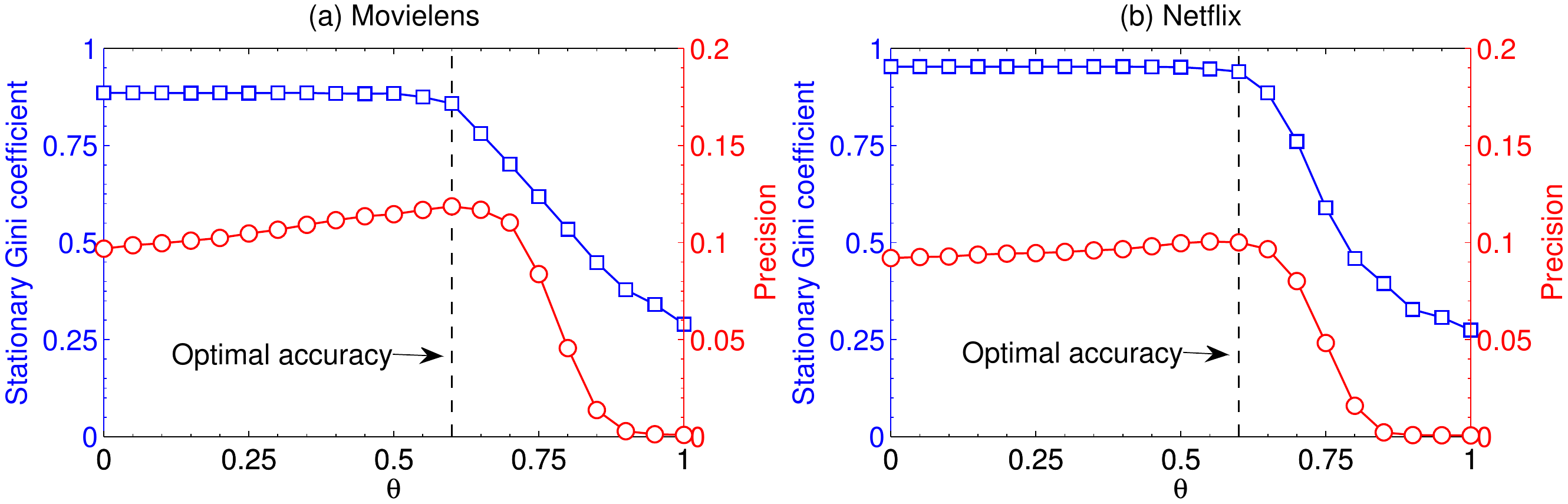}
\caption{Results of the rewiring process with the initial setting corresponding to two different real datasets, (a) Movielens and (b) Netflix, as a function of the $\theta$ parameter of the HeatS-ProbS recommendation method that is used to compute recommendations that drive the rewiring process. The blue left and red right axes show the stationary Gini coefficient in the long-term limit and the short-term recommendation, respectively. (Reprinted from~\cite{zeng2012reinforcing}.)}
\label{fig:gini}
\end{figure*}

\section{Applications of ranking to evolving social, economic and information networks}
\label{sec:applications}

While complex (temporal) networks constitute a general mathematical framework to describe a wide class of real complex systems, intrinsic differences between the meaning of nodes and their relationships across different datasets make it impossible to devise a unique, all-encompassing metric for node centrality. This has resulted in a plethora of static (section \ref{sec:static}) and time-dependent (sections \ref{sec:time} and \ref{sec:temporal} centrality metrics, each of them based on specific assumptions on what important nodes represent. We are left with a fundamental question: given a complex, possibly time-varying, network, which metric shall we use to quantify node importance in the system?

The answer certainly depends on the goals we aim to achieve with a given metric. On the other hand, for a specific goal, the relative performance of two metrics could be very different for different datasets. This makes it necessary to take a close look at the data, establish suitable and well-defined benchmarking criteria, and proceed to performance evaluation. The goal of this section is to present some case studies where this evaluation procedure can be carried out, and to show the benefits from including the temporal dimension into ranking algorithms and/or into the evaluation procedure.

In this section, we focus on three applications of node centrality metrics: ranking of agents (papers, researchers, and journals) in academic networks (paragraph \ref{sec:academic}), prediction of economic development of countries (paragraph \ref{prediction_gdp}), and link prediction in online systems (paragraph \ref{sec:link_prediction}).

\subsection{Quantifying and predicting impact in academic networks}
\label{sec:academic}
Scientific citation data are a popular object of network study for three main reasons. Firstly, accurate citation records are available from a period that spends over more than hundred years (the often-used APS data start in 1893). Secondly, all researchers have a direct experience with citing and being cited, which makes it easier for them to understand and study this system. Thirdly, researchers are naturally keen on knowing how is their work perceived by the others, for which paper citation count is an obvious but rather simplistic measure. Over time, a whole discipline of \emph{scientometrics} has developed whose goal is to study the ways of quantifying the research impact at the level of individual papers and research journals, as well as individual researchers and research institutions~\cite{bar2008informetrics,mingers2015review}. We review here particularly those impact quantification aspects that are strongly influenced by the evolution (in this case largely growth) of the underlying scientific citation data.

\begin{figure*}
\centering
\includegraphics[scale=0.9]{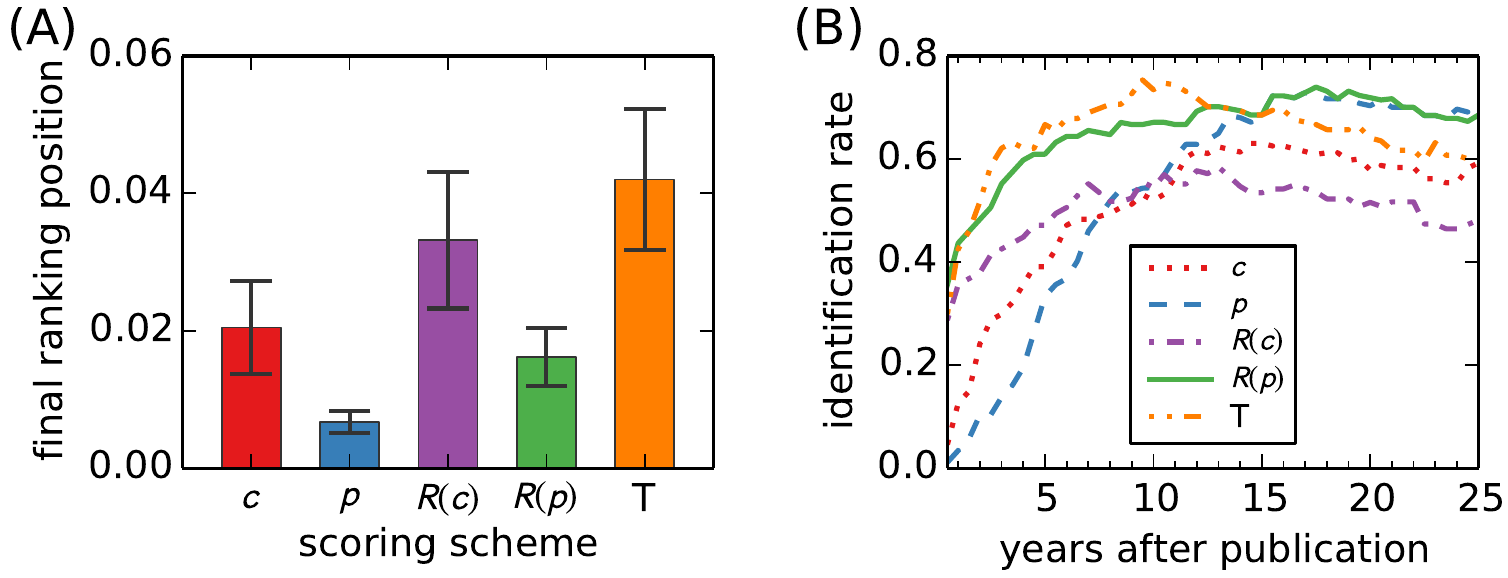}
\caption{Ranking performance comparison for a number of paper centrality metrics: citation count, $c$, PageRank with the teleportation parameter 0.5, $p$, rescaled citation count, $R(c)$, rescaled PageRank, $R(p)$, and CiteRank, $T$. (A) The average ranking position (normalized to the number of papers) computed on the complete APS data spanning from 1893 until 2009. (B) The dependence of the milestone identification rate on the age of milestone papers. (Adapted from~\cite{mariani2016identification}.)}
\label{fig:Rp_identif}
\end{figure*}

\subsubsection{Quantifying the significance of scientific papers}
\label{sec:milestones}
The most straightforward way to estimate the impact of a scientific paper is by computing its citation count (\ie, its in-degree in the citation network). One of the problems of citation count is that it is very broadly distributed~\cite{bar2008informetrics} which makes it difficult to interpret and use it further to, for example, evaluate researchers. For example, if researcher A authored two papers with citation counts 1000 and 10, respectively, and researcher B authored two papers with citation counts 100 and 100, respectively, the average citation count of authored papers points clearly in favor of author A. However, it is long known that the self-reinforcing preferential attachment process strongly contributes to the evolution of paper citation count~\cite{de1965networks,jeong2003measuring,redner2005citation} which suggests that one should approach high citation counts with some caution. Indeed, a recent study shows that the aggregation of logarithmic citation counts leads to better identification of able authors than the aggregation of original citation counts~\cite{medo2016model}. The little reliability of a single highly-cited paper was one of the reasons that motivated the introduction of the popular $h$-index that aims to estimate the productivity and research impact of authors.

To estimate paper impact beyond the mere citation count is therefore an important challenge. The point that first begs an improvement is that while the citation count weights all citations equally, the common sense tells us for that a citation from a highly influential paper should weight more than a citation from an obscure paper. The same logic is at the basis of PageRank~\cite{brin1998anatomy,franceschet2011pagerank} (see paragraph~\ref{sec:pagerank}), that has been therefore applied on the directed network of citations among scientific papers~\cite{chen2007finding,maslov2008promise,gleich2015pagerank}. While PageRank can indeed identify some papers (referred to as ``scientific gems'' by Chen et al.~\cite{chen2007finding}) whose PageRank score is high in comparison with their citation count \cite{chen2007finding}, the interpretation of results is made difficult because of the PageRank's strong bias towards old papers.

The age bias of PageRank in growing networks is natural and well understood~\cite{mariani2015ranking} (see paragraph~\ref{sec:pr_bias} for details). It can be countered by either introducing explicit penalization of old nodes (see the CiteRank algorithm described in Section~\ref{sec:citerank}) or by a posteriori rescaling the resulting PageRank scores as proposed in~\cite{mariani2016identification} (see paragraph~\ref{sec:rescaled} for a detailed description of the rescaling procedure). The latter approach has the advantage of not relying on any model or particular features of the data---it is thus widely applicable without the need for any alterations.

After demonstrating that the proposed rescaling procedure indeed removes the age bias of citation 
impact metrics, Mariani et al. \cite{mariani2016identification} proceeded by validating the resulting 
paper rankings. To this end, they used a small group of 87 milestone papers that have been selected in 
2016 by editors of the prime American Physical Society (APS) journal, Physical Review Letters, on 
the occasion of the journal's 50th anniversary [see the accompanying website \url{http://journals.aps.org/prl/50years/milestones}].
All these papers have been published between 1958 (when Physical Review Letters have been founded) and 2001. The editors choosing the milestone letters thus acted in 2015 with the hindsight of 15 years and more, which adds credibility to their choice.

To compare the rankings of papers by various metric, one can use the input data to produce the rankings and then assess the milestone papers' positions in each of them. Denoting the milestone papers with $i=1,\dots,P$ and the ranking of the milestone paper $i$ by metric $m=1,\dots,M$ as $r_i^{(m)}$, one can for example compute the average ranking of all milestone papers for each metric, $r^{(m)}_{\text{MS}}=\frac1P\sum_i r_i^{(m)}$. A low value of $r^{(m)}_{\text{MS}}$ indicates that metric $m$ assigns the milestone papers good (low) positions in the ranking of all papers. The results shown in Figure~\ref{fig:Rp_identif}a indicate that the original PageRank, followed by its rescaled variant and the citation count, are the best metrics in this respect.

However, using the complete input data to compare the metrics inevitably provides a particular view: it only reflects the ranking of milestone papers after a fixed rather long time period of time (recall that many milestone papers are now more than $30$ years old). Furthermore, a metric with strong bias towards old papers (such as PageRank, for example), is given undue advantage which makes it difficult to compare its result with another metric which has a different level of bias. To overcome these problems, Mariani et al.~\cite{mariani2016identification} provide a ``dynamical'' evaluation where the evolution of each milestone's ranking position is followed as a function of its age (time since publication). The simplest quantity to study in this way is the \emph{identification rate}: the fraction of milestone papers that appear in the top 1\% of the ranking as a function of the time since publication (one can of course choose a different top percentage). This dynamical approach not only provides us with more detailed information but also avoids being influenced by the age distribution of the milestone papers. 

The corresponding results are shown in Figure~\ref{fig:Rp_identif}. There are two points to note: time aware quantities, rescaled PageRank $R(p)$ and CiteRank $T$, rank the milestone papers better than time unaware quantities, citation count $c$ and PageRank $p$, in the first 15 years after publication. Second, network-based quantities, $p$ and $R(p)$, outperform their counterparts based on simply counting the incoming citations, $c$ and $R(c)$, which indicates that taking the complete network structure of scientific citations is indeed beneficial for our ability to identify significant scientific contributions. As shown recently by Ren et al.~\cite{ren2017time}, it is not given that PageRank performs better than citation count: no performance improvement is found, for example, in the movie citation network constructed and analyzed by Wasserman et al.~\cite{wasserman2015cross}. Ren et al.~\cite{ren2017time} argue that the main difference lies in the lack of degree-degree correlations in the movie citation networks, which in turn undermines the basic assumptions of the PageRank algorithm. 

The interactive website \url{sciencenow.info} makes the results obtained with rescaled PageRank and other metrics on regularly updated APS data (as of the time of writing this review, the database includes all papers published until December 2016) publicly available. The very recent paper on the first empirical observation of gravitational waves, for example, reaches a particularly high rescaled PageRank value of 28.6 which is only matched by a handful of papers.\footnote{Recall that the $R(p)$ value quantifies by how many standard deviations a paper outperforms other papers of similar age.}
Despite being very recent, the paper is ranked 17 out of 593,443 papers, as compared to its low rankings by the citation count (rank 1,763) and PageRank (rank 12,277). Among the recent papers that score well by $R(p)$ are recent seminal contributions to the study of graphene (The electronic properties of graphene from 2009) and topological insulators (two papers from 2010 and 2011); both these topics have recently received Nobel Prize in physics. In summary, rescaled PageRank allows one to appraise the significance of published works much earlier than other methods. Thanks to its lack of age bias, the metric can be also used to select both old and recent seminal contributions to a given research field.

The rescaling procedure has also its pitfalls. First, it rewards the papers with immediate impact and, conversely, penalizes the papers that need long time to prove their worthiness, such as the so-called \emph{sleeping beauties} \cite{van2004sleeping, ke2015defining}. 
Second, the early evaluation of papers with rescaled quantities naturally results in an increased rate of false positives---papers that due to a few quickly obtained citations initially reach high values of $R(p)$ and $R(c)$ but these values substantially decrease later\footnote{A spectral-clustering method to classify the papers according to their citation trajectories has been recently proposed by Colavizza and Franceschet \cite{colavizza2016clustering}. According to their terminology, papers with large short-term impact and fast decay of attractiveness for new citations are referred to as \emph{sprinters}, as opposed to \emph{marathoner} papers that exhibit slow decay.}. 
If we take into account that those early citations can be also ``voices of disagreement'', it is clear that early rescaled results need to be used with caution. A detailed study of these issues remains a future challenge.

\subsubsection{Predicting the future success of scientific papers}
\label{sec:papers}
When instead of paper significance, we are only interested in the eventual number of citations we may follow the approach demonstrated by D. Wang et al.~\cite{wang2013quantifying} where the network growth model with preferential attachment and aging~\cite{medo2011temporal} was calibrated on citation data and the aging was found to decay log-normally with time. While the papers differ in the parameters of their log-normal aging decay, these differences do not influence the predicted long-term citation count which is shown to depend only on paper fitness $f_i$ and overall system parameters. If paper fitness is obtained from, for example, a short initial period of the citation count growth, the model thus can be used to predict the eventual citation count. As the authors show, papers with similar estimated fitness indeed achieve much more similar eventual citation count than the papers that had similar citation counts at the moment of estimation. While J. Wang et al.~\cite{wang2014comment} point to a lack of the method's prediction power, D. Wang et al. \cite{wang2014response} explain that this is a consequence of model over-fitting and discuss how it can be avoided.

A recent prediction method proposed by Cao et al.~\cite{cao2016data} predicts future citations of papers by building paper popularity profiles and finding the best-matching profile for any given paper; its results are shown to outperform those obtained by the approach by D. Wang et al.~\cite{wang2013quantifying} (see Figure~\ref{fig:cao_relative_error}). Besides machine-learning approaches being traditionally strong when enough data are available, the method is not built on any specific model of paper popularity growth as it directly uses empirical paper popularity profiles. This can be an important advantage, given that the network growth model used in~\cite{wang2013quantifying,medo2016model} certainly lacks some features of the real citation network. For example, Petersen et al.~\cite{petersen2014reputation} considered the impact of author reputation on the citation dynamics and found that the author's reputation dominates the paper citation growth rate of little cited papers. The crossover citation count between the reputation- and preferential attachment-dominated dynamics is found to be 40 (see~\cite{petersen2014reputation} for details). Golosovsky and Solomon~\cite{golosovsky2017growing} went one step further this and proposed a thoroughly modified citation growth model built on a copying-redirection-triadic closure mechanism. The use of these models in the prediction of paper success has not been tested yet.

\begin{figure*}
\centering
\includegraphics[scale=0.3]{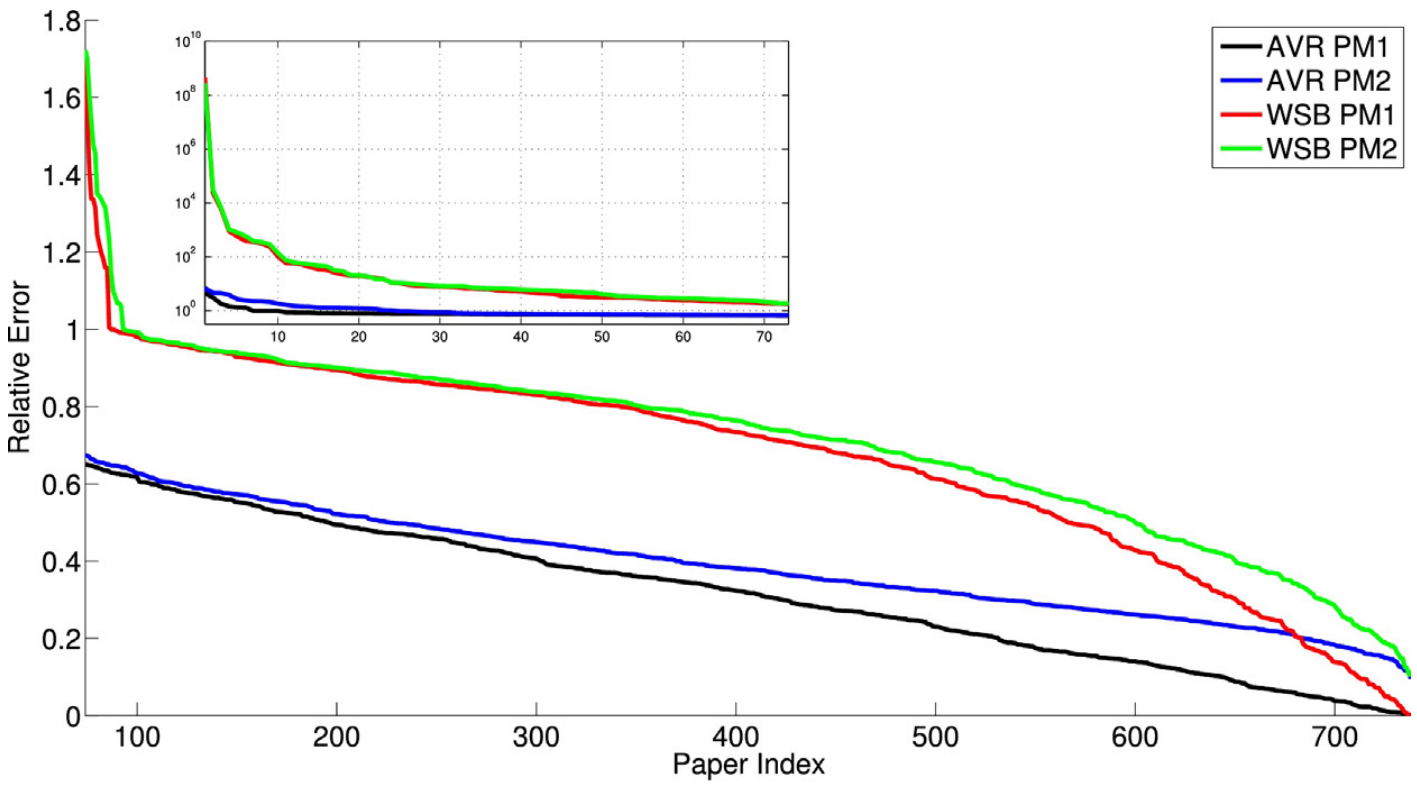}
\caption{The relative citation count prediction error (vertical axis) obtained with the method by D. Wang et al.~\cite{wang2013quantifying}, labeled as WSB, as compared to the basic ``data analytic'' approach by Cao et al.~\cite{cao2016data}, labeled as AVR. Papers published by the Science journal in 2001 were used for the error evaluation. To visualize the error distribution among the papers, the papers are ranked by their relative error for each prediction method separately. Of note is the small group of papers with large relative error shown in the inset. The AVR method suffers from this problem less than the WSB method. (Figure taken from~\cite{cao2016data}.}
\label{fig:cao_relative_error}
\end{figure*}

A comparison of the paper's in-degree and the appearance time, labeled as rescaled in-degree in paragraph~\ref{sec:rescaled}, has been used by Newman~\cite{newman2009first} to find the papers that are likely to prove highly successful in the future (see a detailed discussion in Section~\ref{sec:first-mover}). Five years later~\cite{newman2014prediction}, the previously identified outstanding papers were found to significantly outperform a group of randomly chosen papers as well as a group of randomly chosen papers with the same number of citations.
To specifically address the question of predicting the standing of papers in the citation network consisting solely of the citations made in near future, Ghosh et al. \cite{ghosh2011time} use a model of the reference list creation which is based on following references in existing papers whilst preferring the references to recent papers (see paragraph~\ref{sec:ecm} for details).

An interesting departure from works that build on the citation network among the papers is presented by Sarigol et al.~\cite{sarigol2014predicting} who attempt to predict the paper's success solely on the basis of the paper's authors position in the co-authorship network. On a dataset of more than 100,000 computer science papers published between 1996 and 2007, they show that among the papers authored by an author who is in the top 10\% by (simultaneously) betweenness centrality, degree centrality, $k$-core centrality and eigenvector centrality, 36\% of these papers are eventually among the top 10\% most cited papers five years after publication. By combining the centrality of paper co-authors with a machine-learning approach, this fraction (which represents the prediction's precision) can be further increased to 60\% whilst the corresponding recall is 18\% (\emph{i.e.}, out of the 10\% top cited papers, 18\% are found with the given predictor).

\subsubsection{Predicting the future success of researchers}
Compared with the prediction of the future success of a scientific publication, the prediction of the future success of an individual researcher is much more consequential because it is crucial in distributing the limited research resources (research funds, permanent positions, etc.) among a comparatively large number of researchers. Will a researcher with a highly cited paper produce more papers like that or not? Early on, the predictive power of $h$-index has been studied by Hirsch himself in~\cite{hirsch2007does} with the conclusion that this metric predicts a researcher's success in the future better than other simple metrics (total citation count, citations per paper, and total paper count).

It is probably naive to expect that a single metric, regardless of how well designed, can capture and predict the success in such a complex endeavor as research is. Acuna et al.~\cite{acuna2012future} attempted} to predict the future $h$-index of researchers by combining multiple features using linear regression. For example, the prediction formula for the next year's $h$-index of neuroscientists had the form $h_{+1} = 0.76 + 0.37\sqrt{n} + 0.97h - 0.07y + 0.02j + 0.03q$ where $n$, $h$, $y$, $j$, and $q$ are the number of papers, current $h$-index, number of years since the first paper, number of distinct journals where the author has published, and the number of articles in top journals, respectively. However, this work was later criticized by Penner et al.~\cite{penner2013predictability}. One of the discussed problems is that to predict the $h$-index is relatively simple as this quantity changes slowly and can only grow. The baseline prediction of unchanged $h$-index values is thus likely to be rather accurate. According to Penner et al., one should instead aim to predict the change of the author's $h$-index and evaluate the predictions accordingly; in this way, no excess self-confidence will result from giving ourselves a simple task. Another problem of the original evaluation is that it considered all researchers and thus masked the substantial level of errors specific to young researchers who, at the same time, are the most likely candidates of similar evaluation in evaluations by hiring committees, for example. The young researchers who have published their first first-author paper recently are precisely the target of Zhang et al.~\cite{zhang2016identifying} who use a machine-learning approach to predict which young researchers will be most cited in the near future, and show that their method outperforms a number of benchmark methods.

As recently shown by Sinatra et al.~\cite{sinatra2016quantifying}, the most-cited paper of a researcher is equally likely to appear at the beginning, in the middle, or at the end of the researcher's career. In other words, researchers seem not to become better, or worse, during their careers. Consequently, the authors develop a model which assigns scientists the ability factor $Q$ that multiplicative improves ($Q>1$) or lowers ($Q<1$) the potential impact of their next work. The new $Q$ metric is empirically shown to be stable over time and thus suitable for predicting the authors' future success, as documented by predicting the $h$-index. Note that $Q$ is by definition an age unbiased metric and thus allows to compare scientists of different age without the need for any additional treatment (such as the one presented in Section~\ref{sec:pr_bias} for the PageRank metric).

The literature on this topic is large and we present here only a handful of relevant works. An interested reader is referred to a very recent essay on data-driven predictions of science development~\cite{clauset2017data} summarizes the achieved progress, future opportunities, and the potential impact.

\subsubsection{Ranking scientific journals through non-markovian PageRank}
\label{sec:journals}
In section \ref{sec:temporal}, we have described node centrality metrics for temporal networks.
We have shown that higher-order networks describe and predict actual flows in real networks better than memory-less networks (section \ref{sec:time_respecting_paths}).
In this section, we focus on a particular application of memory-aware centrality metrics, the ranking of scientific journals \cite{rosvall2014memory}.

Weighted variants of the PageRank algorithm~\cite{bergstrom2008eigenfactor, gonzalez2010new} attempt to infer the importance of a journal based on the assumption that a journal is important if its articles are cited relatively many times by papers published in other important journals.
However, such PageRank-based methods are implicitly built on the Markovian assumption, which may provide a poor description of real network traffic (see section \ref{sec:time_respecting_paths}).
As pointed out by \cite{rosvall2014memory}, this is the case also for multi-disciplinary journals like \emph{Science} or \emph{PNAS} which
receive citation flows from journals from various fields. If we neglect memory effects,  multi-disciplinary journals
redistribute the received citation flow toward journals from diverse fields.
This effect causes the flow to repeatedly cross field boundaries, which is arguably unlikely to be done by a real scientific reader and, for this reason, makes the algorithm unreliable as a model for scholars' browsing behavior.

By contrast, in a second-order Markov model, if a multi-disciplinary journal $J$ receives a certain flow from a journal which belongs to field $F$,
then journal $J$ tends to redistribute this flow to journals that belong to the same field $F$ (see panels c-d of Fig. 4 in \cite{rosvall2014memory} for an example).
This results in a sharp reduction of the flow that crosses disciplinary boundaries (referred to as ``flow leak'' in \cite{rosvall2014memory})
and in a clear increase of the within-community flow.
As a result, when including second-order effects, 
multidisciplinary journals (such as \emph{PNAS} and \emph{Science}) 
receive smaller net flow, whereas specialized journals (such as \emph{Ecology} and \emph{Econometrica}) gain flow and thus improve their ranking (see \cite{rosvall2014memory} for the detailed results). 

This property makes also the ranking less subject to manipulation,
which is a worrying problem in quantitative research evaluation \cite{falagas2008top, wallner2009ban}. 
As we have just seen, if a paper written in a journal $J$ specialized in Ecology cites a multi-disciplinary journal,
this credit is likely to be redistributed across journals of diverse fields in a first-order Markov model.
For a memory-less ranking of journals, a specialized journal would be thus tempted to attempt to boost its score through editorial 
policies that discourage citations to multi-disciplinary journals.
By contrast, the second-order Markov model suppresses large part of flow leaks which makes it less convenient this strategic
citation manipulation.
The ranking of journals based on second-order Markov model has also been shown to be more robust with respect
to the selection of included journals in the analysis \cite{bohlin2015robustness},
which is an additional positive property for the second-order model.

In a scientific landscape where metrics of productivity and impact are proliferating \cite{van2010metrics}, assessing the biases and performance~\cite{medo2016model} of widely used metrics and devising improved ones is critical. For example, the widely used impact factor suffers from a number of worrying issues \cite{amin2003impact, plos2006impact, adler2009citation}.
The results presented by Rosvall et al. \cite{rosvall2014memory} indicate that the inclusion of temporal effects holds promise to improve the current rankings of journals, and hopefully future research will continue exploring this direction. 


\subsection{Prediction of economic development of countries}
\label{prediction_gdp}

Predicting the future economic development of countries is of crucial importance in economics. 
The GDP per capita is the traditional monetary variable used by economists to assess a country's development~\cite{kuznets1934national}. Even though it does not account for critical elements such as income equality, environmental impact or social costs~\cite{costanza2014development}, GDP is traditionally regarded as a reasonable proxy for the country's economical wealth, and especially of its industry~\cite{coyle2015gdp}.
Twice a year, in April and in October, the \emph{International Monetary Fund} (IMF) makes projections for the future GDP growth of countries~\cite{IMF}. The web page does not detail the precise procedure, but it indicates that it combines together many factors, such as oil price, food price, and others. 
Development predictions based on complex networks do not aim to beat the projections made by the IMF. They rather aim to extract valuable information from international trade data with a limited number of network-based metrics, attempting thus to isolate the key factors that drive economic development. The respective research area, known as \emph{Economic Complexity}, has attracted considerable attention in recent years both from the economics~\cite{hausmann2011network, felipe2012product} and from the physics community~\cite{tacchella2012new,mariani2015measuring}. 

Two metrics, the Method of Reflections \cite{hidalgo2009building}, and the Fitness-Complexity metric \cite{tacchella2012new}, have been designed to rank countries and  products that they export in the international trade network. The definitions of the two metrics (and some of their variants \cite{mariani2015measuring,wu2016mathematics}) can be found in section \ref{sec:mr} and section \ref{sec:fcm}, respectively.
Both algorithms use a complex network approach to represent the international trade data. One builds a bipartite trade network where the countries are connected with the products that they export. The network is built using the concept of Revealed Compared Advantage, RCA (we refer to \cite{hidalgo2009building} for details), which distinguishes between ``significant'' and ``non-significant'' links in the country-product export network. The resulting network is binary: a country is either considered to be an exporter of a product or it is not. 

Once we have defined network-based metrics to rank countries and products in world trade, a natural question arises: can we use these metric to predict the future development of countries?
This question is not univocal: several prediction-related tasks can be designed and the relative performance of the metrics can be different for different tasks. We shall present two possible approaches below.

\subsubsection{Prediction of GDP through linear regression}
Hidalgo and Hausmann \cite{hidalgo2009building} have employed the standard linear-regression approach to the GDP prediction problem: one designs a linear model where the future GDP, $\log[GDP(t+\Delta t)]$, depends linearly on the current GDP, $\log[GDP(t)]$, and on network-based metrics. The linear-regression equations used by \cite{hidalgo2009building} read
\begin{equation}
\begin{split}
&\log[GDP(t+\Delta t)]-\log[GDP(t)]=a+b_1\, GDP(t) + \\
&b_2 \, k_i^{(n)}+b_3\, k_{i}^{(n+1)},
\end{split}
\end{equation}
where $k_{i}^{n}$ denotes the $n$th-order country Method-of-Reflection score (hereafter MR-score) as determined with Eq.~\eqref{eq:MR}, while $a, b_1, b_2, b_3$ are coefficients to be determined with a fit to the data. 
The results by \cite{hidalgo2009building,hausmann2014atlas} show that the Method-of-Reflection score contributes to the variance of economic growth significantly more than measures of governance and institutional quality, education system quality, and standard competitiveness indexes such as the World Economic Forum Global Competitiveness Index. We refer the reader to \cite{hausmann2014atlas} for the detailed results, and to \url{http://atlas.media.mit.edu/en/} for a web platform with rankings by the MR and visualizations of the international trade data.

\subsubsection{Prediction of GDP through the method of analogues}
The linear-regression approach by \cite{hidalgo2009building,hausmann2014atlas} has been criticized by Cristelli et al. \cite{cristelli2015heterogeneous} who point out that the linear-regression approach assumes that there is a \emph{uniform} trend to uncover. In other words, it assumes that all the countries respond to a variation in their score in the same way, \ie, with the same regression coefficients. 
Cristelli et al. \cite{cristelli2015heterogeneous} use the country fitness score defined by Eq.~\eqref{eq:FC} to show that instead, the dynamics of countries in the Fitness-GDP plane is very heterogeneous, and the predictability of countries' future position is only possible in a limited region of the plane. This undermines the very basic assumption of the linear regression technique \cite{cristelli2015heterogeneous}, and calls for a different prediction analysis tool.

Cristelli et al. \cite{cristelli2015heterogeneous} borrow a well-known method from the literature on dynamical systems \cite{lorenz1969atmospheric,lorenz1969three, wolf1985determining,cencini2010chaos}, called \emph{method of analogues}, to study and predict the dynamics of countries in the Fitness-GDP plane.
The basic idea of the method is that we can try to predict the future history of the system by using our knowledge on the past history of the system.
The main goal of the method is to reveal simple patterns and attempt prediction in systems for which we do not know the laws of evolution.
While~\cite{cristelli2015heterogeneous} only considers the Fitness-GDP plane, with Fitness determined through Eq.~\eqref{eq:FC}, we consider here the score-GDP plane for three different scores: fitness, degree, and the Method-of-Reflections (MR) score\footnote{To obtain the MR score, we perform two iterations of Eq.~\eqref{eq:MR}. We have checked that the performance of the Method of Reflections does not improve by increasing the number of iterations.}.

In the following, we analyze the international trade BACI dataset (1998-2014) composed of $N=261$ countries and $M=1241$ products\footnote{In international trade datasets, different classifications of products can be found. In the BACI dataset used here, products are classified according to the Harmonized System 2007 scheme, which classifies hierarchically the products by assigning them a six-digits code. For the analysis presented in this paragraph, we aggregated the products to the fourth digit.}; from this dataset, we use the procedure introduced by Hidalgo and Hausmann \cite{hidalgo2009building} to construct, year by year, the country-product network that connects the countries with the products they export.
The first insights into the heterogeneity of the country dynamics in the score-GDP plane can be gained through the following procedure:
\begin{itemize}
\item Split the Fitness-GDP plane (in a logarithmic scale) in equally-sized boxes.
\item The box in which each country is at a given year is noted, as well as the one at which it is ten years later.
\item The total number of countries in box $i$ at year $t$ is $N^i_t$, and the number of different boxes
occupied after ten years by countries originating from box $i$ is labeled $n^i_t)$.
\item For each box, the dispersion $\mathcal{C}^i_t= (n^i_t-1)/(N^i_t-1)$ is computed.
\end{itemize}
Here a box whose all countries will occur in the same box after 10 years has zero dispersion, and a box whose $N^i_t$ countries will occur in $N^i_t$ different boxes after 10 years has the dispersion of one. Therefore, for a given box, a small value of $\mathcal{C}^i_t$ means that the future evolution of countries that are located at that box is more predictable than the evolution of countries located at boxes with large dispersion values.

\begin{figure}
\centering
\includegraphics[width=0.55\columnwidth]{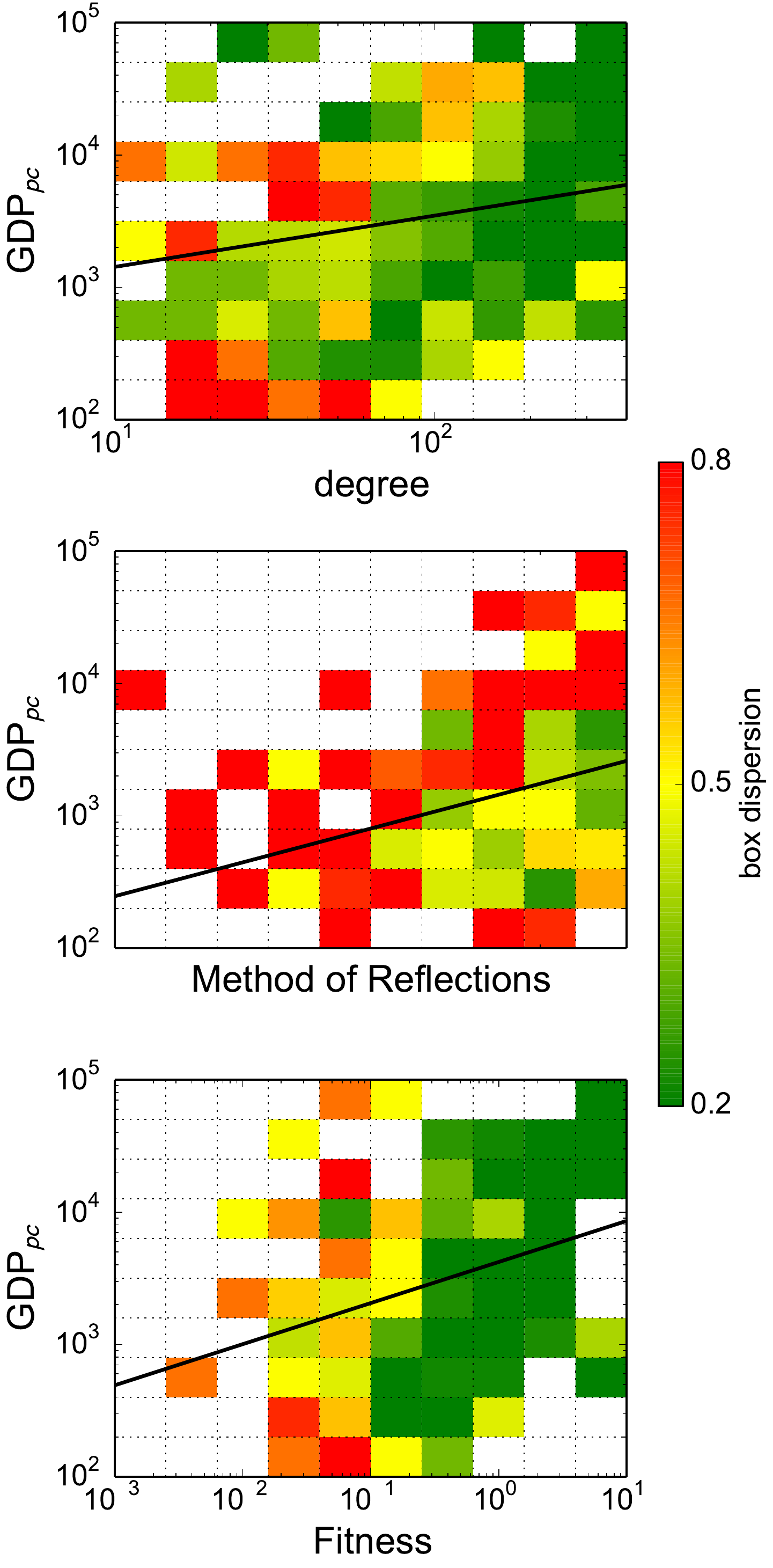}
\caption{Box dispersion in the score-GDP plane for three different country-scoring metrics: degree, Method of Reflections, and Fitness. The results were obtained on the BACI international trade data ranging from 1998 to 2014 \cite{gaulier2010baci} (see the main text for details).}
\label{fig:analogs}
\end{figure}

In agreement with the results in \cite{cristelli2015heterogeneous}, we observe (Fig.~\ref{fig:analogs}) an extended ``low-dispersion'' (\ie, predictable) region in the Fitness-GDP plane, which corresponds essentially to high-fitness countries. This means that for high-fitness countries, fitness is a reliable variable to explain the future evolution of the system. By contrast, for low-fitness countries, fitness fails to capture the future dynamics of the system and additional variables (\eg, institutional or education quality indicators) may be necessary to improve the prediction performance. Cristelli et al. \cite{cristelli2015heterogeneous} point out that this strongly heterogeneous dynamics goes undetected with standard forecasting tools such as linear regressions. Interestingly, an extended low-dispersion region is also found in the degree-GDP plane, whereas the low-dispersion region is comparatively small in the MR-score-GDP plane.

The complex information provided by the dispersion values in a two-dimensional plane can be reduced by computing a weighted mean of dispersion, which weights the contribution of each box with the number of events that fall into that box. The obtained values of the weighted mean of dispersion are $0.41$, $0.71$, and $0.35$ for the degree, Method of Reflections, and Fitness, respectively. These numbers suggest that while both degree and Fitness significantly outperform the predictive power of the Method of Reflections, Fitness is the best metric by some margin. The results are qualitatively the same when a different number of boxes is used.
These findings indicate that fitness and degree can be used to make predictions in the score-GDP plane, whereas the Method of Reflections cannot be used to make reliable predictions with the method of analogues. The simple aggregation method used here to quantify the prediction performance of metrics can be in the future improved to better understand the methods' strong and weak points.
An illustrative video of the evolution of countries in the Fitness-GDP plane, available at 
\url{http://www.nature.com/news/physicists-make-weather-forecasts-for-economies-1.16963}, provides a general perspective on the two above-described economic-complexity approaches to GDP prediction.

\begin{figure*}
\centering
\includegraphics[scale=0.6]{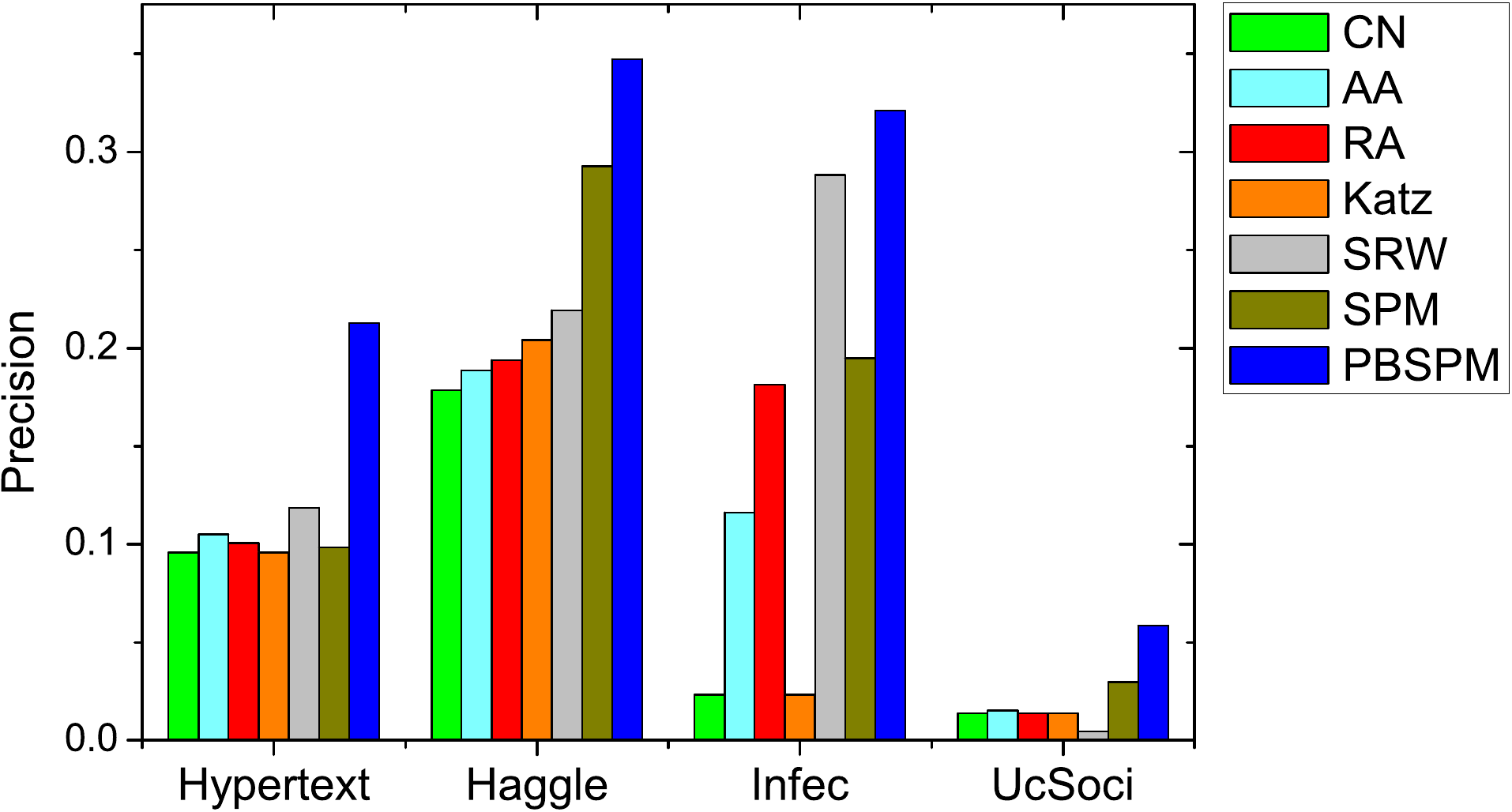}
\caption{
Precision of traditional methods vs. time-aware link prediction based on structural perturbation (PBSPM) for different networks. In the prediction, $90\%$ historical links of the whole network are utilized to predict $10\%$ future links. The six state-of-the-art methods~\cite{lu2011link,lu2015toward} are Common neighbors (CN), Adamic–Adar Index (AA), Resource Allocation Index (RA), Katz Index, Superposed Random Walk (SRW) and Structural Perturbation Method (SPM).}
\label{fig:link_pred_temp}
\end{figure*}

\subsection{Link prediction with temporal information}
\label{sec:link_prediction}
Link prediction is an important problem in network analysis~\cite{hart2006complete,lu2011link}, which aims to identify the potential missing links in a network. This is particularly useful in settings where to obtain information about the network is costly, such as in gene interaction experiments in biology~\cite{barzel2013network,menche2015uncovering}, and it is thus useful to prioritize the experimentation by first studying the node pairs that are the most likely to be actually connected with a link. The problem has its relevance also in the study of social networks~\cite{liben2007link,scott2012social}, for example, where it can tell us about their evolution in the future. There is also an interesting convoluted setting where information diffuses over the social network and this diffusion can influence the process of social link creation~\cite{li2016exploiting}. A common approach to link prediction is based on choosing a suitable node similarity metric, computing the similarity for all yet unconnected node pairs, and finally assuming that the links are most likely to exist among the nodes that are the most similar by the chosen metric. In evolving networks, node similarity should take into account not only the current network topology but also time information (such as the node and link creation time). The simple common neighbor similarity of nodes $i$ and $j$ can be computed as $s_{ij}=\sum_{\alpha}A_{i\alpha}\,A_{j\alpha}$, where $\mathsf{A}$ is the network's adjacency matrix. See~\cite{lu2011link} for more information on node similarity estimation and link prediction in general.

Usual node similarity metrics ignore the time information: the old and new links are assumed to have the same importance, and the same holds for nodes. There are various ways how time can be included in the computation of node similarity. The simple common neighbors metric, for example, can be generalized by assuming that the weight of common neighbors decays exponentially with time, yielding $s_{ij}(t)=\sum_{n\in\Gamma(i, t)\cap\Gamma(j, t)} \ee^{-\lambda (t-t_{ij})}$ where $\Gamma(i,t)$ is the set of neighbors of node $i$ at time $t$, $t_{ij}$ is the time when edge between $i$ and $j$ has been established, and $\lambda$ is a tunable decay parameter. Liu et al.~\cite{liu2009link} introduce time in the resource allocation dynamics on bipartite networks (first described in~\cite{zhou2007bipartite}). Other originally static similarity metrics can be also modified in a similar way~\cite{tylenda2009towards}. Note that in constructing time-reflecting similarity metrics, there exist a number of problems: How to characterize the time-decay parameters, whether edges with different timestamps play different roles in the dynamics, whether and how to include users' potential preference for fresher or older objects, and so on~\cite{dhote2013survey,munasinghe2011time}.

We illustrate now how time can be easily introduced in a recently proposed link prediction method based on a perturbed matrix~\cite{lu2015toward}; this time-aware generalization has been recently presented in~\cite{wang2017link}. The original method assumes an undirected network of $N$ nodes. A randomly chosen fraction $p^H$ of its links constitute a perturbation set with the adjacency matrix $\Delta\mat{A}$, and the remaining links correspond to the adjacency matrix $\mat{A}^R$; the complete network's adjacency matrix is $\mat{A}=\mat{A}^R+\Delta\mat{A}$. Since $\mat{A}^R$ is real and symmetric, it has a set of $N$ orthonormal eigenvectors $\vek{x}_k$ corresponding to eigenvalues $\lambda_k$ and can be written as $\mat{A}^R=\sum_{k=1}^N \lambda_k \vek{x}_k\vek{x}_k\TT$. The complete adjacency matrix, $\mat{A}$, has generally different (perturbed) eigenvalues and eigenvectors that can be written as $\lambda_k+\Delta\lambda_k$ and $\vek{x}_k+\Delta\vek{x}_k$, respectively. Assuming that the perturbation is small, one can find $\Delta\lambda_k = (\vek{x}_k\TT \Delta\mat{A}\vek{x}_k)/(\vek{x}_k\TT\vek{x}_k)$ and $\mat{A}$ can be approximated as
\begin{equation}
\label{eq:perturbationoriginal}
\tilde{\mat{A}} = \sum_{k=1}^N (\lambda_k + \Delta\lambda_k) \vek{x}_k\vek{x}_k\TT,
\end{equation}
where we kept the perturbed eigenvectors unchanged. This matrix can be considered as a linear approximation (extrapolation) of the network's adjacency matrix based on $\mat{A}^R$. As discussed in~\cite{lu2015toward}, the elements of $\tilde{\mat{A}}$ can be considered as link score values in the link prediction problem.

However, the original method divides training and probe sets randomly regardless of time, implying an unattainable scenario that future edges are utilized to predict future edges. It is more appropriate to choose the most recent links to constitute the perturbation set (again, the fraction $p^H$ of all links is chosen). To evaluate link prediction in a system where time plays a strong role, one must not choose the probe at random, but instead divide the network by time so that strictly the future links are there to be predicted (see a similar discussion in Sec.~\ref{sec:rec-time}). To reflect the temporal dynamics of the network's growth, and thus appropriately address the time-based probe division, Wang et al. introduce the ``activity'' of node $i$ as $s_i = k_i^H / k_i$ where $k_i\TT$ and $k_i$ are the degree of node $i$ in the perturbation set and the complete data~\cite{wang2017link}, respectively. Node activity $s_i$ aims the capture the trend of $i$'s popularity growth: whether it is on a rise or it already slows down. In addition, active nodes are more likely to be connected by a link than little active ones.

Here elements of the eigenvector $\vek{x}_k$ can be modified by including node activity as
\begin{equation}
\label{eq:advancedattractiveness}
x_{k,i}'= x_{k,i}(1+\alpha s_i)
\end{equation}
where $\alpha$ is a tunable parameter (by setting $\alpha=0$, node activity is ignored and the original eigenvectors are recovered). The modified eigenvectors can be plugged in~\eqref{eq:perturbationoriginal}, yielding the time-aware perturbed matrix
\begin{equation}
\label{eq:perturbationmodify}
\tilde{\mat{A}}' = \sum_{k=1}^{N}(\lambda_k+\Delta \lambda_k)\vek{x}_k'\vek{x}_k'{}\TT.
\end{equation}
Similarly as before, the elements of $\tilde{\mat{A}}'$ can be interpreted as link prediction scores for the respective links.

In~\cite{wang2017link}, a number of different link prediction methods 
is compared on four distinct datasets: Hypertext data \cite{isella2011s}, 
Haggle data~\cite{chaintreau2007impact}, Infec data~\cite{isella2011s} and 
UcSoci data~\cite{opsahl2009clustering}). Figure~\ref{fig:link_pred_temp} summarizes the results 
obtained by the authors and shows that taking the time evolution of node popularity into account dramatically improves the prediction performance.
Apart from the methods introduced above, several further 
time-aware approaches to link prediction have been 
proposed recently, such as similarity fusion of multilayer networks \cite{moradabadi2016link}, 
real-world information flow feature in link prediction \cite{li2016exploiting}, and others.

\section{Perspectives and conclusions}
\label{sec:perspectives}

After reviewing the recent progress in the actively studied problem of node ranking in complex networks, the reader may arguably feel overwhelmed by the sheer number and variability of the available static and time-aware metrics. Indeed, the understanding which metric to use in which situation is often lacking and one relies on ``proofs'' of metric validity that are often anecdotal, based on a qualitative comparison between the rankings provided by different metrics \cite{zweig2011good}. This issue is quite worrying as a wrongly used metric can have immediate real consequences as is the case, for example, for quantitative evaluation of scientific impact, where the used metrics have consequences on funding decisions and academic careers. Waltman \cite{waltman2016review} points out that since the information that can be extracted from citation networks is inherently limited, researchers should avoid introducing new metrics unless they bring a clear added value to existing metrics. We can only add that this recommendation applies equally to other fields where metrics tend to multiply.

Coming back to the issue of metric validation, we believe that extensive and thorough performance evaluation of (both static and time-aware) network ranking metrics is an important direction future research.
Of course, any ``performance'' evaluation depends on which task we want to achieve with the metric. We outline possible methods to validate centrality metrics in paragraph \ref{sec:benchmarking}. 
Whichever validation procedure we choose for the metrics, we cannot neglect the possible bias of the metrics and their impact on the system's evolution.
In this review, we focused on the bias by age of static algorithms and possible ways to counteract it. However, the temporal dimension is not the only possible source of bias for ranking algorithms. We discuss other possible sources of information bias and their relevance in our highly-connected world in paragraph \ref{sec:impact}.

\subsection{Which metric to use? Benchmarking the ranking algorithms}
\label{sec:benchmarking}
Let us briefly discuss three basic ways that can be used to validate network-based ranking techniques.

\paragraph{Prediction of future edges}
While prediction is vital in physics where wrong predictions on real-world behavior can rule out competing theories, 
it has not yet gained a central role in social and economic sciences \cite{hofman2017prediction}.
Even though many metrics in the last years have been evaluated according to their predictive power, 
as outlined in several paragraphs of section \ref{sec:applications}, wide differences among the adopted predictive evaluation procedures make it often impossible to assess the relative usefulness of the metrics. In agreement with \cite{hofman2017prediction}, we believe that an increased standardization of prediction evaluation would foster a better understanding of the predictive power of the different metrics. 
With the growing importance of data-driven prediction \cite{clauset2017data,hofman2017prediction, subrahmanian2017predicting}, 
benchmarking ranking algorithms through their predictive power may deepen our understanding about which metrics better reproduce the
actual nodes' behavior and eventually lead to useful applications, perhaps in combination with data-mining techniques \cite{zanin2016combining}.

\paragraph{Using external information to evaluate the metrics}
External information can be beneficial both to compare different rankings and to interpret which properties of the system they actually reflect.
For example, network centrality metrics can be evaluated according to their ability to \emph{identify expert-selected significant nodes}. 
Benchmarks of this kind include (but are not limited to) identification of expert-selected movies \cite{wasserman2015cross, ren2017time}, 
identification of awarded conference papers \cite{sidiropoulos2006generalized, dunaiski2016evaluating} or of editor-selected milestone papers \cite{mariani2016identification}, identification of researchers awarded with international prizes \cite{radicchi2009diffusion,fiala2012time}.

The rationale behind this kind of benchmarking is that if a metric well captures the importance of the nodes, it should be able to rank at the top the nodes whose outstanding importance is undeniable. 
Results on different datasets \cite{mariani2016identification,ren2017time} show that there is no unique metric that outperforms the others in all the studied problems.
An extensive thorough investigation of which network properties make different metrics fail or succeed in a given task is a needed step in future research. First steps have been done in \cite{ren2017time}, which introduces a time-aware null model to reveal the role of network structural correlations in determining the relative performance of local and eigenvector-based metrics in identifying significant nodes in citation networks.

Expert judgment is not the only source of external information that can be used to validate the metrics. For example, Smith et al. \cite{smith2014poverty} and Mao et al. \cite{mao2015quantifying} have applied network centrality metrics to mobile-phone communication networks and compared the region-level scores produced by the metrics with the income levels of the region. The effectiveness of centrality metrics can be also validated according to some prior knowledge of the nodes' function in the network. For example, the scores by network centrality metrics have been compared
with known neuronal activity in neural network \cite{fletcher2016structure}.

External information is beneficial not only to evaluate the metrics but also to move toward a better understanding of what the metrics actually represent.
For instance, the long-standing assumption that citation counts are proxies for scientific
impact has been recently investigated in a social experiment \cite{radicchi2016quantifying}.
Radicchi et al. \cite{radicchi2016quantifying} asked $2000$ researchers to perform pairwise comparisons between papers, and then aggregated the results over the researchers, quantifying thereby the papers' perceived impact independently of their citation counts.
The results of this experiment show that when authors are asked to judge their published papers, perceived impact and citation count correlate well, which indicates that the assumption that citation count represents paper impact is in agreement with experts' judgment. 

\paragraph{Using model data to evaluate the metrics}
One can also use models of network growth or network diffusion to evaluate the ability of the metrics to rank the nodes according to their importance in the model.
This strategy opens different alternative possibilities. 
One can use models of network growth \cite{medo2011temporal, medo2016model} where the nodes are endowed with some fitness parameter, grow synthetic networks of a given size and, a posteriori, evaluate the metrics' ability to uncover the nodes' latent fitness \cite{mariani2015ranking,medo2016model} (see also section \ref{sec:failure}).
Another possibility is to use real data and test the metrics' ability to unearth the intrinsic spreading ability of the nodes according to some model of diffusion on the network.
An extensive analysis has been carried out recently in the review article by Lu et al. \cite{lu2016vital} for static metrics and the classical SIR network diffusion model \cite{newman2010networks}. In particular, Lu et al. \cite{lu2016vital} use real data to compare different metrics with respect to their ability to reflect nodes' spreading power as determined by classical network diffusion models.
Analogous studies are still missing for temporal networks. 
For example, it would be interesting to compare different (static and time-aware) metrics with respect to their ability to identify influential spreaders in suitable temporal-network spreading models \cite{van2013non, rosvall2014memory, koher2016infections} similarly to what has been already done for static metrics.
The relative performance of the metrics in identifying of influential spreaders in temporal networks can provide us with important insights on which metrics can be consistently more useful than others in real-world applications.


\subsection{Systemic impact of ranking algorithms: avoiding information bias}
\label{sec:impact}
Besides metrics' performance, it is often of critical importance to detect the possible biases of the metrics and, wherever possible, suppress them.
For example, in the context of quantitative analysis of academic data, the bias of paper- and researcher-level quantitative metrics by age and scientific field is long known \cite{schubert1986relative, vinkler1986evaluation}, and many methods have been proposed in the scientometrics literature to counteract it (see \cite{radicchi2008universality,zhang2014comparison,vaccario2017quantifying}, among others).
In this review, we have mostly focused on the bias of ranking by node age (section \ref{sec:failure}) and possible methods to reduce or suppress it (section \ref{sec:time}).
On the other hand, other types of bias are present in large information systems. 
In particular, possible biases of ranking that come from social effects \cite{scholtes2014social} are largely unexplored in the literature. 

Several shortcomings of automated ranking techniques can stem from social effects. 
For example, there is the risk that online users are only exposed to content similar 
to what they already know and that confirms their points of view, creating
a ``filter bubble'' \cite{pariser2011filter} of information. This is particularly dangerous for our
society since online information is often not reliable -- in a recent report, the World 
Economic Forum has indicated massive digital misinformation as one of the main risks for our
interconnected society [see \url{http://reports.weforum.org/global-risks-2013/risk-case-1/digital-wildfires-in-a-hyperconnected-world/}].
This issue may be exacerbated by the fact that users more prone to consume unreliable 
content tend to cluster together, creating groups of highly-polarized clusters referred to as ``echo chambers''~\cite{del2016echo,del2016spreading}.
Designing statistically-grounded procedures to detect these biases and effective methods 
to suppress them, thereby preserving information diversity and improving information quality, are vital topics for future research.


The impact of centrality metrics and ranking on the behavior of agents in socio-economics systems deserves attention as well.
Just to mention a few examples, online users are heavily influenced by recommender systems when choosing which
items to purchase \cite{lu2012recommender,piramuthu2012input}, ranking can affect the decision of undecided
electors in political elections \cite{ruusuvirta2009online}, the impact factor of scientific journals
heavily affects scholars' research activity \cite{peoples2016twitter}.
Careful considerations of possible future effects are thus needed before adopting a metric for some purpose.

To properly assess the potential impact of ranking on systems' evolution, we would need an accurate model of the nodes' behavior.
Homogeneous models of network growth assume that all nodes are driven by the same mechanism when 
choosing which node to connect to. Examples of homogeneous models include models based on the preferential 
attachment mechanism \cite{fortunato2006scale,medo2011temporal,wang2013quantifying,medo2014statistical}, models where 
nodes tend to connect with the most central nodes and to delete their connections with the least central 
nodes \cite{konig2011network,konig2014nestedness}, among others. While such homogeneous models can be used
to generate benchmark networks to test the performance of network-based algorithms \cite{mariani2015ranking,medo2016model}, 
they neglect the multi-faceted nature of node behavior in real systems and they might provide us with an oversimplified 
description of real systems' evolution. 
Models that account for heterogeneous node behavior have already been 
proposed \cite{sendina2016assortativity,medo2016identification, tomasello2017data}, 
yet we are still at a preliminary stage in this research direction for network modeling and, for this reason, we still lack clear
guidelines on how best to predict the impact of a given centrality metric on the properties of a given network.

\subsection{Summary and conclusions}
Ranking in complex networks plays a crucial role in many real-worlds applications -- we have seen many examples of such applications throughout this review (see also the review by Lu et al. \cite{lu2016vital}).
These applications cover problems in a broad range of research areas, including economics \cite{hidalgo2009building, battiston2012debtrank, tacchella2012new}, neuroscience \cite{lohmann2010eigenvector, fletcher2016structure} and social sciences \cite{katz1953new,bonacich2001eigenvector}, among others.
From a methodological point of view, we have mostly focused on methods inspired by statistical physics, and we have studied three broad classes of node-level ranking methods:
static algorithms (section \ref{sec:static}), time-dependent algorithms based on a time-aggregated network representation of the data (section \ref{sec:time}), and algorithms based on a temporal-network representation of the data (section \ref{sec:temporal}). We have also discussed examples of edge-level ranking algorithms and their application to the problem of link prediction in online systems (paragraph \ref{sec:link_prediction}).

In their recent review on community detection algorithms \cite{fortunato2016community}, Fortunato and Hric write that ``as long as there will be networks, there will be people looking for communities in them''. We might as well say that ``as long as there will be networks, there will be people ranking their nodes''. The broad range of applications of network centrality metrics and the increasing availability of high-quality datasets suggest that research on ranking algorithms will not slow down in the forthcoming years. We expect scholars to become more and more sensitive to the problem of understanding and assessing the performance of the metrics for a given task.
While we hope that the discussion provided in the previous two paragraphs will be useful as a guideline, our discussion does not pretend to be complete and several other dimensions can be added to the evaluation of the methods.

We conclude by stressing the essential role of the \emph{temporal dimension} in the design and evaluation of ranking algorithms for evolving networks. 
The main message of our work is that disregarding the temporal dimension can lead to sub-optimal or even misleading results. The number of methods presented in this review demonstrates that there are many ways to effectively include time in the computation of a ranking.
We hope that this work will be useful for scholars from all research fields where networks and ranking are tools of primary importance.

\section*{Acknowledgements}
We wish to thank Alexander Vidmer for providing us with data and tools used for the analysis presented in paragraph 7.2, and for his early contribution to the text of that paragraph. \\
We would also like to thank all those researchers with whom we have had inspiring and motivating discussions about the topics presented in this 
review, in particular: Giulio Cimini, Matthieu Cristelli, Alberto Hernando de Castro, Flavio Iannelli, Francois Lafond,
Luciano Pietronero, Zhuo-Ming Ren, Andrea Tacchella, Claudio J. Tessone, Giacomo Vaccario, Zong-Wen Wei, An Zeng. \\ 
HL and MYZ acknowledge financial support from the National Natural Science Foundation of China (Grant No. 11547040), Guangdong Province Natural Science Foundation (Grant No. 2016A030310051), Shenzhen Fundamental Research Foundation (Grant No. JCYJ20150625101524056, JCYJ20160520162743717, \\JCYJ20150529164656096), Project SZU R/D Fund (Grant No. 2016047), CCF-Tencent (Grant No. AGR20160201), Natural Science Foundation of SZU (Grant No. 2016-24). MSM and MM and YCZ acknowledge financial support from the EU FET-Open Grant No. 611272 (project Growthcom) and the Swiss National Science Foundation Grant No. 200020-143272.

\bibliographystyle{elsarticle-num}

\end{document}